\documentclass[pra,pra,twocolumn,superscriptaddress,showpacs]{revtex4-1}
\usepackage[pdftex]{graphicx}
\usepackage{amsmath,amsfonts,amsbsy,amssymb,amsthm}
\usepackage{mathrsfs,bbm}
\usepackage{epstopdf}
\usepackage[colorlinks=true,
            urlcolor=blue,
            citecolor=red,
            pdfstartview=FitH,
            pdfpagemode=None]{hyperref}

\newcommand{\ket}[1]{\lvert #1 \rangle}
\newcommand{\bra}[1]{\langle #1 \rvert}

\newcommand{\expect}[1]{\big\langle #1 \big\rangle}
\newcommand{\expects}[1]{\langle #1 \rangle}

\newcommand{\Tr}[1]{\operatorname{Tr}\bigl[#1\bigr]}

\newcommand{\nn}{\nonumber}
\newcommand{\grad}{\nabla}

\newcommand{\erf}[1]{Eq. (\ref{#1})}

\newcommand{\mbf}[1]{\mathbf{#1}}
\newcommand{\rp}{\mathbf{r}_\perp}
\newcommand{\eff}{\text{eff}}
\newcommand{\peakprobe}{\mathcal{E}_0}
\newcommand{\varFz}{\big( \Delta F_z^{00} \big)^2}
\newcommand{\varFzText}{( \Delta F_z^{00} )^2}

\begin{document}
\title{ Three-dimensional light-matter interface for collective spin squeezing in atomic ensembles }
\author{Ben Q. Baragiola}\email{quinn.phys@gmail.com}
	\affiliation{Center for Quantum Information and Control, University of New Mexico, Albuquerque, NM 87131, USA} 
\author{Leigh M. Norris}
	\affiliation{Center for Quantum Information and Control, University of New Mexico, Albuquerque, NM 87131, USA}
\author{Enrique Monta\~{n}o}
	\affiliation{Center for Quantum Information and Control, University of Arizona, Tucson, AZ 85721, USA}
\author{Pascal G. Mickelson}
	\affiliation{Center for Quantum Information and Control, University of Arizona, Tucson, AZ 85721, USA}
\author{Poul S. Jessen}
	\affiliation{Center for Quantum Information and Control, University of Arizona, Tucson, AZ 85721, USA}
\author{Ivan H. Deutsch}
	\affiliation{Center for Quantum Information and Control, University of New Mexico, Albuquerque, NM 87131, USA}
\date{8 December 2013}

\begin{abstract}
	We study the three-dimensional nature of the quantum interface between an ensemble of cold, trapped atomic spins and a paraxial laser beam, coupled through a dispersive interaction. To achieve strong entanglement between the collective atomic spin and the photons, one must match the spatial mode of the collective radiation of the ensemble with the mode of the laser beam while minimizing the effects of decoherence due to optical pumping.  For ensembles coupling to a probe field that varies over the extent of the cloud, the set of atoms that indistinguishably radiates into a desired mode of the field defines an inhomogeneous spin wave.  Strong coupling of a spin wave to the probe mode is not characterized by a single parameter, the optical density, but by a collection of different effective atom numbers that characterize the coherence and decoherence of the system.  To model the dynamics of  the system, we develop a full stochastic master equation, including coherent collective scattering into paraxial modes, decoherence by local inhomogeneous diffuse scattering, and backaction due to continuous measurement of the light entangled with the spin waves.  This formalism is used to study the squeezing of a spin wave via continuous quantum nondemolition (QND) measurement.  We find that the greatest squeezing occurs in parameter regimes where spatial inhomogeneities are significant, far from the limit in which the interface is well approximated by a one-dimensional, homogeneous model.
		
\end{abstract}


\maketitle

\section{Introduction }	

Cold atomic ensembles interacting with electromagnetic fields are powerful tools in quantum information science with applications that include quantum memory \cite{FLukinMemory, Polzik2004, ChoiKimbleMemory}, quantum communication \cite{DLCZ, MatKuzNetwork},  continuous variable quantum computing \cite{BraunReview}, and metrology \cite{AppelPolzikMet, VulMet}.  At the heart of these protocols is the strong coupling between a quantum mode of the field and an effective collective spin of the ensemble. This coupling can generate entanglement between atoms and photons, such that measurement of the light yields strong quantum backaction on the atoms.  Photons can also enable a quantum data bus for entangling atoms with one another.  Enhancing the atom-light interface is thus essential for improving the performance of quantum technologies and for reaching new regimes where a quantum advantage becomes manifest.  This can be achieved through confined modes such as in optical cavities~\cite{Kimble2005, VulMet, Thompson2011} or waveguides in optical nanostructures~\cite{Rauschenbeutel2010, Waks2012, Kimble2013}. 

Strong atom-photon coupling can also occur in free space in the interaction between light and an extended ensemble of atoms.  This occurs when photons are scattered collectively by the ensemble, and interference enhances the radiation into the probe mode relative to diffuse scattering into $4\pi$ steradians~\cite{Vuletic2011, Kaiser2013}.   Early experiments demonstrated such strong coupling and entanglement in high pressure vapor cells where a one-dimensional description of plane wave modes and uniform atomic density is applicable \cite{Kuzmich1999}.  This theory accurately describes a variety of experiments including the entanglement of macroscopic ensembles in remote vapor cells \cite{Polzik2001} and quantum memory for continuous variables~\cite{Polzik2004}.  

	\begin{figure}
    		\includegraphics[width=1\hsize]{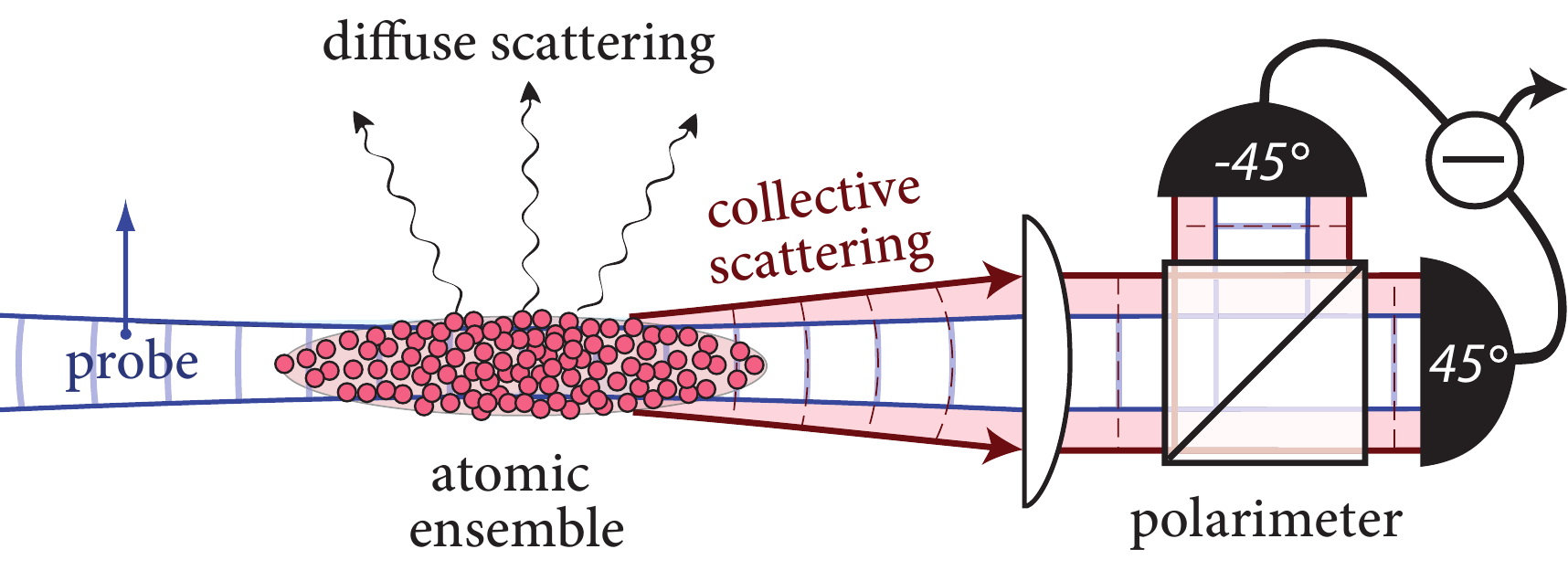}
       		 \caption{Schematic of a linearly polarized laser probe with a Gaussian spatial mode (blue) interacting dispersively with a cold, trapped atomic ensemble.  The light that is collectively scattered by the average density distribution of the ensemble defines the radiated mode (red). The interference of the radiated and probe modes leads to a rotation of the field polarization according to the Faraday effect.  When measured in a polarimeter, this can be used to generate spin squeezing in the ensemble.  The spatial overlap of the collectively scattered field and the probe, a measure of the strong atom-light coupling, depends highly on geometry.  In addition, diffusely scattered photons due to local density fluctuations in the ensemble lead to  decoherence in the collective spin variables. }  \label{Fig::Schematic}     	
	\end{figure}

More recently, experiments have employed ensembles of ultracold atoms in pencil-shaped dipole traps probed by focused laser beams \cite{KamMul12, KosMit10}. When the radiation pattern of the atomic ensemble is effectively matched with the paraxial mode of the probe, the atomic dipoles are indistinguishable and the scattering is cooperative. Such geometries have the potential to strongly enhance the atom-photon quantum interface, but their description is more complex, requiring a full treatment of diffraction, inhomogeneous coupling, radiation patterns, and decoherence.   Harnessing the advantages of these atomic ensembles thus requires a three-dimensional quantum theory of the underlying interaction, including both coherent coupling and quantum noise.

In the last decade there has been significant progress in developing a three-dimensional quantum description of the atom-light interface. A rigorous field-theoretic treatment separates the mean-field classical effects from the quantum fluctuations and noise, including the spatial inhomogeneities of atoms and light modes~\cite{SorSor08}.   Models that include spatial modes have been developed in a variety of contexts~\cite{Kuzmich2004, Windpassinger2008, Koschorreck2009, SauSta10}. Applications include remote entanglement via collective Raman scattering in a DLCZ-type protocol \cite{DuaZol02, SorSor09} and for multimode quantum memories \cite{ZueGro11}.  From such studies, it is clear that one-dimensional models not only fail to describe the relevant coherent and incoherent effects, but they also do not take advantage of the resources associated with spatial modes~\cite{GroSor12, HigBuc12}.  

In this paper we revisit the three-dimensional atom-light interface with particular emphasis on spin squeezing through QND measurement of the collective spin via the Faraday effect \cite{KuzBig00,KosMitSq, Takano2009}, shown schematically in Fig.~\ref{Fig::Schematic}.  In this protocol, the key interaction is the off-resonant scattering of horizontally polarized photons into vertical polarization.  Measurement in a balanced polarimeter corresponds to a homodyne measurement of these scattered photons.  The degree of scattering into the local oscillator, defined by the paraxial laser mode, determines the measurement strength and the resulting backaction that generates squeezing.  

Central to this problem are the spatial modes of the light and the collective spin waves of the atomic ensemble.  In the one-dimensional model, one collective parameter defines the strong-coupling regime of the atom-photon interface, the optical density on resonance, OD $= \eta \sigma_0 L = N \sigma_0 /A$, where $\eta$ is the atomic density, $\sigma_0$ is the resonant scattering cross section, $L$ is the length of the vapor cell, and $N$ is the number of atoms in the volume $V=AL$ for a uniform beam of area $A$.  In contrast, in a fully three-dimensional model, where the atomic density, $\eta(\mbf{r})$, and paraxial beam intensity distribution, $\beta(\mbf{r})$, are not uniform, there is a collection of parameters that dictate the strong-coupling regime.  Different effective atom numbers, $N_{\rm eff}^{(K)} =\int d^3\mbf{r} \, \eta (\mbf{r}) \left[\beta(\mbf{r})\right]^K$, govern different physical effects. For example, $N_{\rm eff}^{(1)}$ determines the mean Faraday signal in the polarimeter, while $N_{\rm eff}^{(2)}$ determines the size of the  measurement uncertainty from spin projection noise.  

The entangling strength of the atom-light interface is determined by the size of the spin projection uncertainty compared to the quantum uncertainty in the measured light quadratures (shot noise).  This collective interaction is proportional to an effective optical density, $OD_\eff = N^{(2)}_\eff \sigma_0/A$.  In contrast, decoherence acts locally on the atoms in a noncollective manner, and the noise injected into the system due to optical pumping and spin flips is governed by other parameters.   A proper accounting of the balance between the coherent coupling and decoherence is especially challenging given the tensor nature of the atom-photon interaction of real alkali atoms.  Previous treatments of quantum noise in a multimode Faraday-based atom-light interface have been carried out in a one-dimensional model~\cite{KupPol05, VasSor12}.  Our goal is to extend this to the three-dimensional case. 

In this work, we derive a stochastic master equation describing the dynamics of the collective atomic state conditioned on balanced polarimetry measurements, including the effects of measurement backaction, collective decoherence from unmeasured paraxial light, and local decoherence from diffuse photon scattering that gives rise to optical pumping. While we apply this to study conditional spin squeezing generated by a QND measurement, the formalism we develop is broadly applicable to other protocols where a strong, free-space atom-light interface is essential, and where measurement backaction may be a tool for induced atom-atom interactions.

The remainder of the article is organized as follows.  We lay out the physical model for an ensemble of alkali atoms dispersively interacting with a coherent probe laser in Sec. \ref{Sec::ParaxialScattering}.  We begin with a semiclassical model that can be used to describe the scattered paraxial fields and to identify the collective spin wave that is coupled to the laser mode.  To understand the entangling Faraday interaction in a multimode geometry, we then present a fully quantum mechanical model.  This serves as the cornerstone for a complete description of QND squeezing and allows us to account for the damaging effects of  decoherence.  When the output light is measured continuously, the quantum dynamics, including the combined effects of measurement backaction and decoherence, are described by a stochastic master equation.  We use this fully quantum mechanical atom-light description to study effects of spatial modes on the squeezing of spin waves in Sec. \ref{Sec::SpinSqueezing}.  In particular we use the multimode description to model the dynamics of spin squeezing and to analyze the dependence of peak squeezing on cloud and beam geometry.  We use numerical simulations to help build physical intuition about the three-dimensional atom-light interface and  to investigate how the model can be used to optimize an experimental design.  Finally, we summarize our results and present future directions for this work in Sec.~\ref{Sec::Conclusion}.

\section{Paraxial Atom-Light Interface}  \label{Sec::ParaxialScattering}

When driven by an off-resonant laser field such that the excited state probability is small, atoms elastically scatter electromagnetic waves in a manner equivalent to a set of linearly polarizable particles.  Thus, a great deal of qualitative and quantitative information can be obtained from classical radiation theory.  In a rigorous field-theoretic analysis, S$\o$rensen and  S$\o$rensen showed that the mean-field effect of the light interacting with an atomic ensemble gives rise to an index of refraction of the gas, while fluctuations are due to the random positions of the atoms and the vacuum noise of the light \cite{SorSor08}.  In particular, the index of refraction is due to the spatially-averaged local density of the atoms, while the diffuse scattering into 4$\pi$ arises from the random positions of the point atomic scatterers and is equivalent to decoherence by local spontaneous emission.  This diffuse scattering, which leads to attenuation of the incident wave and optical pumping of the atomic state, is accounted for by an imaginary part of the polarizability according to the optical theorem.  

We can thus break up the problem into two pieces.  First, the mean field effect is described by classical scattering of a laser beam incident on a linearly polarizable dielectric whose shape is determined by the atomic density distribution.  For a paraxial probe beam and an extended cloud, the scattered field is also paraxial, and the solution is easily found by Fraunhofer scattering theory~\cite{Newton1982}.  As we are interested in the Faraday effect, we include the tensor nature of the atomic polarizability.  Scattering of an incident horizontal polarization to an orthogonal vertical polarization is the key effect that we seek to measure in the polarimeter.  Second, to properly account for quantum backaction on the atoms resulting from measurement and to describe the decoherence due to diffuse scattering and optical pumping, we turn to the fully quantum theory.

\subsection{Semiclassical theory} \label{Sec::SemiclassicalTheory}

	Consider the  scattering of an incident paraxial laser beam with frequency $\omega_0$ and complex amplitude, $\mbf{E}_L(\mbf{r}_\perp, z) = \vec{\mathcal{E}}_L (\rp , z) e^{ik_0 z}$, by a particle located at a position $\mbf{r}'$ with dynamical tensor electric-dipole polarizability $\tensor{\alpha}.$  The field envelope has  the standard form $\vec{\mathcal{E}}_L(\mbf{r}_\perp, z) = \vec{\epsilon}\, \peakprobe u_{00}(\rp,z) $, where $\vec{\epsilon}$ is the laser polarization and $u_{00}$ is chosen to be the Gaussian TEM$_{00}$ mode given by 
\begin{equation}
u_{00}(\rp,z)= \frac{w_0}{w(z)}e^{-\frac{\left|\mbf{\rp}\right|^2}{[w(z)]^2}}e^{\frac{ik_0 \left|\mbf{\rp}\right|^2}{2 R(z)}}e^{-i\Phi(z)}.
\end{equation}
The $z$-dependent beam waist, the radius of curvature of the phase fronts, and the Guoy phase are given by
\begin{subequations} \label{Eq::GaussianParameters}
\begin{align}
w(z) &= w_0 \sqrt{1+(z/z_R)^2} ,  \\
R(z) &= z\left(1+(z_R/z)^2\right) ,  \\
\Phi(z) &= \tan^{-1}(z/z_R), 
\end{align}
\end{subequations}
respectively, with beam waist $w_0$ and Rayleigh range $z_R \equiv k_0 w_0^2/2$.  In the first Born approximation, the scattered field amplitude is that radiated by the induced dipole,
	\begin{align} \label{Eq::DipoleScattering}
		\mbf{E}_{\rm scat}(\mbf{r})&= k_0^2 \left[\tensor{\alpha}\cdot \mbf{E}_L(\mbf{r}')\right]_\perp \frac{e^{ik_0 \left|\mbf{r}-\mbf{r}'\right|}}{\left|\mbf{r}-\mbf{r}'\right|} \nn \\
		& \approx k_0^2 \left[ \tensor{\alpha}\cdot \mbf{E}_L(\mbf{r}')\right]_\perp  e^{ik_0(z-z')}\frac{e^{\frac{i k_0 \left|\rp-\rp'\right|^2}{2(z-z')}}}{z-z'} ,
	\end{align}
where the subscript $\perp$ denotes the component of the dipole transverse to the direction of observation.  The last approximation is valid for paraxial points of observation, $z \gg \left|\rp\right|$.  Gaussian-cgs units for the electromagnetic field equations are used throughout. 

Because the dipole radiation is not mode-matched with the Gaussian laser beam, the light is scattered into all paraxial modes as well as off-axis nonparaxial modes.  In the far field, $z \gg z'$, the total field takes the form,
	\begin{align} 
		\mbf{E}_{\rm out}(\mbf{r}) & =\mbf{E}_L(\mbf{r}) + \mbf{E}_{\rm scat}(\mbf{r}) \nn  \\
		& = \big(  \vec{\epsilon} + \vec{\Upsilon} \big) \peakprobe e^{ik_0 z}u_{00}(\rp,z)+ \mbf{E}'_{\rm scat}(\mbf{r}),
	\end{align}
where $\mbf{E}'_{\rm scat}(\mbf{r})$ is the scattered field into all spatial modes other than the probe mode, and as shown in Appendix A, \erf{Eq::ClassicalScattering},
	\begin{align} 
 	\vec{\Upsilon} \peakprobe  e^{ik_0 z} &\equiv \int \frac{d^2\rp}{A} u_{00}^*(\rp,z) \mbf{E}_{\rm scat}(\mbf{r}) \nn \\
		& = i \frac{2\pi k_0}{A}\left(\tensor{\alpha}\cdot \vec{\epsilon}\right)_\perp   \left| u_{00}(\rp',z') \right|^2  \peakprobe  e^{ik_0 (z-z')}
 	\end{align}
 is the field amplitude ``forward scattered'' into the laser mode. $A = \int d^2\rp |u_{00}(\rp,z)|^2 = \pi w_0^2/2$ is the effective beam area. 
  
The key physical effects are seen in these equations.  The component of the radiated field vector $\vec{\Upsilon}$ along the laser polarization $\vec{\epsilon}$ gives rise to the scalar index of refraction and attenuation.  The component of $\vec{\Upsilon}$ orthogonal to $\vec{\epsilon}$ gives rise to a rotation of the polarization on the Poincar\'{e} sphere -- Faraday rotation and birefringence.  For example, suppose the laser is linearly polarized along $x$ ($\vec{\epsilon} =\mbf{e}_x)$.  The total field thus can be written,
 \begin{align} \label{Eq::OutputField}
 	& \mbf{E}_{\rm out} (\mbf{r}) = \mbf{E}'_{\rm scat}(\mbf{r}) +  \\
	&\quad  \left[ \left(1+i\delta \phi-\frac{a}{2}\right) \mbf{e}_x + \left(\frac{\chi + i\beta}{2}\right) \mbf{e}_y \right]  \peakprobe e^{ik_0 z}u_{00}(\rp,z), \nn
\end{align}
 where,
 \begin{align}
 \delta \phi &=(2 \pi k_0/A) \left| u_{00}(\rp',z') \right|^2 \text{Re} \left\{ \alpha_{xx} \right\} \nonumber \\
 a &= (4 \pi k_0/A) \left| u_{00}(\rp',z') \right|^2 \text{Im} \left\{ \alpha_{xx} \right\} \nonumber \\
 \chi &=-(4 \pi k_0/A) \left| u_{00}(\rp',z') \right|^2   \text{Im} \left\{ \alpha_{yx} \right\} \nonumber \\
 \beta &=(4 \pi k_0/A) \left| u_{00}(\rp',z') \right|^2   \text{Re} \left\{ \alpha_{yx} \right\} \nonumber
 \end{align}
 are respectively: $\phi$ is the index of refraction phase shift, $a$ is the Beer's law attenuation coefficient, $\chi$ is the rotation angle of the Stokes vector corresponding to the Faraday effect, and $\beta$ is the corresponding angle for birefringence, with the polarizability matrix elements denoted as $\alpha_{ij} = \mathbf{e}_i\cdot \tensor{\alpha} \cdot \mathbf{e}_j$ in the $x$-$y$ basis.
 
The above description of the atom-field coupling is most easily generalized using the theory of scattering of paraxial waves \cite{MulPol05}; details can be found in Appendix \ref{Appendix::ParaxialScattering}.  The mean field is described by the electric field envelope $\vec{\mathcal{E}}_L (\rp , z,t)= \mathcal{A}(t-z/c) \vec{\mathcal{U}} (\rp, z)$, where $\mathcal{A}(t)$ is the temporal pulse envelope evaluated at the retarded time, and  $\vec{\mathcal{U}} (\rp, z)$ is the spatial envelope satisfying the paraxial wave equation, 
\begin{equation}
\frac{\partial}{\partial z}\vec{\mathcal{U}}(\rp , z) \! = \!\frac{i}{2 k_0}\grad_\perp^2 \vec{\mathcal{U}}(\rp , z) +i2 \pi k_0 \tensor{\chi}(\rp, z) \cdot \, \vec{\mathcal{U}}(\rp , z),
\end{equation}
with spatially averaged dielectric susceptibility $\tensor{\chi}(\rp, z)$.  The scattering solution to this equation is well known \cite{Newton1982}.   In the first Born approximation, i.e. for dilute samples where multiple scattering is negligible, given an incident field $\vec{\mathcal{U}}_{\rm in} (\rp,z)$, the total field is
	\begin{align}
		&\vec{\mathcal{U}}(\rp , z) =  \vec{\mathcal{U}}_{\rm in} (\rp,z) \label{paraxsol} \\
		&\! + i 2 \pi k_0 \! \!  \int_{-\infty}^z \! \!  \! \!  \! \!  dz' \! \!  \int \! \!  d^2 \rp' K(\rp \! \! -\! \rp', z\!-\!z')  \tensor{\chi}( \rp', z') \! \cdot \vec{\mathcal{U}}_{\rm in} (\rp',z') , \nn
	\end{align}
where $K(\rp-\rp', z-z')$ is the paraxial propagator.  This solution is the superposition of incident and reradiated dipole fields.  The solution for a paraxial field scattered from a point dipole at position $\mbf{r}'$, \erf{Eq::DipoleScattering}, is recovered by setting $\tensor{\chi}(\mbf{r}) = \tensor{\alpha} \, \delta^{(3)}(\mbf{r}-\mbf{r}') $.  

The diagonal matrix elements of the susceptibility give rise to the index of refraction and a slight distortion of the wavefront of the beam.  We can neglect this effect for dilute gases, though it is easily accounted for.  The Faraday effect arises from the scattering of initially $x$-polarized light into orthogonal $y$-polarization as discussed above, governed by the off-diagonal element of the dielectric susceptibility matrix, $\chi_{yx}$.  To measure Faraday rotation, one employs a balanced polarimeter at $\pm 45^{\circ}$,  so that the signal $\mathcal{M}$ is proportional to $\mathcal{U}_x^* \mathcal{U}_y +\mathcal{U}_y^* \mathcal{U}_x$, integrated across the detector surface at position $z_D$ in the far field,
	\begin{equation} \label{Eq::ClassicalMeasurementSignal}
		\mathcal{M} \propto \int d^2 \rp  \text{Re}\big\{ \mathcal{U}_x^*(\rp, z_D) \mathcal{U}_y(\rp, z_D) \big\}.
	\end{equation}
The result is an effective homodyne detector for $\mathcal{U}_y$, with $\mathcal{U}_x$ playing the role of the local oscillator.  Using the solution for  $ \mathcal{U}_y(\rp, z_D)$, \erf{paraxsol}, and the properties of the propagator, Eq. (\ref{backprop}),
\begin{widetext}
	\begin{eqnarray} \label{Eq::ClassicalMeasurementSignal2}
		\mathcal{M} & \propto & \int_{-\infty}^{z_D}  dz' \int d^2 \rp'    \left[ \int  d^2 \rp  \, \mathcal{U}_x^* (\rp, z_D) K(\rp-\rp', z_D-z') \right] \text{Re}\big\{ i \chi_{yx}(\rp', z')\big\}  \mathcal{U}_x (\rp',z') \nonumber \\
		&=& -\int d^3 \mbf{r}' \, \text{Im}\big\{ \chi_{yx}(\rp', z')\big\} |\mathcal{U}_x (\rp',z')|^2.
	\end{eqnarray}
	\end{widetext}
The measured signal is thus proportional to the local value of the susceptibility component $\text{Im}\left\{ \chi_{yx}(\rp, z)\right\}$ integrated over the dielectric, weighted by the local field intensity $|\mathcal{U}_x (\rp,z)|^2$.

For an ensemble of dilute cold atoms at fixed positions $\mbf{r}_i$, the dielectric susceptibility of the gas is
	\begin{equation}
		\tensor{\chi}(\mbf{r}) = \sum_i \expects{ \hat{\tensor{\alpha}}\phantom{}^{(i)} } \, \delta^{(3)}(\mbf{r}-\mbf{r}_i),
	\end{equation}
where $\hat{\tensor{\alpha}}\phantom{}^{(i)}$ is the  the dynamic polarizability tensor operator  for the $i^{th}$ atom. We consider here atoms restricted to a subspace defined by a total (hyperfine) angular momentum $f$. In terms of the hyperfine spin operator $\hat{\mbf{f}}$, the polarizability operator can be decomposed into irreducible tensor components \cite{DeuJes09},
	\begin{equation}\label{Eq::IrreducibleDecomp}
		\hat{\alpha}_{ij} \! = \! \alpha_0 \Big[  C^{(0)} \! +i C^{(1)}  \epsilon_{ijk} \hat{f}_k + C^{(2)} \! \Big( \frac{ \hat{f}_i \hat{f}_j +\hat{f}_j \hat{f}_i}{2}- \delta_{ij}\frac {\hat{\mbf{f}}^2}{3} \Big) \Big]
	\end{equation}
where $\alpha_0$ is the characteristic polarizability and $C^{(K)}$ is the coefficient of the irreducible rank-$K$ tensor component. The rank-0 component is a scalar, which does not influence spin and polarization dynamics. The vector (rank-1) component is responsible for the Faraday effect, while the tensor (rank-2) component induces birefringence.  For alkali atoms driven on a fine-structure multiplet, $\alpha_0$ and the $C^{(K)}$ coefficients are given in \cite{DeuJes09}.

The effect of the tensor component complicates both the collective coupling of the atoms to the probe as well as the internal spin dynamics.  In special cases, the deleterious effects of the rank-2 component of the tensor polarizability can be removed via dynamical decoupling~\cite{Koschorreck2010}.  More generally, a large bias field removes the rank-2 component of the interaction that couples the collective spin to the polarization of the probe~\cite{NorDeu12}, leaving only internal spin dynamics that can be compensated.
We thus retain only the vector component of the off-diagonal element of the dielectric susceptibility, $\chi_{yx}$, which describes a pure Faraday interaction.  Substituting $\text{Im} \{\chi_{yx}(\mbf{r}_\perp,z)\} \propto \sum_i  \expects{ \hat{f}^{(i)}_z}\delta^{(3)}(\mbf{r}-\mbf{r}_i) $ into \erf{Eq::ClassicalMeasurementSignal2} yields
	\begin{equation} \label{Eq::SemiclassicalSpinWave}
		\mathcal{M} \propto \sum_i  |\mathcal{U}_x (\mbf{r}_{\perp i}, z_i)|^2 \expect{ \hat{f}^{(i)}_z }.
	\end{equation}
Equation (\ref{Eq::SemiclassicalSpinWave}) is the central result of the semiclassical model.  In a plane wave, homogeneous, one-dimensional description, the measured observable is $\mathcal{M} \propto \sum_i  \expects{ \hat{f}^{(i)}_z }  =  \expects{ \hat{F}_z } $, the symmetric collective spin of the ensemble.  For paraxial beams, the polarimeter measures an effective {\em spin wave} determined by the inhomogeneous weighting of the atomic spin operators by the local intensity of the beam.  The spin wave is stationary because it is coupled to the {\em forward-scattered light}, where the absorbed and emitted modes are the same. Physically, it is this collective observable that radiates indistinguishably into the probe mode and is effectively selected by the homodyne measurement of the polarimeter.

	\begin{figure}
    		\includegraphics[width=1\hsize]{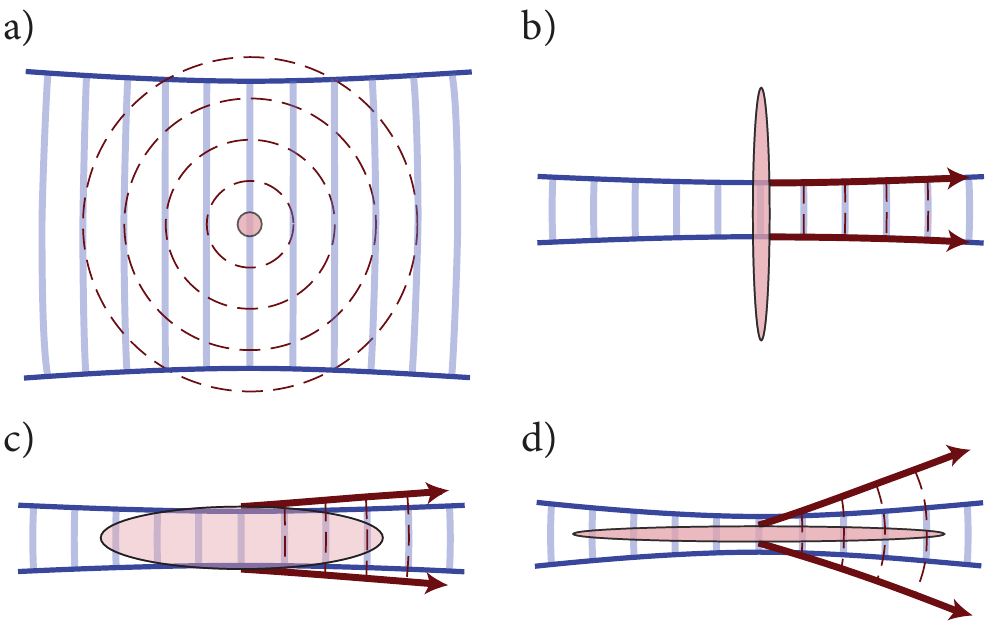}
       		 \caption{ Scattered modes for various atomic cloud and beam geometries.  The probe laser mode is indicated in blue and the mode scattered by a given dielectric distribution is shown in red. a) An  atomic ensemble localized with respect to the probe beam scatters light like a point dipole.  b) A ``pancake"-shaped cloud at a fixed $z$-plane radiates nearly perfectly into the probe mode.  Extended ``pencil"-shaped clouds can radiate into the probe mode well as in c) or poorly as in d) depending on the geometry of the probe. }  \label{Fig::ModeMatching}     	
	\end{figure}

Further intuition can be gained from the semiclassical model.  We recover symmetric atom-light coupling when the field intensity $ |\mathcal{U}_x (\mbf{r}_{\perp}, z)|^2$ is constant over the atomic ensemble.  Geometrically, this is achieved when the beam waist, $w_0$, is much larger than the transverse extent of the cloud and the length of the cloud is short compared to twice the Rayleigh range, $z_R = k_0 w_0^2/ 2$.    The mean-field radiation pattern of such a cloud described by \erf{paraxsol}, however, has poor overlap with the probe as depicted in Fig. \ref{Fig::ModeMatching}(a).  The end result is that the polarimeter detects only a small fraction of the signal photons.  On the other hand, perfect mode matching is achieved for atoms confined as a uniform dielectric sheet at a fixed $z$-plane as seen in Fig \ref{Fig::ModeMatching}(b).  However, for a finite number of atoms, the realizable OD is low in this configuration.  Indeed, a uniform dielectric slab of extent much larger than the beam waist achieves perfect mode matching, but one cannot achieve such an dielectric distribution with high OD using cold atomic gases.   An intermediate ``pencil"-shaped geometry is more realistic, allowing for reasonable mode matching while maintaining a high OD, as in Figs. \ref{Fig::ModeMatching}(c-d). 

In addition to maximizing the signal, we must minimize the sources of noise. There are two fundamental effects: (i) the polarimeter has a finite shot noise resolution; (ii) atoms  scatter photons diffusely into all directions (spontaneous emission).  The latter is accompanied by optical pumping that can both depolarize the spins and inject noise into the measured spin wave.   In order to address these effects, we must turn to the fully quantum theory.

\subsection{Quantum theory} \label{Sec::QuantumTheory}

Following Refs. \cite{Ham06, VasSor12}, we partition the quantized electric field into paraxial modes and nonparaxial, diffuse modes, 
	\begin{align} \label{Eq::ModeDecomp}
		\hat{\mathbf{E}}^{(+)}(\mathbf{r}, t) = \hat{\mathbf{E}}_{\rm para}^{(+)}(\mathbf{r}, t) + \hat{\mathbf{E}}_{\rm diff}^{(+)}(\mathbf{r}, t).
	\end{align}	
This decomposition is motivated by the geometry we consider -- photon scattering of a paraxial laser beam by an extended atomic ensemble.  The mean-field, spatially averaged atomic density, which plays the role of the index of refraction in the classical theory, appears as coherent radiation by a collective atomic observable in the quantum theory.  The coupling of collective atomic observables to paraxial modes thus describes the coherent atom-photon light-shift interaction, mediated by the Hermitian part of the atomic polarizability operator.  

The diffuse modes, in contrast, couple to the density fluctuations in the ensemble due to the discrete atomic positions and thus act locally on each atom \cite{SorSor08}.  In the usual Born-Markov approximation, tracing over these modes leads to decoherence and is described by the anti-Hermitian part of the atomic polarizability \cite{DeuJes09}.    

	In this section we first derive a multimode generalization of the Faraday interaction that coherently entangles the atomic ensemble and the paraxial quantum field.  Then, we employ a master equation to account for the effects of local decoherence (optical pumping) driven by diffuse scattering.  Finally, we present the stochastic master equation describing the conditional collective atomic state given polarimetry measurements of the paraxial field, which we use to analyze spin squeezing in Sec. \ref{Sec::SpinSqueezing}.


	\subsubsection{Paraxial multimode Faraday interaction}

	Quantization of paraxial electromagnetic fields was discussed in \cite{DeuGar91};  relevant extensions to the current problem are summarized in  Appendix \ref{Appendix::ParaxialQuantization}.  We decompose the paraxial field operator into an orthogonal set of tranverse spatial modes, here the Laguerre-Gauss modes $ u_{pl}(\rp,z)$, given in \erf{Eq::LGModes}, which are convenient for cylindrical symmetry.  The positive-frequency component of the electric field restricted to the paraxial subspace is,
	\begin{align}
		\hat{\mbf{E}}&^{(+)}_{\rm para}(\mbf{r},t) = \nn \\
		 & \sum_{p,l,\alpha=x,y} \sqrt{\frac{2 \pi \hbar \omega_0}{c A}}\, \mbf{e}_\alpha \, \hat{a}_{pl,\alpha}(z,t) \,  u_{pl}(\rp,z)  e^{i(k_0 z - \omega_0 t)},
	\end{align}
where the quantization area is chosen as the natural scale of the Gaussian beam, $A =\pi w_0^2/2$.  The traveling wave creation/annihilation operators for each transverse mode freely propagate according the Hamiltonian
\begin{equation}
	\hat{H}_{\text{free}} =   \sum_{p,l, \alpha} \int dz \; \hat{a}^\dag_{pl, \alpha}(z,t) \left( -i  \hbar c \frac{\partial}{\partial z} \right) \hat{a}_{pl, \alpha}(z,t),
\end{equation}
with solution, $\hat{a}_{pl, \alpha}(z,t) = \hat{a}_{pl, \alpha}(z-ct,0)=\hat{a}_{pl, \alpha}(0,t-z/c),$ and free-field commutation relations,
	\begin{equation}
		\left[ \hat{a}_{pl,\alpha}(z,t),\hat{a}^{\dag}_{p'l',\beta}(z',t')\right] =\delta_{p,p'} \delta_{l,l'} \delta_{\alpha, \beta} \, \delta(t-t'-(z-z')/c).
	\end{equation}
We have normalized so that $\hat{a}^{\dag}_{pl,\alpha}(z,t)\hat{a}_{pl,\alpha}(z,t)$ is the local photon flux in transverse mode $pl$ with polarization $\alpha$.

For weak excitation (linear atomic response), the interaction Hamiltonian governing the coupling of the quantized paraxial modes is
	\begin{equation} \label{Eq::ScatteringInteraction}
		\hat{H}_{\text{int}} = - \sum_i \hat{\mathbf{E}}^{(-)}_{\rm para}(\mbf{r}_i, t) \cdot \hat{\tensor{\alpha}}\phantom{}^{(i)} \cdot \hat{\mathbf{E}}^{(+)}_{\rm para}(\mbf{r}_i, t).
	\end{equation}
As before, the index $i$ is summed over atoms in the ensemble at respective positions $\mathbf{r}_i$.  Upon substituting the decomposition of $\hat{\alpha}_{ij}$ into its irreducible components given in Eq. (\ref{Eq::IrreducibleDecomp}), we find scalar (rank-0), vector (rank-1), and tensor (rank-2) contributions to the interaction.  We retain only the vector contribution that leads to the Faraday effect, as the scalar contribution does not entangle photons with the atoms and the tensor contribution can in principle be removed \cite{KosMit10}.  The Faraday interaction is then, 
\begin{widetext}
	\begin{align} \label{Eq::MultimodeFaraday}
		 \hat{H}_{\text{int}} &= - \alpha_0 C^{(1)} \sum_i \hat{\mathbf{E}}^{(+)}_{\rm para}(\mbf{r}_i, t) \times\hat{\mathbf{E}}^{(-)}_{\rm para}(\mbf{r}_i, t) \cdot \hat{ \mbf{f}}^{(i)} \nonumber \\
	 &= \frac{\hbar \chi}{2} \sum_{ i,p,l, p',l' } \Big[ i u^*_{p'l'}(\mbf{r}_{\perp i}, z_i) u_{pl}(\mbf{r}_{\perp i}, z_i)  \, \hat{a}^\dag_{p'l',y}(z_i,t) \hat{a}_{pl,x } (z_i,t) 
	 + \text{H.c.} \Big] \hat{f}_z^{(i)}, 
	\end{align}
\end{widetext}
where
	\begin{equation}
 		\chi =  -C^{(1)} \frac{4 \pi \omega_0}{A c}  \alpha_0 =  -C^{(1)} \left(\frac{\sigma_0}{A}\right)  \left(\frac{\Gamma}{2 \Delta}\right) 
 	\end{equation}
is the Faraday rotation angle, $\sigma_0 = 3 \lambda^2/(2\pi)$ is the resonant cross-section for unit oscillator strength, $\Gamma$ is the atomic linewidth, and $\Delta$ is the detuning from resonance.  For an $S_{1/2}\rightarrow P_J$ transition with $\Delta$ much larger than the excited state hyperfine splitting, $|C^{(1)}| = g_f/3$, where $g_f$ is the Land\'{e} g-factor. We can interpret \erf{Eq::MultimodeFaraday} as a scattering process, whereby an $x$-polarized photon in a given transverse mode, $pl$, is absorbed and  a $y$-polarized photon in the mode $p'l'$ is emitted, and vice versa, as mediated by the collective atomic spin wave.

Here, we consider an initial macroscopic occupation in the laser probe, again taken to be the fundamental Gaussian TEM$_{00}$ mode with $x$-polarization.  In that case the interaction can be linearized by substituting $\hat{a}_{pl,x}(z,t) \rightarrow \sqrt{\dot{N}_L}\delta_{p,0}\delta_{l,0}$, where $\dot{N}_L  = A I_0/(\hbar \omega_0)$ is the photon flux of the laser with peak intensity $I_0$.  The quantum fluctuations in the field of interest are then represented by the $y$-polarized mode,  $\hat{a}_{pl,y}(z,t)$, and the Faraday interaction then takes the form
	\begin{align}\label{Eq::TheFaradayInteraction}
 		\hat{H}_{\text{int}} \!= \! -i  \frac{\hbar \sqrt{\kappa}}{2} \! \sum_{i,p,l} \! \Big[ & \beta^*_{pl}(\mbf{r}_{\perp i}, z_i ) \hat{a}_{pl,y}(z_i,t) \!- \! \mbox{h.c.} \Big] \! \hat{f}_z^{(i)} 
	\end{align}
where the local amplitude for scattering from the fundamental (laser) mode $00$ into mode $pl$ is given by
	\begin{align} \label{Eq::Beta}
		\beta_{pl}(\mbf{r})  \equiv u^*_{pl}(\mbf{r}) u_{00}(\mbf{r}). 
	\end{align}
The interaction has been written in terms of the ``measurement strength" per atom,
	\begin{align} \label{Eq::MeasurementStrengthperAtom}
		\kappa = \chi^2 \dot{N}_L =  \frac{1}{9f^2} \left( \frac{\sigma_0}{A} \right)\gamma_0,
	\end{align}	
which characterizes the rate at which photons are scattered into the paraxial modes, where $\gamma_0 = (\sigma_0 I_0 /\hbar \omega_0) (\Gamma^2/4\Delta^2)$ is the unit-oscillator-strength photon scattering rate at the peak intensity.  

The Heisenberg equation of motion for a $y$-polarized traveling wave mode interacting with the atomic media in the presence of the probe field is
\begin{equation}
\left(\frac{\partial}{\partial t} +c \frac{\partial}{\partial z}\right) \hat{a}_{pl,y}(z,t) = \frac{\sqrt{\kappa}}{2} \sum_i \beta_{pl} (\mbf{r}_{\perp i} z_i) \hat{f}_z^{(i)}  (t) \delta(z-z_i),
\end{equation}
whose solution is
\begin{align}
\hat{a}_{pl,y}(z&,t) =\hat{a}_{pl,y}(0,t-z/c)&\\
&+\frac{\sqrt{\kappa}}{2} \sum_i \beta_{pl} (\mbf{r}_{\perp i} z_i) \hat{f}_z^{(i)} (t-(z-z_i)/c) \Theta(z-z_i),\nonumber
\end{align}
where $ \Theta(z)$ is the Heaviside step function.  Neglecting the time it takes light to propagate across the sample, the mode amplitude at the detector plane, $z_D$, in the far field is
\begin{equation}
\hat{a}_{pl,y}(z_D,t) =\hat{a}_{pl,y}(0,t-z_D/c)+\frac{\sqrt{\kappa}}{2}\hat{F}_z^{pl}, 
\end{equation}
a form familiar from input-output theory \cite{GardZoller}.  The collective atomic spin wave that couples to this paraxial mode is\begin{equation}
\hat{F}_z^{pl}=  \sum_i \beta_{pl} (\mbf{r}_{\perp i}, z_i) \hat{f}_z^{(i)}.
\end{equation}

In the balanced polarimeter, the probe mode acts as a local oscillator so that one measures the Stokes vector component associated with the fundamental spatial mode defined by the laser beam, $(\hat{a}^\dag_{00,x}\hat{a}_{00,y}+\hat{a}^\dag_{00,y}\hat{a}_{00,x})/2 \approx \sqrt{\dot{N}/2}\hat{X}_{00}$, where $\hat{X}_{00}= (\hat{a}_{00,y}+\hat{a}^\dag_{00,y})/\sqrt{2}$ is the mode quadrature.  The measured quadrature at the detector plane, $z_D$, is thus
	\begin{equation}  \label{Eq::XplOut}
		 \hat{X}_{00}(z_D,t) =  \hat{X}_{00}(0,t-z_D/c) +\sqrt{\frac{\kappa}{2}} \hat{F}^{00}_z .
	\end{equation}
Thus, the total polarimeter signal, integrated over a time $T$, is determined by the output operator
	\begin{equation}\label{Eq::Measurement}
		\hat{\mathcal{M}} = \int_0^T  dt\, \hat{X}_{00}(0,t-z_D/c) + T \sqrt{\frac{\kappa}{2}}  \hat{F}^{00}_z
	\end{equation}
where $\hat{F}^{00}_z$ is the fundamental spin wave found in the semiclassical calculation, \erf{Eq::SemiclassicalSpinWave}.  The fully quantum theory explicitly includes the additional vacuum noise entering the polarimeter, $\langle \Delta \hat{X}_{00}(0,t)\Delta \hat{X}_{00}(0,t') \rangle=\delta(t-t')/2$, that leads to a shot-noise (SN) variance of the polarimeter signal, in Eq. (\ref{Eq::Measurement}), $\Delta \mathcal{M}_{\rm SN}^2 = T/2$.   

Of particular interest here is the application to spin squeezing via QND measurement.  In this case, the signal we seek to measure arises from different spin-projections associated with the eigenstates of $\hat{F}_z^{00}$.  Whereas in magnetometry these shot-to-shot variations are known as ``projection noise"  (PN), in the context of creating a spin squeezed state, these variations from the mean value represent the ``signal" one seeks to resolve over the laser shot noise. For the fundamental spin wave measured in the polarimeter, the projection noise variance is
	\begin{align}\label{Eq::variance}
		\left( \Delta F^{00}_z \right)^2_{\rm PN} =  & \sum_{i}  \beta^2_{00}(\mbf{r}_i) \expect{(\Delta \hat{f}_z^{(i)})^2} \\
		& + \sum_{i\neq j}  \beta_{00}(\mbf{r}_i) \beta_{00}(\mbf{r}_j )\expect{\Delta \hat{f}_z^{(i)}\Delta \hat{f}_z^{(j)}}. \nn
	\end{align}
Given an initial spin coherent state polarized orthogonal to $z$,  $\expects{\Delta \hat{f}_z^{(i)} \Delta \hat{f}_z^{(j)}}=(f/2) \delta_{ij}$, and thus,
\begin{equation}\label{Eq::PNvariance}
\left( \Delta F^{00}_z \right)^2_{\rm PN}= \frac{ N^{(2)}_\eff f}{2}.
\end{equation}
Here we define a set of effective atom numbers
	\begin{align} \label{Eq::EffectiveAtomNumbers}
		N^{(K)}_\eff & = \sum_{i} \big[ \beta_{00}(\mbf{r}_i) \big]^K = \sum_{i} |u_{00}(\mbf{r}_i)|^{2K} \\ 
		&\rightarrow \int d^3\mbf{r} \, \eta(\mbf{r}) |u_{00}(\mbf{r}_i)|^{2K},	
	\end{align}
where the sum becomes an integral in the continuum limit.  The atomic density distribution $\eta(\mbf{r})$, is normalized so that $\int d^3\mbf{r} \, \eta(\mbf{r}) = N$, the total atom number.  The effective atom number $N^{(2)}_\eff$ determines  the projection noise contribution to Eq. (\ref{Eq::Measurement}), $\Delta\mathcal{M}_{\rm PN}^2 = \kappa T^2 ( \Delta F_z^{00} )^2 _{\rm PN}/2 =\kappa T^2   N^{(2)}_\eff f/4$.

The coupling strength $\xi$ that sets the degree of entanglement one can attain between the atoms and photons is the ratio of the projection noise variance to the shot noise resolution \cite{DeuJes09}.  Using Eqs. (\ref{Eq::MeasurementStrengthperAtom}) and (\ref{Eq::PNvariance}) we find
	\begin{equation}\label{Eq::CouplingStrength}
		\xi = \frac{\Delta \mathcal{M}_{\rm PN}^2}{\Delta \mathcal{M}_{\rm SN}^2}= \left( \Delta F_z^{00}\right)^2  \kappa T  =  \mbox{OD}_\eff \, \frac{\gamma_s T}{18f},
	\end{equation}	
where we have defined the effective optical density for the laser mode probing the spin wave on a unit-oscillator-strength transition, 
	\begin{align} \label{Eq::ODeff}
		\mbox{OD}_\eff = N^{(2)}_\eff \frac{\sigma_0}{A}.
	\end{align}  	
The key to achieving a large OD$_\eff$ is choosing an atomic and beam geometry that addresses a large number of atoms and maximizes $N_\eff^{(2)}$ while keeping the mode area $A$ small. It should be noted that whereas in the one-dimensional case the optical density is associated with both the coupling strength and the Beer's law attenuation of the probe, in the three-dimensional case different parameters are associated with each of these effects. Because the attenuation coefficient in \erf{Eq::OutputField} is proportional to the local intensity of the field, the total attenuation depends upon the effective atom number $N^{(1)}_\eff $. 

While \erf{Eq::CouplingStrength} implies an ever increasing coupling strength with integration time $T$, we have neglected so far the decoherence that limits the total useful  integration time and the strength of the atom-light interface.  In the following section we treat these effects from a first-principles master equation, including spatial variations in the scattering rate which drives local decoherence.

		\subsubsection{Local decoherence and optical pumping }

The discrete random atomic positions are associated with the density fluctuations that give rise to diffuse scattering into 4$\pi$ steradians \cite{SorSor08}.  We consider light far detuned from any atomic resonance in a highly transparent regime, and thus we can safely neglect the small attenuation of the laser probe associated with this absorption.  The scattering processes, however, cause decoherence of the spin wave due to optical pumping.  This local decoherence breaks the collective symmetry of the problem and adds additional noise, which is detected in the polarimeter and competes with squeezing. 

To treat the decoherence due to diffuse scattering, we employ a master equation,
\begin{equation}
\frac{d \hat{\rho}}{dt} = -\frac{i}{\hbar} [\hat{H}_{\rm int}, \hat{\rho}] + \frac{d \hat{\rho}}{dt} \Big|_{\rm diff},
\end{equation} 
where $\hat{H}_{\rm int}$ is the multimode Faraday interaction given in \erf{Eq::TheFaradayInteraction}.  The key feature of this equation is that the paraxial modes couple to collective spin waves, while the diffuse scattering couples to localized atoms and induces optical pumping according to
	\begin{equation} \label{Eq::GenME}
		\frac{d \hat{\rho}}{dt} \Big|_{\rm diff} = \sum_i \gamma_s(\mathbf{r}_i) \mathcal{D}_i [\hat{\rho}].
	\end{equation}
The map $\mathcal{D}_i$ acts on the $i^{th}$ atom, proportional to the local scattering rate,
\begin{align} \label{Eq::LocalScatRate}
		\gamma_s(\mathbf{r}_i) = I(\mathbf{r}_i) \frac{\sigma_0}{\hbar \omega} \frac{\Gamma^2}{4 \Delta^2} = \gamma_{0}\beta_{00}(\mathbf{r}_i).
	\end{align}
Here  $I(\mathbf{r}_i) = I_0 \beta_{00}(\mbf{r}_i)$ is the local intensity at the position of the atom and $\gamma_{0}$ is the peak scattering rate. We consider here a probe driving an $S_{1/2} \rightarrow P_J$ transition in an alkali atom, with a detuning that is small compared to the ground state hyperfine splitting but large compared to any hyperfine splitting in the excited state.  In this case, the light coherently couples substantially only to atoms in a given ground-electronic hyperfine manifold $f$ and the master equation is restricted to this subspace.  As shown in Appendix \ref{Appendix::MultimodeJumpOperators}, with an $x$-polarized probe and applying a large bias magnetic field along the $z$-direction, the local decoherence in the master equation due to optical pumping is given by the map
	\begin{align} \label{Eq::DiffuseME}
		\mathcal{D}_i [\hat{\rho}]  = \! -\frac{2}{9}\hat{\rho} + \frac{g_f^2}{9} \Big[ \hat{f}^{(i)}_z\hat{ \rho} \hat{f}^{(i)}_z + \frac{1}{2} \big( \hat{f}^{(i)}_x \hat{\rho} \hat{f}^{(i)}_x + \hat{f}^{(i)}_y \hat{\rho} \hat{f}^{(i)}_y \big)   \Big].
	\end{align}
The first term on the right-hand side of \erf{Eq::DiffuseME} describes the decay of correlations due to optical pumping, while the second term represents a feeding due to ``transfer of coherences'' that can reduce this decay rate \cite{CohenTannoudji}. Note that for $f>1/2$, this master equation is not trace preserving, since atoms can be optically pumped to the other ground hyperfine manifold where they are lost to any further measurement.

Given the master equation, we can find the effect of diffuse scattering on atomic correlations.   Consider a inhomogeneous collective operator of the form $\hat{X} = \sum_i \beta_{00}(\mbf{r}_i) \hat{x}^{(i)}$.  Because $\hat{X}$ is a weighted sum over single atom operators, the equation of motion for its expectation value depends upon the evolution of the single atom density operator, $\hat{\rho}^{(i)}$. By summing over a single index $i$ in Eq. (\ref{Eq::GenME}) we obtain
\begin{align}\label{eq::rhoi}
\frac{d\hat{\rho}^{(i)}}{dt}\Big|_{\rm diff}=\gamma_s(\mbf{r}_i) \mathcal{D}_i[\hat{\rho}^{(i)}],
\end{align}
from which the evolution of $\expects{\hat{X}}$ is given by 
\begin{align}\label{Eq::1stOrderEvol} 
	\frac{d}{dt}\expects{\hat{X}}\Big|_{\rm diff} & =\sum_i\ \beta_{00}(\mbf{r}_i) \mbox{Tr} \bigg[ \hat{x}^{(i)}\frac{d\hat{\rho}^{(i)}}{dt}\Big|_{\rm diff}\bigg]  \nn \\
		&= \sum_i \gamma_s(\mbf{r}_i) \beta_{00}(\mbf{r}_i) \expect{\mathcal{D}_i[\hat{x}^{(i)}]}.
\end{align}
For inhomogeneous collective operators that depend on pairs of atoms,
	\begin{align}
		\hat{O}  =  \sum_{i \neq j} \beta_{00}(\mbf{r}_i)  \beta_{00}(\mbf{r}_j)  \hat{x}^{(i)} \hat{y}^{(j)},
	\end{align}
we require the joint density operator of the $i^{th}$ and $j^{th}$ atoms, $\hat{\rho}^{(i,j)}$, with equation of motion 
\begin{align}\label{Eq::rhoij}
	\frac{d}{dt}\hat{\rho}^{(i,j)}\Big|_{\rm diff} =\gamma_s(\mbf{r}_i)\mathcal{D}_i[\hat{\rho}^{(i,j)}]+ \gamma_s(\mbf{r}_j) \mathcal{D}_j[\hat{\rho}^{(i,j)}].
\end{align}
The evolution of $\expects{\hat{O}}$ due to diffuse scattering is then
\begin{align} \label{Eq::2ObservableEOM}
		\frac{d}{dt} \expects{\hat{O}} \Big|_{\rm diff} & =  \sum_{i \neq j}\Big\{  \gamma_s(\mbf{r}_i) \beta_{00}(\mbf{r}_i)  \beta_{00}(\mbf{r}_j)  \expect{ \mathcal{D}_i [\hat{x}^{(i)}] \hat{y}^{(j)}} \nn \\
		&+  \gamma_s(\mbf{r}_j) \beta_{00}(\mbf{r}_i)  \beta_{00}(\mbf{r}_j) \expect{ \hat{x}^{(i)} \mathcal{D}_j[ \hat{y}^{(j)}] } \Big\}. 
	\end{align}	

The degree of squeezing that one can ultimately produce is determined by a balance between QND measurement backaction on the spin wave mediated by the collective radiation and the damage to that observable caused by diffuse scattering. To properly treat this we must include the effects of measurement on the atoms, as discussed in the next section.

	\subsubsection{The conditional stochastic master equation} \label{Sec::HomodyneSME}

The Faraday Hamiltonian, \erf{Eq::TheFaradayInteraction}, is an entangling interaction between the atomic spin waves and the paraxial modes of the field.  When the light is measured in the polarimeter, quantum backaction leads to stochastic evolution of the atomic state, conditioned on the measurement result.  A complete description of the dynamics is then described by a stochastic master equation (SME), with decoherence from unmeasured light and squeezing due to information gained from the continuous measurement record.  In a balanced polarimeter, the measurement signal is proportional to the interference of the probe and scattered fields integrated over the detector faces, as in \erf{Eq::ClassicalMeasurementSignal}. Due to the orthogonality of the spatial modes, \erf{Eq::TransverseOrthogonality}, such a measurement selects only paraxial light that is scattered into the mode of the probe, $u_{00}$.  The result is a continuous measurement of the quadrature $\hat{X}_{00}$. 
		
We derive the SME for the atoms following \cite{JacSte06, WieMilBook}, with details presented in Appendix \ref{Appendix::SMEDerivation}.  Measurement of the quadrature $\hat{X}_{00}$ by the homodyne polarimeter generates a differential stochastic measurement record
 \begin{equation}
 dy_{00} =\expect{\hat{F}_z^{00}} dt + \frac{1}{ \sqrt{\kappa}}dW,
 \end{equation}
where $dW$ is a Weiner interval in the It\={o} calculus and $\kappa$ is the measurement strength given in \erf{Eq::MeasurementStrengthperAtom}.  Assuming unit measurement efficiency, the evolution of the ensemble conditioned upon the measurement record $y_{00}$ is given by
	\begin{align} \label{ApEq::HomodyneSMEtext}
		d \hat{\rho}&= \sqrt{ \frac{\kappa}{4} } \mathcal{H}_{00}[\hat{\rho}] \, dW + \frac{\kappa}{4} \sum_{p,l} \mathcal{L}_{pl}[\hat{\rho}] \, dt.
  	\end{align}
The effects of measurement backaction on the ensemble are taken into account by the superoperator $\mathcal{H}_{00}[\hat{\rho}] $, where
\begin{align}  \label{Eq::HSuperoperator}
		\mathcal{H}_{pl}[\hat{\rho}] = \hat{F}^{pl}_z \hat{\rho} + \hat{\rho} \hat{F}^{pl \dagger}_z - \Tr{ \big(\hat{F}^{pl}_z + \hat{F}^{pl\dagger}_z \big) \hat{\rho} } \hat{\rho}.
	\end{align}
The Lindblad dissipator, 
	\begin{align} \label{Eq::LSuperoperator}
		\mathcal{L}_{pl}[\hat{\rho}] = \hat{F}_z^{pl} \hat{\rho} \hat{F}_z^{pl\dagger} - \frac{1}{2} \hat{F}_z^{pl\dagger} \hat{F}_z^{pl} \hat{\rho}  - \frac{1}{2} \hat{\rho} \hat{F}_z^{pl\dagger} \hat{F}_z^{pl},
	\end{align}
describes the effect on the atomic ensemble arising from collective radiation into all paraxial modes of the field.

Including local decoherence from diffuse scattering, \erf{Eq::DiffuseME}, the full stochastic master equation for homodyne polarimetry measurements of the $00$-mode is 
	\begin{align} \label{Eq::HomodyneSME}
	d \hat{\rho}   =&  - \frac{i}{\hbar} [\hat{H}, \hat{\rho}]dt + \sqrt{ \frac{\kappa}{4} } \mathcal{H}_{00}[\hat{\rho}] \, dW \\ 
	&+ \frac{\kappa}{4} \sum_{p,l} \mathcal{L}_{pl}[ \hat{\rho}]\, dt 
		 + \sum_i \gamma_s(\mathbf{r}_i) \mathcal{D}_i [\hat{\rho}] \, dt. \nn
	\end{align}  
This SME is a complete description of the evolution of the collective atomic state, accounting for the three-dimensional nature of the atom-photon modes, decoherence, and measurement backaction.  We see that through its  interaction with the probe, the atomic ensemble undergoes an additional form of \emph{collective} decoherence, \erf{Eq::LSuperoperator}, corresponding to light radiated into paraxial modes $pl \neq 00$ that ultimately goes unmeasured.  Thus we have arrived at the same conclusion as in Ref. \cite{DuaZol02}.  That is, decoherence arises through two distinct processes - first, the inherent mode-mismatch that gives rise to collectively scattered light in spatial modes other than the probe mode and second, the diffuse scattering of photons that acts locally on atoms in the ensemble.

\section{QND squeezing of spin waves} \label{Sec::SpinSqueezing}

\subsection{Quantifying squeezing of the spin waves}	
One typically quantifies the amount of squeezing created in a QND measurement according to the Wineland squeezing parameter~\cite{Wineland94},
\begin{equation} \label{Eq::StandardParam}
\zeta =\left(\frac{\Delta \phi}{\Delta \phi_{\text{SCS}}}\right)^2,
\end{equation}
where $\Delta \phi$ is the projection-noise limited resolution when measuring an angle of rotation for a generic spin $J$ of the given input state, and  $\Delta \phi_{\text{SCS}}$ is the corresponding resolution when the input is a spin coherent state (SCS).  For a mean value $J_\parallel \equiv |\expects{\hat{\mbf{J}}}|$, and variance $\Delta J_\perp^2$ orthogonal to the mean, the projection-noise resolution is $\Delta \phi =\Delta J_\perp/J_\parallel$.  With  $\Delta \phi_{\rm SCS} =1/\sqrt{2J}$,  the Wineland squeezing parameter is then
\begin{equation}
\zeta = 2J\frac{\Delta J_\perp^2}{J_\parallel^2}.
\label{simple_squeezing}
\end{equation}

For the spin waves of the inhomogeneous ensemble under consideration here, we must tie the squeezing parameter directly to the measured quantities. For an initial mean spin polarization along $x$ and a small rotation around $y,$ the polarimeter signal will be determined by the mean spin wave component $\expects{\hat{F}_x^{00}} = \sum_i \beta_{00} (\mbf{r}_i) \expects{\hat{f}^{(i)}_x}$, and the projection noise contribution to the resolution of the measurement will be given by   $\Delta F^{00}_z$, defined in Eq. (\ref{Eq::variance}).  The projection-noise limited resolution of this rotation is therefore  $\Delta \phi^{00} = \Delta F_z^{00}/\expects{\hat{F}_x^{00}}$.  Furthermore, given a SCS initially polarized along x, the mean spin of interest is $\expects{\hat{F}_x^{00}}_{\text{SCS}}= N^{(1)}_\eff f$  where the effective atom number contributing to this signal is given in \erf{Eq::EffectiveAtomNumbers}, while the projection noise in the spin coherent state is $( \Delta F_z^{00})^2_{\text{SCS}}=N^{(2)}_\eff f/2$.  The projection noise limited resolution for a SCS preparation is thus, $( \Delta \phi^{00}_\text{SCS})^2=N_\eff^{(2)}/[2f  (N^{(1)}_\eff )^2 ]$, and will depend on the shape of the atomic cloud and beam geometry.  Putting this together, we define the squeezing parameter for the measured spin wave to be
	\begin{equation} \label{Eq::SqueezingParam}
		\zeta \equiv \left(\frac{\Delta \phi^{00}}{\Delta \phi^{00}_{\text{SCS}}}\right)^2 = 2f \frac{ \big(N^{(1)}_\eff \big)^2 }{N_\eff^{(2)}} \frac{\big(\Delta F_z^{00}\big)^2}{\expect{\hat{F}_x^{00}}^2}.
	\end{equation}
This parameter quantifies the degree of ``quantum backaction," on a spin coherent state, accounting for the change in projection noise due to QND measurement as well as the damage done to both the mean spin polarization and variance due to optical pumping.  

In a real-world metrological application such as an optically probed atomic magnetometer~\cite{BudRom07, SewMit12}, spin rotations are measured by passing the probe through the atom sample and measuring the resulting Faraday rotation in a polarimeter.  In addition to spin projection noise, the measurement resolution is then subject also to ``technical noise," including probe shot noise, detector electronic noise, and atom number fluctuations.  Under those circumstances, optimizing the squeezing parameter  as defined in Eq. (\ref{Eq::SqueezingParam}) is distinct from optimizing the magnetometer sensitivity.

\subsection{The dynamical evolution of squeezing}

	To determine the squeezing as function of time, we employ the SME in \erf{Eq::HomodyneSME} to track $( \Delta{F}_z^{00})^2$ and $\expects{\hat{F}_x^{00}}$.  For ensembles with large numbers of atoms, we can work in the central-limit approximation where fluctuations in the spin waves are treated as Gaussian random variables~\cite{PolzikRMP, VasSor12}.  Following  \cite{JacSte06}, the SME then couples solely means and covariances. The moments of the fundamental spin wave that characterize the spin squeezing parameter then evolve according to
	\begin{subequations}  \label{Eq::FullSpinWaveEOM}
	\begin{align}
		d\varFz =&  -\kappa  \big[ \left(\Delta F_z^{00} \right)^2 \big]^2 dt +\varFz\Big|_{\rm diff}dt , \label{Eq::Var00Evolution}\\
	d \expect{\hat{F}^{00}_x} = & \sqrt{ \frac{\kappa}{4} } \big  \langle\mathcal{H}_{00}\big[\hat{F}^{00}_x\big]\big  \rangle \, dW \label{Eq::Mean00Evolution} \\
	& +\sum_{p,l}\frac{\kappa}{4} \big  \langle \mathcal{L}_{pl}\big[ \hat{F}^{00}_x \big] \big \rangle dt +\expect{\hat{F}^{00}_x} \Big|_{\rm diff} dt. \nn
	\end{align}
	\end{subequations}
Because we assume the fundamental mode is measured with unit efficiency, diffuse scattering by local spontaneous emission is the only process contributing to the decoherence of the variance $\varFzText $.  Collective radiation into other transverse modes commutes with $\hat{F}_z^{00}$ and does not contribute to any decay or noise injection into the fundamental variance.  In contrast, the mean spin $\expects{\hat{F}^{00}_x}$ decoheres due to both diffuse scattering and collective scattering into other unmeasured paraxial modes.  It also evolves stochastically due to the continuous measurement of $\hat{F}_z^{00}$.  However, the contributions to the dynamics from both collective scattering and continuous measurement are small in comparison to diffuse scattering and can be neglected when the radiation pattern of the cloud is well matched to that of the probe.  

We consider the moment evolution, \erf{Eq::FullSpinWaveEOM}, with the initial condition that the ensemble is in a SCS polarized along $x$. The initial mean spin and variance are $\expects{ \hat{F}_{x}^{00} (t_0) }= N_{\rm eff}^{(1)} f$ and $( \Delta F_z^{00} (t_0) )^2 =  N_{\rm eff}^{(2)}f/2$ respectively.  Along with the cross-sectional area of the probe laser,  $N_{\rm eff}^{(2)}$ specifies the effective optical density, OD$_{\rm eff}$ defined in \erf{Eq::ODeff}.  The  OD$_{\rm eff}$ is the critical geometric parameter for determining how the atomic density distribution influences collective scattering into the probe mode and ultimately leads to spin squeezing. Both of these effective atom numbers are determined solely by the cloud shape and beam geometry, and can be found from the semiclassical model in Sec. \ref{Sec::SemiclassicalTheory}.  

For times short compared to the photon scattering rate, where decoherence is negligible, the mean spin is essentially constant and the spin variance is affected only by measurement backaction. The solution to \erf{Eq::FullSpinWaveEOM} takes the familiar form \cite{PolzikRMP} 
	\begin{align} 
		\big( \Delta F_z^{00} (t) \big)^2 & = \frac{\big( \Delta F_z^{00} (t_0) \big)^2 }{1+ \big( \Delta F_z^{00} (t_0) \big)^2 \,\kappa t} = \frac{ \big( \Delta F_z^{00} (t_0) \big)^2 }{1+\xi} \\
		& \Rightarrow \zeta = \frac{1}{1+\xi}, \label{Eq::ShortTimeSolution}
	\end{align}
where $\xi$ is the integrated coupling strength in Eq. (\ref{Eq::CouplingStrength}).  In \erf{Eq::ShortTimeSolution}, the squeezing parameter decreases as OD$_\eff^{-1}$ for $\xi\gg1$.

For longer times, decoherence due to diffuse photon scattering must be included.  The mean spin will depolarize according to Eq. (\ref{Eq::1stOrderEvol}),
\begin{equation}\label{Eq::MeanSpinArb}
\frac{d}{dt} \expect{\hat{F}^{00}_x} = \sum_i \gamma_s(\mbf{r}_i)\beta_{00}(\mbf{r}_i) \big\langle \mathcal{D}_i \big[\hat{f}_x^{(i)}\big] \big\rangle.
\end{equation}
The variance involves both single atom and pairwise atomic correlations,
	\begin{align}  \label{Eq::VarianceDecompositionApprox}
		\frac{d}{dt}\varFz = & \sum_i \big[ \beta_{00}(\mbf{r}_i) \big]^2 \frac{d}{dt}\big\langle(\Delta \hat{f}_z^{(i)})^2\big\rangle \\
		& +  \sum_{i\neq j} \beta_{00}(\mbf{r}_i) \beta_{00}(\mbf{r}_j)  \frac{d}{dt}\big\langle \Delta \hat{f}_z^{(i)}\Delta \hat{f}_z^{(j)} \big\rangle, \nn
	\end{align}
where the first term is the spin projection noise of the uncorrelated spins and the second term contains the correlations that generate spin squeezing.   Following Eqs. (\ref{Eq::1stOrderEvol}) and  (\ref{Eq::2ObservableEOM}), these correlation functions decay due to diffuse scattering according to
	\begin{subequations} \label{Eq::Correlations}
	\begin{align}
& \frac{d}{dt}\sum_{i} \big\langle (\Delta \hat{f}_z^{(i)})^2  \big\rangle  \big|_{\rm diff} = \nn \\  & \quad   \sum_{i}  \gamma_s(\mbf{r}_i) \Big\{  \big\langle\mathcal{D}_i \big[ \hat{f}_z^{(i)2} \big] \big\rangle  -2 \big\langle \mathcal{D}_i \big[ \hat{f}_z^{(i)} \big] \big\rangle \big\langle \hat{f}_z^{(i)} \big\rangle  \Big\} \label{eq::NoiseTerm2}, \\
		& \frac{d}{dt}\sum_{i\neq j} \big\langle \Delta \hat{f}_z^{(i)}\Delta \hat{f}_z^{(j)}   \big\rangle \big|_{\rm diff} = \nn \\ 
		& \! \sum_{i\neq j} \! \Big\{ \! \gamma_s(\mbf{r}_i) \big\langle \Delta\mathcal{D}_i \big[ \hat{f}_z^{(i)} \big]\Delta \hat{f}_z^{(j)}\big\rangle  \! +\!  \gamma_s(\mbf{r}_j) \big\langle \Delta \hat{f}_z^{(i)}\Delta \mathcal{D}_j \big[\hat{f}_z^{(j)}\big] \big\rangle \! \Big\} \label{eq::correlationDecay2}.
	\end{align}
	\end{subequations}
The local decoherence acts via the map $\mathcal{D}_i$ in \erf{Eq::DiffuseME}.

\subsubsection{Spin-1/2 ensembles}

We first restrict our attention to ensembles of spin $f=1/2$ atoms to focus on spatial effects without the complications that arise for ensembles with larger-spin. Using the fact that the local scattering rate is proportional to the probe intensity, $\gamma_s(\mathbf{r}) = \gamma_{0}\beta_{00}(\mathbf{r})$, the mean spin evolution of \erf{Eq::Mean00Evolution} is 
\begin{equation} \label{Eq::Mean00GammaBeta}
	\frac{d}{dt} \expect{\hat{F}^{00}_x} = -\frac{\gamma_{0}}{3} \sum_i \big[ \beta_{00}(\mbf{r}_i) \big]^2 \big\langle \hat{f}_x^{(i)} \big\rangle.
\end{equation}
The local decoherence does not respect the orthogonality of the transverse paraxial modes and we will see that the diffuse scattering acts to couple the fundamental spin wave to higher order spin waves.  
 
Because the transverse modes are orthogonal in a plane at a fixed $z$,  we can  derive a set of coupled equations by decomposing products of the spatial weighting coefficients, \erf{Eq::Beta}, in the basis of mode functions as follows:
	\begin{equation}
		\big[ \beta_{00}(\mbf{r}_\perp, z) \big]^2 =|u_{00}(\mathbf{r}_{\perp}, z)|^4  =  \sum_{p,l} c^{00}_{pl}(z) \beta_{pl}(\mathbf{r}_\perp, z) ,
	\end{equation}
with $z$-dependent projection coefficients,
\begin{equation}
c^{00}_{pl}(z) \equiv \frac{1}{A} \int d^2 \mathbf{r}_\perp \big[ u_{00}(\mathbf{r}_\perp, z) \big]^2 u^*_{00}(\mathbf{r}_\perp, z) u_{pl}(\mathbf{r}_\perp, z).
	\end{equation}
Performing the sum over atoms within each coarse-grained slice, it follows that \erf{Eq::Mean00GammaBeta} can be expressed as	\begin{equation}  \label{Eq::MeanZsliced}
		\frac{d}{dt} \big\langle \hat{F}^{00}_{x} \big\rangle =  - \frac{ \gamma_{0}}{3} \sum_{k, p,l} c^{00}_{pl} (z_k) \big\langle \hat{F}_x^{pl}(z_k) \big\rangle,
	\end{equation}
where we have defined spin waves in coarse-grained slices of thickness $\delta z$ around $z_k$, 
	\begin{equation}  \label{Eq::CoarseGrainedSpinWave}
		\hat{F}^{pl}_z (z_k) \equiv \sum_{i_k} \beta_{pl}(\mbf{r}_{\perp {i_k}}, z_{i_k}) \hat{f}_z^{(i_k)},
	\end{equation}
with the index $i_k$ labeling atoms in the slice $z_k$.  The total spin wave for a given transverse mode is $\hat{F}^{pl}_z = \sum_k \hat{F}^{pl}_z (z_k)$. The mean spin in the fundamental mode $\expects{\hat{F}_x^{00}}$ is thus coupled to other spin waves $\expect{\hat{F}_x^{pl}}$ within each slice $z_k$.  Generally, the expected value of the $pl$-spin wave in the slice $z_k$ evolves according to
	\begin{align}  \label{Eq::MeanZsliced}
		\frac{d}{dt} \expect{\hat{F}^{pl}_{x}(z_k)} =  - \frac{ \gamma_{0}}{3} \sum_{p',l'} c^{pl}_{p'l'} (z_k) \expect{\hat{F}_x^{p'l'}(z_k)},
	\end{align}
with projection coefficients $c^{pl}_{p'l'} (z_k)$ given in \erf{Eq::ProjCoeff}.  Details of this derivation are found in Appendix \ref{Appendix::SpinCovariances}. The initial conditions, \erf{Eq::meanSlice}, account for the matching between the probe mode and cloud geometry.  By projecting onto the spin waves, we obtain a hierarchy of coupled equations.  Numerically, we truncate once the desired convergence is achieved.

	The effect of diffuse scattering on the evolution of the collective spin variance follows in an analogous manner.  For spin-1/2, $\expects{\Delta \hat{f}^2_z }=1/4$ for all atoms.  The map for local decoherence, $\Delta \mathcal{D}_i\big[\hat{f}^{(i)}_z\big] = -2 \Delta \hat{f}^{(i)}_z/9$, corresponds to decay of spin-spin correlations with no feeding of coherences.  The evolution of the fundamental spin wave variance, \erf{Eq::Var00Evolution}, simplifies to
	\begin{widetext}
	\begin{align} \label{Eq::SpinHalfVariance}
		& \frac{d}{dt}  \varFz =  -\kappa \Big[ \big( \Delta F_z^{00} \big)^2 \Big]^2   -\frac{2\gamma_{0}}{9}  \sum_{i,j} \big[ \beta_{00}(\mathbf{r}_i)  +\beta_{00}(\mathbf{r}_j) \big] \beta_{00}(\mbf{r}_i)\beta_{00}(\mbf{r}_j) \expect{\Delta \hat{f}_z^{(i)}\Delta \hat{f}_z^{(j)}} + \frac{\gamma_{0}}{9} \sum_i \big[ \beta_{00}(\mbf{r}_i) \big]^3, 
	\end{align}
	\end{widetext}
where again we have used \erf{Eq::LocalScatRate}.  The first term describes squeezing of the variance due to measurement backaction, the second represents decay of correlations due to diffuse scattering, and the third is the noise injected into the variance from spin flips (optical pumping).   Following the same procedure as above, the decay terms are projected onto higher order spin waves, 
\begin{align}  \label{Eq::ProjectedSpin1/2VarianceEOM}
		& \frac{d}{dt} \varFz =  -\kappa \Big[ \big( \Delta F_z^{00} \big)^2 \Big]^2 \\ 
		& -\frac{4 \gamma_0}{9} \sum_{p,l} \sum_{k,k'} c^{00}_{pl}(z_k) \big\langle \Delta \hat{F}_z^{00}(z_k) \Delta \hat{F}_z^{pl}(z_{k'})\big\rangle  + \frac{ \gamma_{0}}{9}N^{(3)}_\eff. \nn
	\end{align}
Here, $N^{(3)}_\eff$ is the effective atom number governing the injection of noise through optical pumping, defined in \erf{Eq::EffectiveAtomNumbers}.  Equation (\ref{Eq::ProjectedSpin1/2VarianceEOM}) is a covariance description of the dynamics, similar to that commonly employed for spin squeezing \cite{MadMol04}, but which also accounts for local decoherence from first principles.  To solve for the fundamental variance, we must track the evolution of the covariances between coarse-grained slices and between transverse modes, $\langle \Delta \hat{F}_z^{pl}(z_k) \Delta \hat{F}_z^{p'l'}(z_{k'})\rangle = \expects{\hat{F}_z^{pl}(z_k)\, \hat{F}_z^{p'l'}(z_{k'}) } -\expects{\hat{F}_z^{pl}(z_k)}\expects{ \hat{F}_z^{p'l'}(z_{k'}) }$.  Equations of motion for these covariances follow readily from the SME. 
A detailed derivation is given in Appendix \ref{Appendix::SpinCovariances}.

\subsubsection{Spin-f alkali atom ensembles} \label{Sec::SpinfEnsembles}

The constituent atoms in many spin squeezing experiments are alkali metal atoms whose ground state structure is more complex than spin-1/2.  For example, in $^{133}$Cs, the ground electronic subspace is defined by two hyperfine manifolds with total spin angular momentum $f = \{3,4\}$.  Owing to the large ground-state hyperfine splitting (9.2 GHz in $^{133}$Cs), a single hyperfine manifold $f$ is addressed by the coherent interaction with the probe laser.

Though ensembles of higher spin atoms can be squeezed by the same QND measurement process, spin size affects both the coherent squeezing dynamics and decoherence.  Recall that the strength of the Faraday interaction is quantified by the coupling strength $\xi$, Eq. (\ref{Eq::CouplingStrength}). Because $\xi\propto 1/f^2$, the atom-light coupling decreases with increasing spin size. This decreased coupling strength is partially offset by an increased robustness to the effects of optical pumping. When $f>1/2$, optical pumping events can be broadly divided into two categories: (i) ``loss" that occurs when an atom is pumped from the $f$ manifold into the other ground hyperfine manifold and (ii) ``spin flips" that leave the atom in the $f$ manifold. Because atoms lost to the other ground manifold are no longer resonant with the probe, loss events decrease the mean spin $\langle\hat{F}_x^{00}\rangle$, though they contribute no excess noise to $(\Delta F_z^{00})^2$. ``Spin flips" are responsible for both a decrease in $\langle\hat{F}_x^{00}\rangle$ and a noise injection into $(\Delta F_z^{00})^2$. For the SCS preparation, the deleterious effects of spin flips are mitigated by ``transfers of coherence" between pairs of magnetic sublevels that reduce the rate of decay of correlations~\cite{CohenTannoudji}. While the interplay between these effects is complex, the rate of spin flips remains a good indicator of an ensemble's robustness to optical pumping. For an ensemble of spin-$f$ alkalis prepared in a SCS, the spin flip rate is $\gamma_s(\mathbf{r})/(12f)$, thus decreasing for larger spin size.  

Due to these decoherence processes, the dynamics of the squeezing parameter is substantially more complicated for larger spin atoms. For spin-$f$, we obtain the evolution of the mean value of a spin wave in slice $z_k$ by projecting onto the different spin waves in a manner analogous to \erf{Eq::MeanZsliced},
	\begin{align}  \label{Eq::MeanSpins}
		\frac{d}{dt} & \expect{\hat{F}^{pl}_{x}(z_k)}  = - \frac{2 \gamma_{0}}{9}  \sum_{p',l'} c^{pl}_{p'l'} (z_k) \expect{\hat{F}_x^{p'l'}(z_k)} \\
		& \quad + \frac{g_F^2 \gamma_0}{9}  \sum_{p',l'} \sum_{i_k} c^{pl}_{p'l'} (z_k) \beta_{p'l'} (\mathbf{r}_{\perp i_k},z_k )\mathcal{C}_{i_k}[\hat{f}_x^{(i_k)}]. \nn
	\end{align}
 Here, we have defined a local superoperator that arises solely from the ``feeding" terms in the master equation:
	\begin{align}
		\mathcal{C}_i[\hat{X}] \equiv \hat{f}_z^{(i)}\hat{X}\hat{f}_z^{(i)} + \tfrac{1}{2} \big( \hat{f}_x^{(i)}\hat{X}\hat{f}_x^{(i)} + \hat{f}_y^{(i)} \hat{X}\hat{f}_y^{(i)} \big).
	\end{align}
Similarly, we find equations of motion for the fundamental spin wave variance,
	\begin{align} \label{Eq::GenCovarianceEvolution}
		& \frac{d}{dt}  \varFz =- \kappa \big[\left(\Delta F_z^{00} \right)^2\big]^2 \\
		& -\frac{4\gamma_{0}}{9} \sum_{p,l} \sum_{k',k}c^{00}_{pl}(z_k)\big\langle \Delta \hat{F}_{z}^{00}(z_{k'}) \Delta \hat{F}_{z}^{pl}(z_k) \big\rangle \nn \\
		&  +  \frac{g_f^2 \gamma_{0}}{9} \sum_{p,l}\sum_{k',k,i_k} c^{00}_{pl}(z_k)\beta_{pl}(\mathbf{r}_{i_k})  \Big\{ \big\langle \Delta \hat{F}_{z}^{00}(z_{k'}) \Delta \mathcal{C}_{i_k}[ \hat{f}_{z}^{(i_k)} ] \big\rangle \nn \\
		&\quad \quad \quad +\big\langle \Delta \mathcal{C}_{i_k}[ \hat{f}_{z}^{(i_k)} ] \Delta \hat{F}_{z}^{00}(z_{k'}) \big\rangle \Big\} \nn \\
		&+\gamma_{0}\sum_{k,i_k} \big[ \beta_{00}(\mathbf{r}_{i_k}) \big]^3 \bigg\{ \frac{2}{9} \expect{(\hat{f}_z^{(i_k)})^2} \nn \\
		& \quad \quad \quad  + \frac{g_f^2}{9} \left( \big\langle \mathcal{C}_{i_k}[\hat{f}_z^{(i_k)2}]\big\rangle-\big\langle \big\{\hat{f}_z^{(i_k)},\:\mathcal{C}_{i_k}[\hat{f}_z^{(i_k)} ] \big\}_+\big\rangle \right) \bigg\} , \nn
	\end{align}
where $\{ \hat{X}, \hat{Y}\}_+$ denotes the anti-commutator. As for the case of spin-1/2, we have an infinite hierarchy of equations that couple spin wave operators in the different $z_k$-slices. In general, the feeding terms in \erf{Eq::GenCovarianceEvolution} couple to covariances outside the set  $\expects{\Delta\hat{F}_z^{pl}(z_k) \Delta\hat{F}_z^{p'l'}(z_{k'})}$. This expands considerably the number of equations that must be solved to reach convergence. Solving these equations, furthermore, requires different methods than the spin-1/2 case. A detailed treatment of the spin-$f$ case will be provided in future work.

\subsection{Results}

Using our formalism we can calculate the dynamics of QND measurement and the peak achievable squeezing in the presence of decoherence.  We now consider the fundamental effects of geometry and the optimization of experimentally relevant quantities to maximize spin squeezing.  Most of our results are shown for the simplest case of spin-1/2 atoms in order to focus on the effects of spatial modes and spin waves.  We also consider some preliminary calculations for spin-$f>1/2$; more complete studies will be presented elsewhere.

\subsubsection{Geometric effects of local decoherence for a fixed rate of squeezing}\label{Sec::GeomEffects}	

The geometry of the atom-laser system plays two distinct roles in determining the amount of achievable squeezing. First,  OD$_{\rm eff} \propto N_{\rm eff}^{(2)}$, Eq. (\ref{Eq::ODeff}), is a purely geometrical quantity, derivable from the semiclassical model  (see also \cite{MulPol05}).   OD$_{\rm eff}$  sets the measurement strength, $\xi$, that characterizes the amount of light that is collectively scattered into the spatial mode of the probe.  Second, because of the inhomogeneous intensity profile of the laser mode, the rate of diffuse photon scattering that causes the local decoherence, and ultimately caps the amount of squeezing that is generated, varies across the cloud.    Further complications arise from the fact that optical pumping both injects noise into the spin wave variance and causes a decay of the mean spin.  

To gain physical insight, in this section we fix the OD$_\eff$ as we vary the geometry in order to isolate the effects of local decoherence as they relate specifically to the squeezing parameter, \erf{Eq::SqueezingParam}. For simulations, we choose the ensemble to be a cylindrically symmetric Gaussian cloud with average density
	\begin{align} \label{Eq::AtomicDistribution}
		\eta(\mathbf{r}) = \eta_0 \exp \left( - 2\frac{\rho^2}{\sigma_\perp^2}  - 2\frac{z^2}{\sigma_z^2} \right),
	\end{align}
where $\sigma_\perp^2$ and $\sigma_z^2$ are the transverse and longitudinal $1/e^2$ variances, $\eta_0$ is the peak density, and $\int d^3 \mathbf{r} \eta(\mathbf{r}) = N$, the total atom number.  To characterize the geometry of the atomic distribution we use the \emph{aspect ratio}, defined as AR $\equiv \sigma_z/\sigma_\perp$.  A longitudinally-extended, pencil-shaped cloud commonly employed in cold, dipole-trapped atomic ensemble experiments has an AR $\gg 1$; a pancake-shaped cloud that is much wider than it is long has an AR $\ll 1$. Note, we vary $\eta_0$ as a function of the cloud geometry such that OD$_\eff \propto N_\eff^{(2)}/A$ remains constant.

\begin{figure}[b]
        		\includegraphics[width=1\hsize]{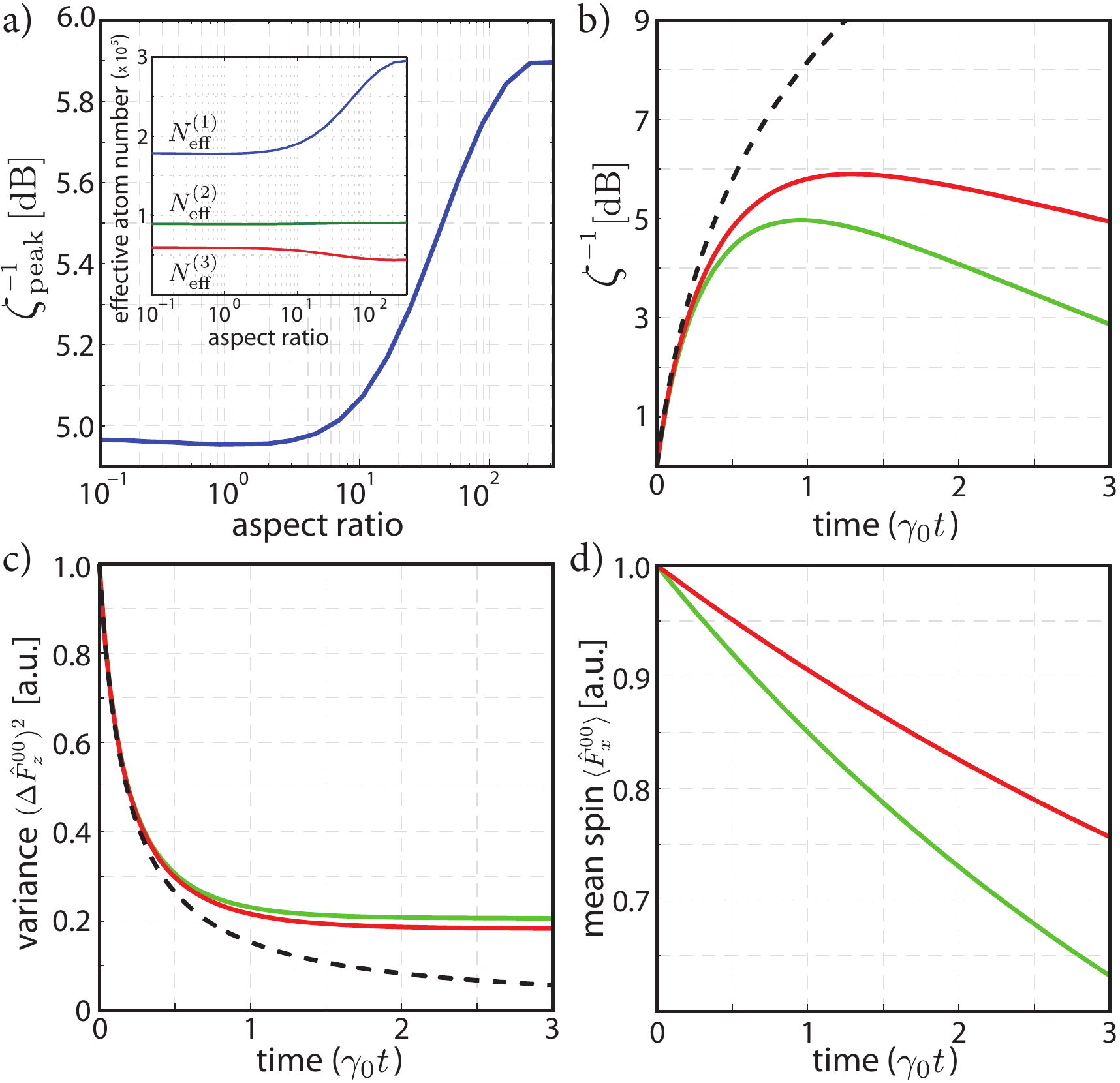}
       		 \caption{Squeezing dynamics for a fixed OD$_\eff = 50$ and different atomic cloud geometries.  The laser probe is a TEM$_{00}$ mode with beam waist $w_0 = 20$ $\mu$m.   a) Peak squeezing, denoted as the inverse of the squeezing parameter, $\zeta^{-1}$ in dB, as a function of aspect ratio of the cloud.  The inset shows effective atom numbers as a function of aspect ratio; $N_\eff^{(2)}$ is constant by design.  b) Comparison of squeezing dynamics for clouds with AR= 0.1 (green line) and AR $= 316$ (red line).  The behavior in the absence of decoherence, \erf{Eq::ShortTimeSolution} (black dashed line), is plotted at the same OD$_\eff$, showing  agreement for short times. c) Dynamics of the spin wave variance for the two clouds, normalized by dividing each by its initial variance, $N^{(2)}_{\rm eff}/4$.  d) Dynamics of the mean spin for the two clouds, normalized by dividing each by its initial mean spin, $N^{(1)}_{\rm eff}/2$.  For fixed OD$_\eff$, the superior squeezing of the pencil-shaped cloud over the pancake-shaped cloud is attributed to slower decay of the mean spin.}  \label{Fig::SpinHalfCompareSqueezing}
   	\end{figure}

To find the peak squeezing, we perform numerical simulations by integrating the evolution of the collective mean spin and variance, Eqs.  (\ref{Eq::Mean00GammaBeta}) and (\ref{Eq::SpinHalfVariance}), and then calculating the spin squeezing as a function of time.   Figure \ref{Fig::SpinHalfCompareSqueezing} shows the resulting spin squeezing for different cloud geometries for a fixed beam waist, $w_0 = 20$ $\mu$m.  The effective optical density is held constant, OD$_\eff = 50$, which guarantees identical squeezing in the absence of decoherence for any geometry. Fig. \ref{Fig::SpinHalfCompareSqueezing}(a) shows the peak squeezing as a function of the AR.  An increase in peak squeezing accompanies an increasing aspect ratio, indicating that decoherence is less detrimental to longitudinally extended clouds.  The dynamics of the squeezing parameter are plotted in Fig. \ref{Fig::SpinHalfCompareSqueezing}(b) for the opposing cases of a pancake-shaped cloud with AR = 0.1 and a pencil-shaped cloud with AR = 316.  For comparison, the short-time solution \erf{Eq::ShortTimeSolution} is shown, which describes the squeezing for either cloud in absence of decoherence.  

	To understand these results we separately examine the dynamical evolution of the projection noise variance and the mean spin in Figs. \ref{Fig::SpinHalfCompareSqueezing}(c-d), both of which contribute to the squeezing parameter.  The effects of decoherence lead to different steady state values of the fundamental spin wave variance in Fig. \ref{Fig::SpinHalfCompareSqueezing}(c)  because the noise injection due to optical-pumping-induced spin flips, set by $N_{\rm eff}^{(3)}$, is slightly smaller for the pencil than for the pancake (see subplot in Fig. \ref{Fig::SpinHalfCompareSqueezing}(a)).  More importantly, the decay rate of the mean spin is a strong function of the AR, as seen in  \ref{Fig::SpinHalfCompareSqueezing}(d).  For a fixed OD$_\eff$, under consideration here, different cloud geometries correspond to different  $N_{\rm eff}^{(1)}$, which determines the mean spin of the ensemble addressed by the beam.  The pencil geometry address a larger $N_{\rm eff}^{(1)}$ when compared to the pancake geometry, as seen in the subplot of \ref{Fig::SpinHalfCompareSqueezing}(a).  In addition, for the pencil geometry $N_{\rm eff}^{(1)}$  also decays more favorably.  This occurs because for a fixed OD$_\eff$, in the pencil geometry a large fraction of the atoms are spread far from the beam waist where rates of optical pumping are lower.  For the pancake geometry, to achieve the same OD$_\eff$, more of the atoms the we address are concentrated in the high intensity region and more quickly depolarize.

	\begin{figure*}
    		    		\includegraphics[width=1\hsize]{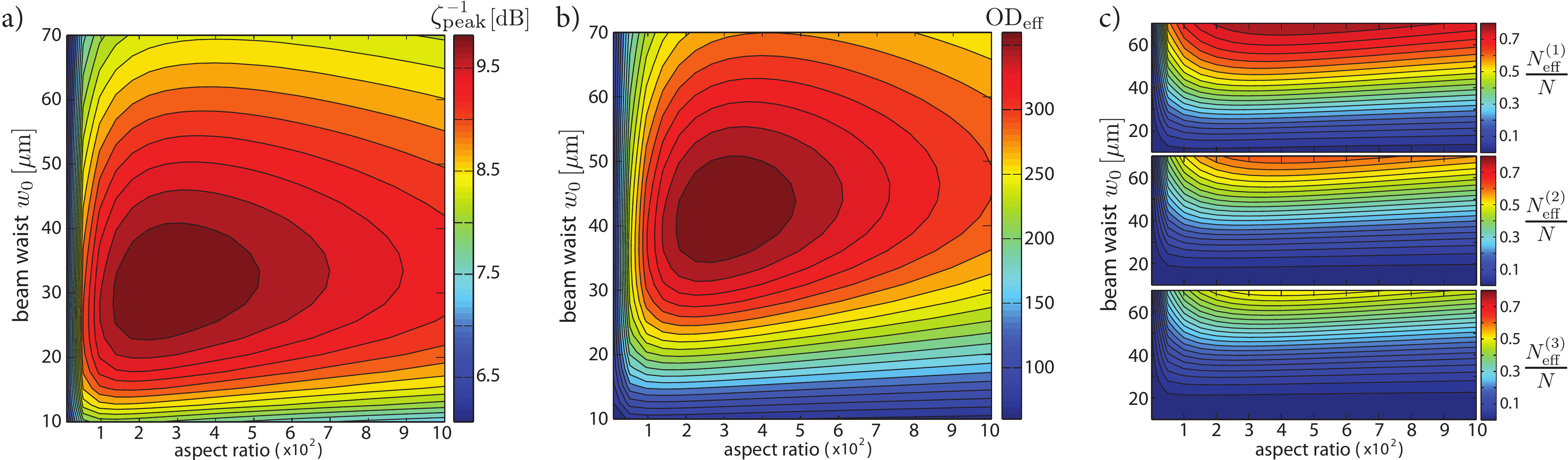}
       		 \caption{Squeezing for different cloud geometries with Gaussian atomic density distribution, \erf{Eq::AtomicDistribution}, and a fixed total atom number $N = 9.84 \times 10^6$.  a) Contours of peak squeezing, $\zeta^{-1}$ in dB, as a function of cloud aspect ratio and laser probe beam waist. b) Contours of the coupling strength, OD$_{\rm eff}$.  The difference between the optimal coupling strength and the resulting squeezing depends on the balance between coherent interactions and decoherence, characterized by different effective atom numbers,  $N_{\rm eff}^{(1)}$, $N_{\rm eff}^{(2)}$, and $N_{\rm eff}^{(3)}$, shown in (c). }  \label{Fig::SpinHalfScaling}
    		\end{figure*}

	\subsubsection{Optimizing geometry for fixed atom number}
	
	We gain further insight into the nature of the atom-light interface by keeping the atom number $N$ fixed and optimizing the cloud dimensions for peak squeezing.  We fix the peak density at $\eta_0 = 5 \times 10^{11}$ cm$^{-3}$, typical of dipole-trapped atoms, and  keep the total atom number constant, $N = 9.8 \times 10^6$, for a fixed cloud volume.  In Fig. \ref{Fig::SpinHalfScaling}(a), we plot contours of peak squeezing as a function of aspect ratio and beam waist.  The optimal peak squeezing, $\zeta_{\rm opt}^{-1} = 10.0$ dB, is found for AR $= 256$ at a beam waist of $w^{\rm opt}_0=31$ $\mu$m.  At the optimal geometry, the cloud length extends over several Rayleigh ranges, $\sigma_z/z^{\rm opt}_R = 2.42$, and the transverse width of the cloud is slightly larger than the beam waist, $\sigma_\perp/w^{\rm opt}_0 = 1.09$. 
	
	To further understand the region of peak squeezing, in Fig. \ref{Fig::SpinHalfScaling}(b), we plot contours of OD$_{\rm eff}$.  Comparison of Figs. \ref{Fig::SpinHalfScaling}(a-b) shows that the optimal peak squeezing occurs in a parameter region where OD$_{\rm eff}$ is high, as expected.  However, the optimal peak squeezing arises from a balance between high OD$_{\rm eff}$ with low noise injection into the spin wave variance and low decay of the mean spin.   Figure \ref{Fig::SpinHalfScaling}(c) shows the fraction of total atoms contributing to the mean spin, $N_{\rm eff}^{(1)}/N$, to the effective optical density, $N_{\rm eff}^{(2)}/ N$, and to the noise injection $N_{\rm eff}^{(3)}/ N$.  As the cloud becomes too long and narrow, there does not exist a beam waist that can address a sufficiently large number of atoms while keeping a high OD$_{\rm eff}$.  Said another way, when the cloud becomes too long, the diffraction of scattered light is too large to effectively mode match with the probe field, as seen in Fig. \ref{Fig::ModeMatching}(c).  Similarly, too small a waist leaves many atoms outside the Rayleigh range and too large a waist increases the beam area, thus decreasing OD$_{\rm eff}$, both manifestations of poor mode matching of the probe and the scattered field from the atom cloud.

	\begin{figure}[b]
    		\includegraphics[width=1\hsize]{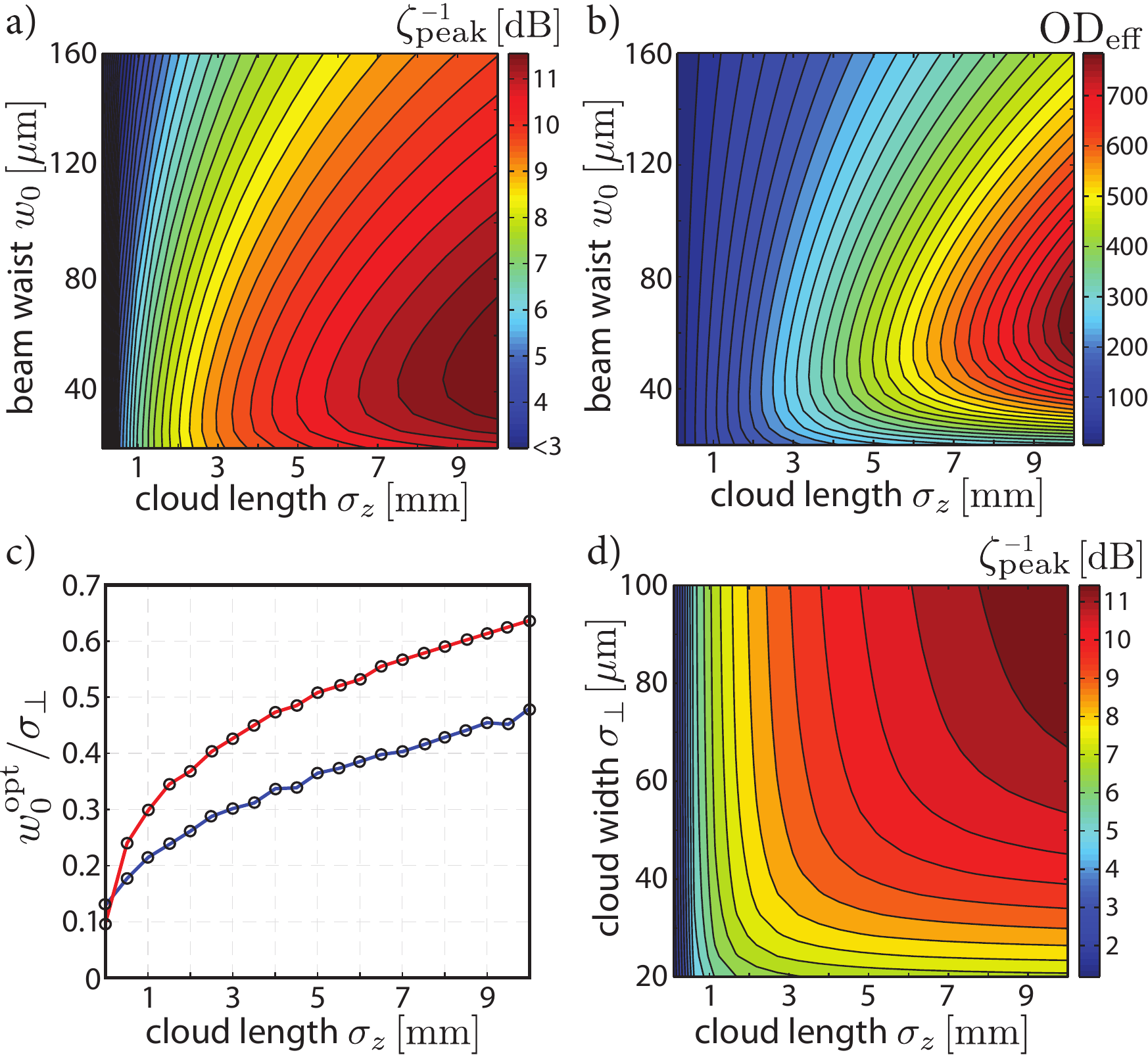}
       		 \caption{ Squeezing for a fixed peak density $\eta_0 = 5 \times 10^{11}$ cm$^{-3}$ and variable atom number that fills a dipole trap for cold atoms.  In a), b), and c) the transverse cloud width is fixed at $\sigma_\perp = 100$ $\mu$m and cloud length is taken to be variable.  a) Contours of peak squeezing, $\zeta^{-1}$ in dB.  b) Contours of OD$_{\rm eff}$.  c) Optimal beam waist for maximizing OD$_{\rm eff}$ (red) and for maximizing peak squeezing (blue).  For a given atomic geometry, the beam waist that optimizes the OD$_{\rm eff}$ is not the same as that which optimizes peak squeezing.  d) Peak squeezing as a function of cloud size for the optimal beam waist at each point.}  \label{Fig::Fixed_SigmaT}     	
	\end{figure}

	\subsubsection{Optimizing the beam waist for a fixed atomic cloud geometry}  

	With a better understanding of how cloud geometry influences decoherence, we study the optimization of squeezing in a situation typical of experiments with dipole-trapped cold atoms, where both the trap dimensions and beam waist $w_0$ can be tuned, the peak atomic density $\eta_0$ is fixed, and the total atom number is variable depending on the trap volume.  For each cloud geometry there exists a beam waist that maximizes OD$_\eff$.  This is seen in Fig. \ref{Fig::Fixed_SigmaT}(b) where contours of OD$_{\rm eff}$ are shown for a cloud with a fixed transverse width of $\sigma_\perp = 100$ $\mu$m as the cloud length $\sigma_z$ and beam waist $w_0$ are varied.

	Contours for peak squeezing are shown in Fig. \ref{Fig::Fixed_SigmaT}(a).  Comparison with \ref{Fig::Fixed_SigmaT}(b) demonstrates that for a given cloud geometry, the peak squeezing is achieved with a smaller beam waist than that which optimizes OD$_\eff$. This is seen most clearly in Fig. \ref{Fig::Fixed_SigmaT}(c), where we compare the optimal beam waist for maximizing OD$_\eff$ to the beam waist that maximizes peak squeezing. Optimal squeezing occurs at smaller beam waists where the region of the beam with greatest intensity, the Rayleigh range, is smaller. Because the scattering rate $\gamma_s(\mathbf{r})$ is proportional to the local intensity, atoms outside the Rayleigh range experience a decreased rate of optical pumping.  Although a smaller Rayleigh range implies a decreased OD$_\eff$ and $N_{\rm eff}^{(1)}$ as well, the reduction of the decoherence rate dominates in this regime. This is a direct analogy to  Sec. \ref{Sec::GeomEffects}, in which pencil-shaped clouds with higher mean spins were more robust to decay due to a large number of atoms farther away from the beam waist.  Finally, in Fig. \ref{Fig::Fixed_SigmaT}(d) we plot contours of peak squeezing for different geometries at the optimal beam waist for each point.

	\subsubsection{Relation to the symmetric one-dimensional model} \label{SubSec::1Dmodel}
	
	Spin squeezing by QND measurement is traditionally modeled using a one-dimensional description of the atom-light interface where the ensemble is symmetrically coupled to plane waves with no spatial variations \cite{PolzikRMP}.  When accounting only for squeezing due to collective scattering and QND measurement, the full three-dimensional system can be effectively described by such a model, with the symmetric OD replaced by OD$_\eff$. When decoherence from local diffuse scattering is included, however, such models become insufficient. In addition, a symmetric description does not account for the difference between the effective atom number contributing to the spin wave variance, $N_{\rm eff}^{(2)}$,  that contributing to the mean spin, $N_{\rm eff}^{(1)}$,  and that contributing to noise injection by spin flips, $N_{\rm eff}^{(3)}$.

	To better understand the limit in which we recover the simple symmetric description, consider a symmetric 1D model where an ensemble of spin-1/2 atoms is coupled to a uniform plane wave and scatters collectively into this mode and locally into diffuse modes.  In this case a single atom number suffices; every atom contributes equally to the optical density, to the mean spin, and to the injection of noise, $ N_{\rm eff}^{(1)} = N_{\rm eff}^{(2)} = N_{\rm eff}^{(3)}= N$.  The equation of motion for the spin wave variance \erf{Eq::SpinHalfVariance} becomes \cite{MadMol04, DeuJes09, TraDeu10, NorDeu12}
	\begin{align} \label{Eq::VarSpin1/2Symmetric}
		\frac{d}{dt}  \Delta F^2_z   & = - \kappa \big(  \Delta F^2_z \big)^2  - \frac{4}{9}\gamma_0\,  \Delta F^2_z + \frac{1 }{9}\gamma_0 \, N,
	\end{align}  
where $\gamma_0$ is the scattering rate and $\kappa = (\sigma_0/A)(4\gamma_0/9)$ is the measurement strength corresponding to the rate of scattering into the probe mode per atom, Eq. (\ref{Eq::MeasurementStrengthperAtom}).  This equation can be solved analytically, yielding\begin{align}
	 \Delta F^2_z(t) = \frac{N}{4} \, \frac{\sqrt{ \mbox{OD}+1} + \tanh \left[ \frac{(1+\text{OD})}{  \sqrt{\text{OD}}} \frac{2}{9} \gamma_0 t \right] }{\sqrt{\mbox{OD}+1} + \big(\frac{ \text{OD} }{2}+1\big)\tanh\left[\frac{(1+\text{OD})}{  \sqrt{\text{OD}}} \frac{2}{9} \gamma_0 t \right]}.
\end{align}
In the limit of short times, $ \gamma_0 t  \ll 1$, we recover the  expression for QND squeezing in the absence of decoherence, \erf{Eq::ShortTimeSolution},  $ \Delta F^2_z(t) \approx (N/4)[1+\xi(t)]^{-1}$, where $\xi (t) = \mbox{OD} \, \gamma_0 t/9$.  In the opposite limit of long times, $\gamma_0 t \rightarrow \infty$, and large optical density, OD $\gg 1$, we find the expected scaling $ \Delta F^2_z(t\rightarrow \infty) \propto \mbox{OD}^{-1/2}$\cite{PolzikRMP}. 

	We can compare the symmetric 1D model to a limiting case of the full three-dimensional model developed here.  When the transverse extent of the cloud is much smaller than the beam waist and the longitudinal extent is well within the Rayleigh range, then spatial variations of the field across the cloud are minimal and $\beta_{00}(\mathbf{r}_i) \rightarrow 1$.  Although this limiting case replicates the squeezing expected from the symmetric 1D model, it is in fact far from a single-mode description.  As discussed in Sec. \ref{Sec::QuantumTheory}, this geometry radiates paraxial light into \emph{many} of transverse modes defined relative to the beam, and the associated spin waves couple together through diffuse scattering, \erf{Eq::ProjectedSpin1/2VarianceEOM}.   After numerically solving these coupled equations according to the procedure outlined above, we recover the same results Eq. (\ref{Eq::VarSpin1/2Symmetric}), as is manifest from Eq. (\ref{Eq::SpinHalfVariance}) in the limit that $\beta_{00}(\mbf{r}_i) \rightarrow 1$ over the extent of the atom cloud.

	\begin{figure}
    		\includegraphics[width=.9\hsize]{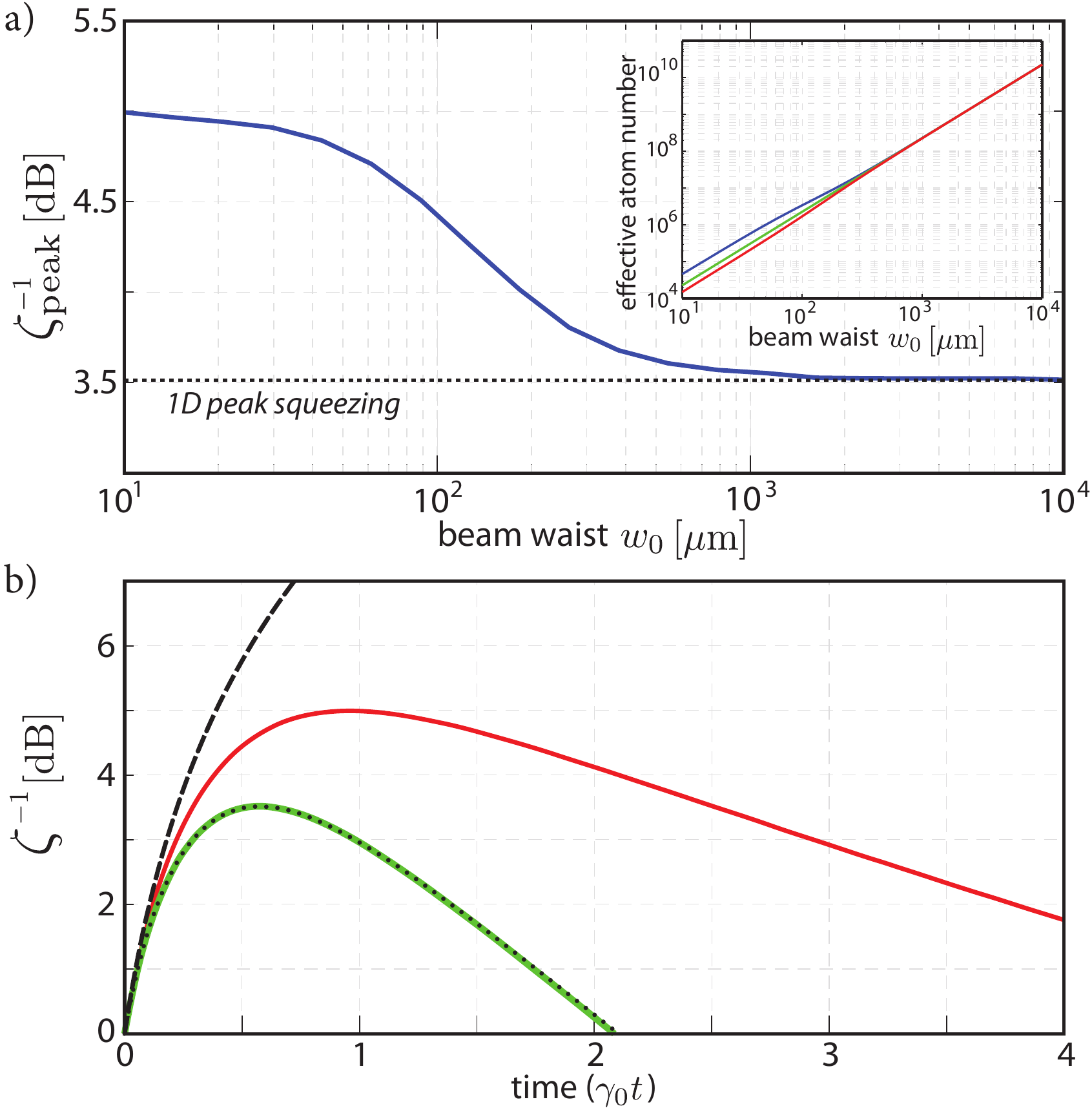} 
       		 \caption{ Comparison between the symmetric 1D model and the three-dimensional spin wave model.  a) Peak squeezing, $\zeta^{-1}$ in dB, for a spherical cloud with OD$_{\rm eff} = 50$ as the beam waist is increased.  Inset shows the convergence of $N_{\rm eff}^{(1)}$ (blue), $N_{\rm eff}^{(2)}$ (green), and $N_{\rm eff}^{(3)}$ (red) as $w_0$ increases.  b) Comparison of squeezing dynamics for the extremal waists from a): the smallest $w_0 = 10$ $\mu$m (red line) and the largest $w_0 = 10^4$ $\mu$m (green line).  For comparison, the symmetric 1D case using \erf{Eq::VarSpin1/2Symmetric} is plotted with decoherence (black dotted line) and without (black dashed line).} \label{Fig::PlaneWaveComp}
	\end{figure}

	We investigate this limit numerically in Fig. \ref{Fig::PlaneWaveComp} for a spherical cloud ($\sigma_\perp, \sigma_z = 100$ $\mu$m) probed by beams of increasing waist $w_0$.  In each case OD$_{\rm eff} = 50$, such that in the absence of decoherence the different geometries would achieve identical squeezing.  In Fig. \ref{Fig::PlaneWaveComp}(a) we see that as the beam waist is increased, the peak squeezing approaches that of the symmetric 1D model.  The inset shows the convergence of the effective atom numbers as the beam waist increases.  Figure \ref{Fig::PlaneWaveComp}(b) shows the dynamics of the squeezing parameter $\zeta^{-1}(t)$ for the spherical cloud at both extremes in Fig. \ref{Fig::PlaneWaveComp}(a).  For comparison, the squeezing parameter for the symmetric 1D model is plotted both with and without decoherence.  The difference between the models is substantial -- the optimal peak squeezing for the symmetric 1D model and full model are $\zeta^{-1}_{\rm peak} = \{ 3.52 \text{ dB}, 4.99 \text{ dB} \}$, respectively.  This difference can be understood in terms of the effective atom numbers.  The advantage for spin squeezing in the three-dimensional model comes from the fact that $N_{\rm eff}^{(1)} \geq  N_{\rm eff}^{(2)} \geq N_{\rm eff}^{(3)}$ due to different dependence on the spatial weightings $\beta_{00}(\mathbf{r})$, while for the symmetric 1D case they are equal.  For the three-dimensional model, not only can the effective number of atoms contributing to the noise injection be smaller than that contributing to the OD$_{\rm eff}$, but the effective number of atoms contributing to the mean spin, and thus the signal, is larger than both.  Inspecting Fig. \ref{Fig::SpinHalfCompareSqueezing}(d) we see an additional advantage for the three-dimensional model -- when geometry is properly chosen, the mean spin decays at a much reduced rate.

\subsubsection{Spin $f>1/2$ atoms}	
\begin{figure}[b]
    		\includegraphics[width=1\hsize]{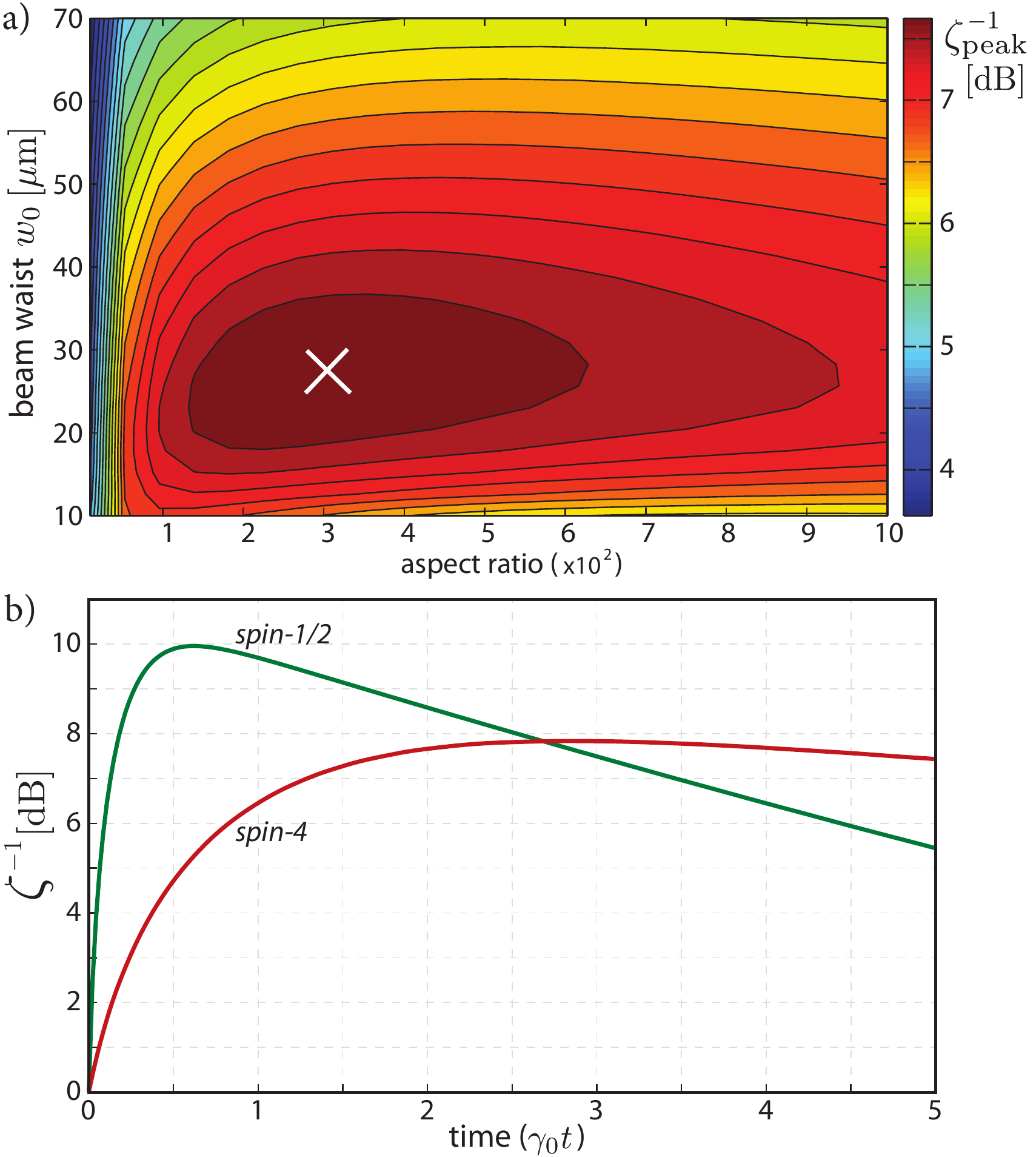}
       		 \caption{  Squeezing of a spin-4 ensemble.  a) Contours of peak squeezing for an ensemble of spin-4 atoms for the same fixed atom number as in Fig. \ref{Fig::SpinHalfScaling} as function of beam waist and cloud aspect ratio.  The optimal squeezing, $\zeta^{-1} = 7.8$ dB, is achieved for a geometry AR = 300, $w_0 = 28$ $\mu$m. b) Dynamics of the inverse of the squeezing parameter for the optimal geometry in  a) for spin-4 (red line) and for spin-1/2 (green line) } \label{Fig::Spin4Scaling}
	\end{figure}

 We present here some initial results that illustrate the differences between spin-1/2 and larger spin ensembles. Consider an ensemble of $^{133}$Cs atoms prepared in the $f=4$ ground hyperfine manifold. Figure \ref{Fig::Spin4Scaling}(a) shows contours of peak squeezing as a function of aspect ratio using the atomic density given by \erf{Eq::AtomicDistribution}.  Note that the peak squeezing is substantially smaller than the peak squeezing for spin-1/2 (compare to Fig. \ref{Fig::SpinHalfScaling}(a)) and the optimal aspect ratio and beam waist are different for spin-4 alkali atoms than for spin-1/2 atoms. This can be attributed to a reduction of the coupling strength $\xi$ that is not compensated by an equal reduction in the decoherence rate.  In principle, the coupling strength can be increased by internal spin control of the $f=4$ hyperfine spin \cite{NorDeu12}.  Figure \ref{Fig::Spin4Scaling}(b) compares the squeezing dynamics for spin-4 and spin-1/2 ensembles at the same geometry and illustrates not only the disparity in peak squeezing but also in the time at which it occurs.  These effects arise from the subtle interplay between the rates of depolarization and injected noise due to spin flips when applied to the spatial modes of larger spin ensembles.

\section{Summary and Outlook} \label{Sec::Conclusion}

The entangling power of the quantum interface between photons and an ensemble of cold atoms is at the heart of a variety of quantum information processing tasks.  When considering extended atomic clouds in dipole traps, one must consider a full three-dimensional description of the electromagnetic modes and atomic density distribution in order to optimize this entangling power.  Inhomogeneous coupling between atoms and photons is essential to maximize the strength of the quantum interface, but this comes with substantial complexity in the theoretical description.  The model presented in this work addressed these issues and examined regimes of optimal performance.

We have studied the strength of the atom-photon interface in a traveling wave configuration in the context of spin squeezing via QND measurement.  We developed a description in terms of quantized paraxial modes of the field in order to model the inhomogeneous atom-light coupling across the atomic ensemble, which leads to two distinct effects.  First, the collective coupling describes a generalization of the Faraday interaction that entangles the quantized Stokes vector of the laser field with a spin wave defined by the weighted ensemble of atoms that indistinguishably radiates into the  mode of the probe.  The spin wave that is squeezed is defined by the probe mode we measure in a balanced polarimeter.  Second, diffusely scattered photons lead to optical pumping and decoherence across the ensemble at a rate proportional to the local probe intensity. The delicate balance of these two effects favors certain geometries for spin squeezing.

We have numerically investigated the ultimate limits to spin squeezing based on a stochastic master equation, including the effects of QND measurement backaction and decoherence by photon scattering into unmeasured modes.  Unlike the usual one-dimensional description in which the amount of squeezing is set by a single parameter, the optical density, we find that due to  inhomogeneous coupling, multiple parameters are required. Of particular importance in a metrological context are the mean collective spin and  the projection noise variance, determined by effective atom numbers $N^{(1)}_\eff$ and $N^{(2)}_\eff$ respectively.  Optimal geometries maximize the effective optical density, OD$_\eff$, proportional to $N^{(2)}_\eff$, while minimizing the depolarization of  $N^{(1)}_\eff$  and injected noise into the spin wave by optical pumping.  We found that optimal mode matching occurs for geometries where a large number of atoms are addressed by a beam with a small transverse area, yielding a high OD$_{\rm eff}$, but also where the depolarization rate due to optical pumping is relatively small.  This geometry corresponds to a longitudinally extended, pencil-shaped cloud, with  a probe beam chosen to optimize the tradeoffs between OD$_\eff$ and decoherence.  Such a geometry is far from the regime describing squeezing of a symmetric atomic spin ensemble, as is typically assumed.  One recovers the symmetric description only for ensembles confined with extents much smaller than the beam waist and Rayleigh range, which yield much smaller OD$_\eff$.  

While the three-dimensional model developed in this work was specifically tailored to study the problem of spin squeezing by QND measurement, it can be extended to other protocols involving the quantum interface between photons and free-space atomic ensembles.
Mode-matching and spatial effects are important for other spin squeezing protocols including the double-pass counter-twisting interaction \cite{TakTak05,TraDeu10} or the recently proposed planar squeezing protocol \cite{PueMit13}. Understanding spatial effects in order to identify regimes of strong coupling is also essential for quantum memories and repeaters in free-space atomic ensembles.  In addition, recent work has considered ensembles of higher-spin alkali atoms, in which control over the rich internal hyperfine structure can enhance the entangling strength of the atom-light interface \cite{NorDeu12}. Quantifying the gains achievable though such control techniques requires a realistic description of the inhomogeneous interaction between light and atoms. Finally, a multimode description of the entangling Hamiltonian offers the possibility to exploit spatial modes and their associated spin waves as a resource.   The creation of entanglement between spin waves could lead to novel states with potential application in continuous variable quantum computation and communication \cite{CV2007}.

\emph{Acknowledgments}.  We thank Robert Cook for helpful discussions and insights.  This work was supported by NSF Grants PHY-0969391, 0969997, 1212445, 1306171, 1307520, and AFOSR Grant No. Y600242.

\bibliography{3DInterfaceBib.bib}

\begin{thebibliography}{55}%
\makeatletter
\providecommand \@ifxundefined [1]{%
 \@ifx{#1\undefined}
}%
\providecommand \@ifnum [1]{%
 \ifnum #1\expandafter \@firstoftwo
 \else \expandafter \@secondoftwo
 \fi
}%
\providecommand \@ifx [1]{%
 \ifx #1\expandafter \@firstoftwo
 \else \expandafter \@secondoftwo
 \fi
}%
\providecommand \natexlab [1]{#1}%
\providecommand \enquote  [1]{``#1''}%
\providecommand \bibnamefont  [1]{#1}%
\providecommand \bibfnamefont [1]{#1}%
\providecommand \citenamefont [1]{#1}%
\providecommand \href@noop [0]{\@secondoftwo}%
\providecommand \href [0]{\begingroup \@sanitize@url \@href}%
\providecommand \@href[1]{\@@startlink{#1}\@@href}%
\providecommand \@@href[1]{\endgroup#1\@@endlink}%
\providecommand \@sanitize@url [0]{\catcode `\\12\catcode `\$12\catcode
  `\&12\catcode `\#12\catcode `\^12\catcode `\_12\catcode `\%12\relax}%
\providecommand \@@startlink[1]{}%
\providecommand \@@endlink[0]{}%
\providecommand \url  [0]{\begingroup\@sanitize@url \@url }%
\providecommand \@url [1]{\endgroup\@href {#1}{\urlprefix }}%
\providecommand \urlprefix  [0]{URL }%
\providecommand \Eprint [0]{\href }%
\providecommand \doibase [0]{http://dx.doi.org/}%
\providecommand \selectlanguage [0]{\@gobble}%
\providecommand \bibinfo  [0]{\@secondoftwo}%
\providecommand \bibfield  [0]{\@secondoftwo}%
\providecommand \translation [1]{[#1]}%
\providecommand \BibitemOpen [0]{}%
\providecommand \bibitemStop [0]{}%
\providecommand \bibitemNoStop [0]{.\EOS\space}%
\providecommand \EOS [0]{\spacefactor3000\relax}%
\providecommand \BibitemShut  [1]{\csname bibitem#1\endcsname}%
\let\auto@bib@innerbib\@empty
\bibitem [{\citenamefont {Fleischhauer}\ and\ \citenamefont
  {Lukin}(2002)}]{FLukinMemory}%
  \BibitemOpen
  \bibfield  {author} {\bibinfo {author} {\bibfnamefont {M.}~\bibnamefont
  {Fleischhauer}}\ and\ \bibinfo {author} {\bibfnamefont {M.~D.}\ \bibnamefont
  {Lukin}},\ }\href {http://link.aps.org/doi/10.1103/PhysRevA.65.022314}
  {\bibfield  {journal} {\bibinfo  {journal} {Phys. Rev. A}\ }\textbf {\bibinfo
  {volume} {65}},\ \bibinfo {pages} {022314} (\bibinfo {year}
  {2002})}\BibitemShut {NoStop}%
\bibitem [{\citenamefont {Julsgaard}\ \emph {et~al.}(2004)\citenamefont
  {Julsgaard}, \citenamefont {Sherson}, \citenamefont {Cirac}, \citenamefont
  {Fiurasek},\ and\ \citenamefont {Polzik}}]{Polzik2004}%
  \BibitemOpen
  \bibfield  {author} {\bibinfo {author} {\bibfnamefont {B.}~\bibnamefont
  {Julsgaard}}, \bibinfo {author} {\bibfnamefont {J.}~\bibnamefont {Sherson}},
  \bibinfo {author} {\bibfnamefont {J.~I.}\ \bibnamefont {Cirac}}, \bibinfo
  {author} {\bibfnamefont {J.}~\bibnamefont {Fiurasek}}, \ and\ \bibinfo
  {author} {\bibfnamefont {E.~S.}\ \bibnamefont {Polzik}},\ }\href
  {http://dx.doi.org/10.1038/nature03064} {\bibfield  {journal} {\bibinfo
  {journal} {Nature}\ }\textbf {\bibinfo {volume} {432}},\ \bibinfo {pages}
  {482} (\bibinfo {year} {2004})}\BibitemShut {NoStop}%
\bibitem [{\citenamefont {Choi}\ \emph {et~al.}(2008)\citenamefont {Choi},
  \citenamefont {Deng}, \citenamefont {Laurat},\ and\ \citenamefont
  {Kimble}}]{ChoiKimbleMemory}%
  \BibitemOpen
  \bibfield  {author} {\bibinfo {author} {\bibfnamefont {K.~S.}\ \bibnamefont
  {Choi}}, \bibinfo {author} {\bibfnamefont {H.}~\bibnamefont {Deng}}, \bibinfo
  {author} {\bibfnamefont {J.}~\bibnamefont {Laurat}}, \ and\ \bibinfo {author}
  {\bibfnamefont {H.~J.}\ \bibnamefont {Kimble}},\ }\href
  {http://dx.doi.org/10.1038/nature06670} {\bibfield  {journal} {\bibinfo
  {journal} {Nature}\ }\textbf {\bibinfo {volume} {452}},\ \bibinfo {pages}
  {67} (\bibinfo {year} {2008})}\BibitemShut {NoStop}%
\bibitem [{\citenamefont {Duan}\ \emph {et~al.}(2001)\citenamefont {Duan},
  \citenamefont {Lukin}, \citenamefont {Cirac},\ and\ \citenamefont
  {Zoller}}]{DLCZ}%
  \BibitemOpen
  \bibfield  {author} {\bibinfo {author} {\bibfnamefont {L.~M.}\ \bibnamefont
  {Duan}}, \bibinfo {author} {\bibfnamefont {M.~D.}\ \bibnamefont {Lukin}},
  \bibinfo {author} {\bibfnamefont {J.}~\bibnamefont {Cirac}}, \ and\ \bibinfo
  {author} {\bibfnamefont {P.}~\bibnamefont {Zoller}},\ }\href
  {http://dx.doi.org/10.1038/35106500} {\bibfield  {journal} {\bibinfo
  {journal} {Nature}\ }\textbf {\bibinfo {volume} {414}},\ \bibinfo {pages}
  {413} (\bibinfo {year} {2001})}\BibitemShut {NoStop}%
\bibitem [{\citenamefont {Matsukevich}\ and\ \citenamefont
  {Kuzmich}(2004)}]{MatKuzNetwork}%
  \BibitemOpen
  \bibfield  {author} {\bibinfo {author} {\bibfnamefont {D.~N.}\ \bibnamefont
  {Matsukevich}}\ and\ \bibinfo {author} {\bibfnamefont {A.}~\bibnamefont
  {Kuzmich}},\ }\href {http://www.sciencemag.org/content/306/5696/663.abstract}
  {\bibfield  {journal} {\bibinfo  {journal} {Science}\ }\textbf {\bibinfo
  {volume} {306}},\ \bibinfo {pages} {663} (\bibinfo {year}
  {2004})}\BibitemShut {NoStop}%
\bibitem [{\citenamefont {Braunstein}\ and\ \citenamefont {van
  Loock}(2005)}]{BraunReview}%
  \BibitemOpen
  \bibfield  {author} {\bibinfo {author} {\bibfnamefont {S.~L.}\ \bibnamefont
  {Braunstein}}\ and\ \bibinfo {author} {\bibfnamefont {P.}~\bibnamefont {van
  Loock}},\ }\href {http://link.aps.org/doi/10.1103/RevModPhys.77.513}
  {\bibfield  {journal} {\bibinfo  {journal} {Rev. Mod. Phys.}\ }\textbf
  {\bibinfo {volume} {77}},\ \bibinfo {pages} {513} (\bibinfo {year}
  {2005})}\BibitemShut {NoStop}%
\bibitem [{\citenamefont {Appel}\ \emph {et~al.}(2009)\citenamefont {Appel},
  \citenamefont {Windpassinger}, \citenamefont {Oblak}, \citenamefont {Hoff},
  \citenamefont {Kjaergaard},\ and\ \citenamefont {Polzik}}]{AppelPolzikMet}%
  \BibitemOpen
  \bibfield  {author} {\bibinfo {author} {\bibfnamefont {J.}~\bibnamefont
  {Appel}}, \bibinfo {author} {\bibfnamefont {P.~J.}\ \bibnamefont
  {Windpassinger}}, \bibinfo {author} {\bibfnamefont {D.}~\bibnamefont
  {Oblak}}, \bibinfo {author} {\bibfnamefont {U.~B.}\ \bibnamefont {Hoff}},
  \bibinfo {author} {\bibfnamefont {N.}~\bibnamefont {Kjaergaard}}, \ and\
  \bibinfo {author} {\bibfnamefont {E.~S.}\ \bibnamefont {Polzik}},\ }\href
  {http://www.pnas.org/content/106/27/10960.abstract} {\bibfield  {journal}
  {\bibinfo  {journal} {Proc. Natl. Acad. Sci. U.S.A.}\ }\textbf {\bibinfo
  {volume} {106}},\ \bibinfo {pages} {10960} (\bibinfo {year}
  {2009})}\BibitemShut {NoStop}%
\bibitem [{\citenamefont {Leroux}\ \emph {et~al.}(2010)\citenamefont {Leroux},
  \citenamefont {Schleier-Smith},\ and\ \citenamefont {Vuletic}}]{VulMet}%
  \BibitemOpen
  \bibfield  {author} {\bibinfo {author} {\bibfnamefont {I.~D.}\ \bibnamefont
  {Leroux}}, \bibinfo {author} {\bibfnamefont {M.~H.}\ \bibnamefont
  {Schleier-Smith}}, \ and\ \bibinfo {author} {\bibfnamefont {V.}~\bibnamefont
  {Vuletic}},\ }\href {http://link.aps.org/doi/10.1103/PhysRevLett.104.073602}
  {\bibfield  {journal} {\bibinfo  {journal} {Phys. Rev. Lett}\ }\textbf
  {\bibinfo {volume} {104}},\ \bibinfo {pages} {073602} (\bibinfo {year}
  {2010})}\BibitemShut {NoStop}%
\bibitem [{\citenamefont {Miller}\ \emph {et~al.}(2005)\citenamefont {Miller},
  \citenamefont {Northup}, \citenamefont {Boca}, \citenamefont {Boozer},\ and\
  \citenamefont {Kimble}}]{Kimble2005}%
  \BibitemOpen
  \bibfield  {author} {\bibinfo {author} {\bibfnamefont {R.}~\bibnamefont
  {Miller}}, \bibinfo {author} {\bibfnamefont {K.}~\bibnamefont {Northup},
  \bibfnamefont {T.E.~Birnbaum}}, \bibinfo {author} {\bibfnamefont
  {A.}~\bibnamefont {Boca}}, \bibinfo {author} {\bibfnamefont {A.}~\bibnamefont
  {Boozer}}, \ and\ \bibinfo {author} {\bibfnamefont {H.}~\bibnamefont
  {Kimble}},\ }\href {http://stacks.iop.org/0953-4075/38/i=9/a=007} {\bibfield
  {journal} {\bibinfo  {journal} {J. Phys. B: At. Mol. Opt. Phys.}\ }\textbf
  {\bibinfo {volume} {38}},\ \bibinfo {pages} {551} (\bibinfo {year}
  {2005})}\BibitemShut {NoStop}%
\bibitem [{\citenamefont {Chen}\ \emph {et~al.}(2011)\citenamefont {Chen},
  \citenamefont {Bohnet}, \citenamefont {Sankar}, \citenamefont {Dai},\ and\
  \citenamefont {Thompson}}]{Thompson2011}%
  \BibitemOpen
  \bibfield  {author} {\bibinfo {author} {\bibfnamefont {Z.}~\bibnamefont
  {Chen}}, \bibinfo {author} {\bibfnamefont {J.~G.}\ \bibnamefont {Bohnet}},
  \bibinfo {author} {\bibfnamefont {S.~R.}\ \bibnamefont {Sankar}}, \bibinfo
  {author} {\bibfnamefont {J.}~\bibnamefont {Dai}}, \ and\ \bibinfo {author}
  {\bibfnamefont {J.~K.}\ \bibnamefont {Thompson}},\ }\href {\doibase
  10.1103/PhysRevLett.106.133601} {\bibfield  {journal} {\bibinfo  {journal}
  {Phys. Rev. Lett.}\ }\textbf {\bibinfo {volume} {106}},\ \bibinfo {pages}
  {133601} (\bibinfo {year} {2011})}\BibitemShut {NoStop}%
\bibitem [{\citenamefont {Vetsch}\ \emph {et~al.}(2010)\citenamefont {Vetsch},
  \citenamefont {Reitz}, \citenamefont {Sagu\'e}, \citenamefont {Schmidt},
  \citenamefont {Dawkins},\ and\ \citenamefont
  {Rauschenbeutel}}]{Rauschenbeutel2010}%
  \BibitemOpen
  \bibfield  {author} {\bibinfo {author} {\bibfnamefont {E.}~\bibnamefont
  {Vetsch}}, \bibinfo {author} {\bibfnamefont {D.}~\bibnamefont {Reitz}},
  \bibinfo {author} {\bibfnamefont {G.}~\bibnamefont {Sagu\'e}}, \bibinfo
  {author} {\bibfnamefont {R.}~\bibnamefont {Schmidt}}, \bibinfo {author}
  {\bibfnamefont {S.~T.}\ \bibnamefont {Dawkins}}, \ and\ \bibinfo {author}
  {\bibfnamefont {A.}~\bibnamefont {Rauschenbeutel}},\ }\href {\doibase
  10.1103/PhysRevLett.104.203603} {\bibfield  {journal} {\bibinfo  {journal}
  {Phys. Rev. Lett.}\ }\textbf {\bibinfo {volume} {104}},\ \bibinfo {pages}
  {203603} (\bibinfo {year} {2010})}\BibitemShut {NoStop}%
\bibitem [{\citenamefont {Bose}\ \emph {et~al.}(2012)\citenamefont {Bose},
  \citenamefont {Sridharan}, \citenamefont {Kim}, \citenamefont {Solomon},\
  and\ \citenamefont {Waks}}]{Waks2012}%
  \BibitemOpen
  \bibfield  {author} {\bibinfo {author} {\bibfnamefont {R.}~\bibnamefont
  {Bose}}, \bibinfo {author} {\bibfnamefont {D.}~\bibnamefont {Sridharan}},
  \bibinfo {author} {\bibfnamefont {H.}~\bibnamefont {Kim}}, \bibinfo {author}
  {\bibfnamefont {G.~S.}\ \bibnamefont {Solomon}}, \ and\ \bibinfo {author}
  {\bibfnamefont {E.}~\bibnamefont {Waks}},\ }\href {\doibase
  10.1103/PhysRevLett.108.227402} {\bibfield  {journal} {\bibinfo  {journal}
  {Phys. Rev. Lett.}\ }\textbf {\bibinfo {volume} {108}},\ \bibinfo {pages}
  {227402} (\bibinfo {year} {2012})}\BibitemShut {NoStop}%
\bibitem [{\citenamefont {Hung}\ \emph {et~al.}(2013)\citenamefont {Hung},
  \citenamefont {Meenehan}, \citenamefont {Chang}, \citenamefont {Painter},\
  and\ \citenamefont {Kimble}}]{Kimble2013}%
  \BibitemOpen
  \bibfield  {author} {\bibinfo {author} {\bibfnamefont {C.-L.}\ \bibnamefont
  {Hung}}, \bibinfo {author} {\bibfnamefont {S.}~\bibnamefont {Meenehan}},
  \bibinfo {author} {\bibfnamefont {D.}~\bibnamefont {Chang}}, \bibinfo
  {author} {\bibfnamefont {O.}~\bibnamefont {Painter}}, \ and\ \bibinfo
  {author} {\bibfnamefont {H.}~\bibnamefont {Kimble}},\ }\href
  {http://stacks.iop.org/1367-2630/15/i=8/a=083026} {\bibfield  {journal}
  {\bibinfo  {journal} {N. J. Phys.}\ }\textbf {\bibinfo {volume} {15}},\
  \bibinfo {pages} {083026} (\bibinfo {year} {2013})}\BibitemShut {NoStop}%
\bibitem [{\citenamefont {Tanji-Suzuki}\ \emph {et~al.}(2011)\citenamefont
  {Tanji-Suzuki}, \citenamefont {Leroux}, \citenamefont {Schleier-Smith},
  \citenamefont {Cetina}, \citenamefont {Grier}, \citenamefont {Simon},\ and\
  \citenamefont {Vuleti{\'c}}}]{Vuletic2011}%
  \BibitemOpen
  \bibfield  {author} {\bibinfo {author} {\bibfnamefont {H.}~\bibnamefont
  {Tanji-Suzuki}}, \bibinfo {author} {\bibfnamefont {I.~D.}\ \bibnamefont
  {Leroux}}, \bibinfo {author} {\bibfnamefont {M.~H.}\ \bibnamefont
  {Schleier-Smith}}, \bibinfo {author} {\bibfnamefont {M.}~\bibnamefont
  {Cetina}}, \bibinfo {author} {\bibfnamefont {A.~T.}\ \bibnamefont {Grier}},
  \bibinfo {author} {\bibfnamefont {J.}~\bibnamefont {Simon}}, \ and\ \bibinfo
  {author} {\bibfnamefont {V.}~\bibnamefont {Vuleti{\'c}}},\ }in\ \href
  {\doibase 10.1016/B978-0-12-385508-4.00004-8} {\emph {\bibinfo {booktitle}
  {Advances in Atomic, Molecular, and Optical Physics}}},\ \bibinfo {series}
  {Advances In Atomic, Molecular, and Optical Physics}, Vol.~\bibinfo {volume}
  {60},\ \bibinfo {editor} {edited by\ \bibinfo {editor} {\bibfnamefont
  {P.~B.}\ \bibnamefont {E.~Arimondo}}\ and\ \bibinfo {editor} {\bibfnamefont
  {C.}~\bibnamefont {Lin}}}\ (\bibinfo  {publisher} {Academic Press},\ \bibinfo
  {year} {2011})\ pp.\ \bibinfo {pages} {201 -- 237}\BibitemShut {NoStop}%
\bibitem [{\citenamefont {Bienaim\'{e}}\ \emph {et~al.}(2013)\citenamefont
  {Bienaim\'{e}}, \citenamefont {Bachelard}, \citenamefont {Piovella},\ and\
  \citenamefont {Kaiser}}]{Kaiser2013}%
  \BibitemOpen
  \bibfield  {author} {\bibinfo {author} {\bibfnamefont {T.}~\bibnamefont
  {Bienaim\'{e}}}, \bibinfo {author} {\bibfnamefont {R.}~\bibnamefont
  {Bachelard}}, \bibinfo {author} {\bibfnamefont {N.}~\bibnamefont {Piovella}},
  \ and\ \bibinfo {author} {\bibfnamefont {R.}~\bibnamefont {Kaiser}},\ }\href
  {http://dx.doi.org/10.1002/prop.201200089} {\bibfield  {journal} {\bibinfo
  {journal} {Fortschr. Phys.}\ }\textbf {\bibinfo {volume} {61}},\ \bibinfo
  {pages} {377} (\bibinfo {year} {2013})}\BibitemShut {NoStop}%
\bibitem [{\citenamefont {Kuzmich}\ \emph {et~al.}(1999)\citenamefont
  {Kuzmich}, \citenamefont {Mandel}, \citenamefont {Janis}, \citenamefont
  {Young}, \citenamefont {Ejnisman},\ and\ \citenamefont
  {Bigelow}}]{Kuzmich1999}%
  \BibitemOpen
  \bibfield  {author} {\bibinfo {author} {\bibfnamefont {A.}~\bibnamefont
  {Kuzmich}}, \bibinfo {author} {\bibfnamefont {L.}~\bibnamefont {Mandel}},
  \bibinfo {author} {\bibfnamefont {J.}~\bibnamefont {Janis}}, \bibinfo
  {author} {\bibfnamefont {Y.~E.}\ \bibnamefont {Young}}, \bibinfo {author}
  {\bibfnamefont {R.}~\bibnamefont {Ejnisman}}, \ and\ \bibinfo {author}
  {\bibfnamefont {N.~P.}\ \bibnamefont {Bigelow}},\ }\href {\doibase
  10.1103/PhysRevA.60.2346} {\bibfield  {journal} {\bibinfo  {journal} {Phys.
  Rev. A}\ }\textbf {\bibinfo {volume} {60}},\ \bibinfo {pages} {2346}
  (\bibinfo {year} {1999})}\BibitemShut {NoStop}%
\bibitem [{\citenamefont {Julsgaard}\ \emph {et~al.}(2001)\citenamefont
  {Julsgaard}, \citenamefont {Kozhekin},\ and\ \citenamefont
  {Polzik}}]{Polzik2001}%
  \BibitemOpen
  \bibfield  {author} {\bibinfo {author} {\bibfnamefont {B.}~\bibnamefont
  {Julsgaard}}, \bibinfo {author} {\bibfnamefont {A.}~\bibnamefont {Kozhekin}},
  \ and\ \bibinfo {author} {\bibfnamefont {E.~S.}\ \bibnamefont {Polzik}},\
  }\href {http://dx.doi.org/10.1038/35096524} {\bibfield  {journal} {\bibinfo
  {journal} {Nature}\ }\textbf {\bibinfo {volume} {413}},\ \bibinfo {pages}
  {400} (\bibinfo {year} {2001})}\BibitemShut {NoStop}%
\bibitem [{\citenamefont {Kaminski}\ \emph {et~al.}(2012)\citenamefont
  {Kaminski}, \citenamefont {Kampel}, \citenamefont {Steenstrup}, \citenamefont
  {Griesmaier}, \citenamefont {Polzik},\ and\ \citenamefont
  {M\"uller}}]{KamMul12}%
  \BibitemOpen
  \bibfield  {author} {\bibinfo {author} {\bibfnamefont {F.}~\bibnamefont
  {Kaminski}}, \bibinfo {author} {\bibfnamefont {N.~S.}\ \bibnamefont
  {Kampel}}, \bibinfo {author} {\bibfnamefont {M.~P.~H.}\ \bibnamefont
  {Steenstrup}}, \bibinfo {author} {\bibfnamefont {A.}~\bibnamefont
  {Griesmaier}}, \bibinfo {author} {\bibfnamefont {E.~S.}\ \bibnamefont
  {Polzik}}, \ and\ \bibinfo {author} {\bibfnamefont {J.~H.}\ \bibnamefont
  {M\"uller}},\ }\href {http://dx.doi.org/10.1140/epjd/e2012-30038-0}
  {\bibfield  {journal} {\bibinfo  {journal} {Eur. Phys. J. D}\ }\textbf
  {\bibinfo {volume} {66}},\ \bibinfo {pages} {227} (\bibinfo {year}
  {2012})}\BibitemShut {NoStop}%
\bibitem [{\citenamefont {Koschorreck}\ \emph
  {et~al.}(2010{\natexlab{a}})\citenamefont {Koschorreck}, \citenamefont
  {Napolitano}, \citenamefont {Dubost},\ and\ \citenamefont
  {Mitchell}}]{KosMit10}%
  \BibitemOpen
  \bibfield  {author} {\bibinfo {author} {\bibfnamefont {M.}~\bibnamefont
  {Koschorreck}}, \bibinfo {author} {\bibfnamefont {M.}~\bibnamefont
  {Napolitano}}, \bibinfo {author} {\bibfnamefont {B.}~\bibnamefont {Dubost}},
  \ and\ \bibinfo {author} {\bibfnamefont {M.~W.}\ \bibnamefont {Mitchell}},\
  }\href {\doibase 10.1103/PhysRevLett.105.093602} {\bibfield  {journal}
  {\bibinfo  {journal} {Phys. Rev. Lett.}\ }\textbf {\bibinfo {volume} {105}},\
  \bibinfo {pages} {093602} (\bibinfo {year} {2010}{\natexlab{a}})}\BibitemShut
  {NoStop}%
\bibitem [{\citenamefont {S\o{}rensen}\ and\ \citenamefont
  {S\o{}rensen}(2008)}]{SorSor08}%
  \BibitemOpen
  \bibfield  {author} {\bibinfo {author} {\bibfnamefont {M.~W.}\ \bibnamefont
  {S\o{}rensen}}\ and\ \bibinfo {author} {\bibfnamefont {A.~S.}\ \bibnamefont
  {S\o{}rensen}},\ }\href {http://link.aps.org/doi/10.1103/PhysRevA.77.013826}
  {\bibfield  {journal} {\bibinfo  {journal} {Phys. Rev. A}\ }\textbf {\bibinfo
  {volume} {77}},\ \bibinfo {pages} {013826} (\bibinfo {year}
  {2008})}\BibitemShut {NoStop}%
\bibitem [{\citenamefont {Kuzmich}\ and\ \citenamefont
  {Kennedy}(2004)}]{Kuzmich2004}%
  \BibitemOpen
  \bibfield  {author} {\bibinfo {author} {\bibfnamefont {A.}~\bibnamefont
  {Kuzmich}}\ and\ \bibinfo {author} {\bibfnamefont {T.~A.~B.}\ \bibnamefont
  {Kennedy}},\ }\href {http://link.aps.org/doi/10.1103/PhysRevLett.92.030407}
  {\bibfield  {journal} {\bibinfo  {journal} {Phys. Rev. Lett.}\ }\textbf
  {\bibinfo {volume} {92}},\ \bibinfo {pages} {030407} (\bibinfo {year}
  {2004})}\BibitemShut {NoStop}%
\bibitem [{\citenamefont {Windpassinger}\ \emph {et~al.}(2008)\citenamefont
  {Windpassinger}, \citenamefont {Oblak}, \citenamefont {Hoff}, \citenamefont
  {Appel}, \citenamefont {Kj{\ae}rgaard},\ and\ \citenamefont
  {Polzik}}]{Windpassinger2008}%
  \BibitemOpen
  \bibfield  {author} {\bibinfo {author} {\bibfnamefont {P.~J.}\ \bibnamefont
  {Windpassinger}}, \bibinfo {author} {\bibfnamefont {D.}~\bibnamefont
  {Oblak}}, \bibinfo {author} {\bibfnamefont {U.~B.}\ \bibnamefont {Hoff}},
  \bibinfo {author} {\bibfnamefont {J.}~\bibnamefont {Appel}}, \bibinfo
  {author} {\bibfnamefont {N.}~\bibnamefont {Kj{\ae}rgaard}}, \ and\ \bibinfo
  {author} {\bibfnamefont {E.~S.}\ \bibnamefont {Polzik}},\ }\href
  {http://stacks.iop.org/1367-2630/10/i=5/a=053032} {\bibfield  {journal}
  {\bibinfo  {journal} {N. J. Phys.}\ }\textbf {\bibinfo {volume} {10}},\
  \bibinfo {pages} {053032} (\bibinfo {year} {2008})}\BibitemShut {NoStop}%
\bibitem [{\citenamefont {Koschorreck}\ and\ \citenamefont
  {Mitchell}(2009)}]{Koschorreck2009}%
  \BibitemOpen
  \bibfield  {author} {\bibinfo {author} {\bibfnamefont {M.}~\bibnamefont
  {Koschorreck}}\ and\ \bibinfo {author} {\bibfnamefont {M.~W.}\ \bibnamefont
  {Mitchell}},\ }\href {http://stacks.iop.org/0953-4075/42/i=19/a=195502}
  {\bibfield  {journal} {\bibinfo  {journal} {J. Phys. B}\ }\textbf {\bibinfo
  {volume} {42}},\ \bibinfo {pages} {195502} (\bibinfo {year}
  {2009})}\BibitemShut {NoStop}%
\bibitem [{\citenamefont {Sau}\ \emph {et~al.}(2010)\citenamefont {Sau},
  \citenamefont {Leslie}, \citenamefont {Cohen},\ and\ \citenamefont
  {Stamper-Kurn}}]{SauSta10}%
  \BibitemOpen
  \bibfield  {author} {\bibinfo {author} {\bibfnamefont {J.~D.}\ \bibnamefont
  {Sau}}, \bibinfo {author} {\bibfnamefont {S.~R.}\ \bibnamefont {Leslie}},
  \bibinfo {author} {\bibfnamefont {M.~L.}\ \bibnamefont {Cohen}}, \ and\
  \bibinfo {author} {\bibfnamefont {D.~M.}\ \bibnamefont {Stamper-Kurn}},\
  }\href {http://stacks.iop.org/1367-2630/12/i=8/a=085011} {\bibfield
  {journal} {\bibinfo  {journal} {New Journal of Physics}\ }\textbf {\bibinfo
  {volume} {12}},\ \bibinfo {pages} {085011} (\bibinfo {year}
  {2010})}\BibitemShut {NoStop}%
\bibitem [{\citenamefont {Duan}\ \emph {et~al.}(2002)\citenamefont {Duan},
  \citenamefont {Cirac},\ and\ \citenamefont {Zoller}}]{DuaZol02}%
  \BibitemOpen
  \bibfield  {author} {\bibinfo {author} {\bibfnamefont {L.~M.}\ \bibnamefont
  {Duan}}, \bibinfo {author} {\bibfnamefont {J.~I.}\ \bibnamefont {Cirac}}, \
  and\ \bibinfo {author} {\bibfnamefont {P.}~\bibnamefont {Zoller}},\ }\href
  {http://link.aps.org/doi/10.1103/PhysRevA.66.023818} {\bibfield  {journal}
  {\bibinfo  {journal} {Phys. Rev. A}\ }\textbf {\bibinfo {volume} {66}},\
  \bibinfo {pages} {023818} (\bibinfo {year} {2002})}\BibitemShut {NoStop}%
\bibitem [{\citenamefont {S\o{}rensen}\ and\ \citenamefont
  {S\o{}rensen}(2009)}]{SorSor09}%
  \BibitemOpen
  \bibfield  {author} {\bibinfo {author} {\bibfnamefont {M.~W.}\ \bibnamefont
  {S\o{}rensen}}\ and\ \bibinfo {author} {\bibfnamefont {A.~S.}\ \bibnamefont
  {S\o{}rensen}},\ }\href {http://link.aps.org/doi/10.1103/PhysRevA.80.033804}
  {\bibfield  {journal} {\bibinfo  {journal} {Phys. Rev. A}\ }\textbf {\bibinfo
  {volume} {80}},\ \bibinfo {pages} {033804} (\bibinfo {year}
  {2009})}\BibitemShut {NoStop}%
\bibitem [{\citenamefont {Zeuthen}\ \emph {et~al.}(2011)\citenamefont
  {Zeuthen}, \citenamefont {Grodecka-Grad},\ and\ \citenamefont
  {S\o{}rensen}}]{ZueGro11}%
  \BibitemOpen
  \bibfield  {author} {\bibinfo {author} {\bibfnamefont {E.}~\bibnamefont
  {Zeuthen}}, \bibinfo {author} {\bibfnamefont {A.}~\bibnamefont
  {Grodecka-Grad}}, \ and\ \bibinfo {author} {\bibfnamefont {A.~S.}\
  \bibnamefont {S\o{}rensen}},\ }\href
  {http://link.aps.org/doi/10.1103/PhysRevA.84.043838} {\bibfield  {journal}
  {\bibinfo  {journal} {Phys. Rev. A}\ }\textbf {\bibinfo {volume} {84}},\
  \bibinfo {pages} {043838} (\bibinfo {year} {2011})}\BibitemShut {NoStop}%
\bibitem [{\citenamefont {Grodecka-Grad}\ \emph {et~al.}(2012)\citenamefont
  {Grodecka-Grad}, \citenamefont {Zeuthen},\ and\ \citenamefont
  {S\o{}rensen}}]{GroSor12}%
  \BibitemOpen
  \bibfield  {author} {\bibinfo {author} {\bibfnamefont {A.}~\bibnamefont
  {Grodecka-Grad}}, \bibinfo {author} {\bibfnamefont {E.}~\bibnamefont
  {Zeuthen}}, \ and\ \bibinfo {author} {\bibfnamefont {A.~S.}\ \bibnamefont
  {S\o{}rensen}},\ }\href
  {http://link.aps.org/doi/10.1103/PhysRevLett.109.133601} {\bibfield
  {journal} {\bibinfo  {journal} {Phys. Rev. Lett.}\ }\textbf {\bibinfo
  {volume} {109}},\ \bibinfo {pages} {133601} (\bibinfo {year}
  {2012})}\BibitemShut {NoStop}%
\bibitem [{\citenamefont {Higginbottom}\ \emph {et~al.}(2012)\citenamefont
  {Higginbottom}, \citenamefont {Sparkes}, \citenamefont {Rancic},
  \citenamefont {Pinel}, \citenamefont {Hosseini}, \citenamefont {Lam},\ and\
  \citenamefont {Buchler}}]{HigBuc12}%
  \BibitemOpen
  \bibfield  {author} {\bibinfo {author} {\bibfnamefont {D.~B.}\ \bibnamefont
  {Higginbottom}}, \bibinfo {author} {\bibfnamefont {B.}~\bibnamefont
  {Sparkes}}, \bibinfo {author} {\bibfnamefont {M.}~\bibnamefont {Rancic}},
  \bibinfo {author} {\bibfnamefont {O.}~\bibnamefont {Pinel}}, \bibinfo
  {author} {\bibfnamefont {M.}~\bibnamefont {Hosseini}}, \bibinfo {author}
  {\bibfnamefont {P.~K.}\ \bibnamefont {Lam}}, \ and\ \bibinfo {author}
  {\bibfnamefont {B.}~\bibnamefont {Buchler}},\ }\href
  {http://link.aps.org/doi/10.1103/PhysRevA.86.023801} {\bibfield  {journal}
  {\bibinfo  {journal} {Phys. Rev. A}\ }\textbf {\bibinfo {volume} {86}},\
  \bibinfo {pages} {023801} (\bibinfo {year} {2012})}\BibitemShut {NoStop}%
\bibitem [{\citenamefont {Kuzmich}\ \emph {et~al.}(2000)\citenamefont
  {Kuzmich}, \citenamefont {Mandel},\ and\ \citenamefont {Bigelow}}]{KuzBig00}%
  \BibitemOpen
  \bibfield  {author} {\bibinfo {author} {\bibfnamefont {A.}~\bibnamefont
  {Kuzmich}}, \bibinfo {author} {\bibfnamefont {L.}~\bibnamefont {Mandel}}, \
  and\ \bibinfo {author} {\bibfnamefont {N.~P.}\ \bibnamefont {Bigelow}},\
  }\href {http://link.aps.org/doi/10.1103/PhysRevLett.85.1594} {\bibfield
  {journal} {\bibinfo  {journal} {Phys. Rev. Lett.}\ }\textbf {\bibinfo
  {volume} {85}},\ \bibinfo {pages} {1594} (\bibinfo {year}
  {2000})}\BibitemShut {NoStop}%
\bibitem [{\citenamefont {Koschorreck}\ \emph
  {et~al.}(2010{\natexlab{b}})\citenamefont {Koschorreck}, \citenamefont
  {Napolitano}, \citenamefont {Dubost},\ and\ \citenamefont
  {Mitchell}}]{KosMitSq}%
  \BibitemOpen
  \bibfield  {author} {\bibinfo {author} {\bibfnamefont {M.}~\bibnamefont
  {Koschorreck}}, \bibinfo {author} {\bibfnamefont {M.}~\bibnamefont
  {Napolitano}}, \bibinfo {author} {\bibfnamefont {B.}~\bibnamefont {Dubost}},
  \ and\ \bibinfo {author} {\bibfnamefont {M.~W.}\ \bibnamefont {Mitchell}},\
  }\href {http://link.aps.org/doi/10.1103/PhysRevLett.104.093602} {\bibfield
  {journal} {\bibinfo  {journal} {Phys. Rev. Lett.}\ }\textbf {\bibinfo
  {volume} {104}},\ \bibinfo {pages} {093602} (\bibinfo {year}
  {2010}{\natexlab{b}})}\BibitemShut {NoStop}%
\bibitem [{\citenamefont {Takano}\ \emph {et~al.}(2009)\citenamefont {Takano},
  \citenamefont {Fuyama}, \citenamefont {Namiki},\ and\ \citenamefont
  {Takahashi}}]{Takano2009}%
  \BibitemOpen
  \bibfield  {author} {\bibinfo {author} {\bibfnamefont {T.}~\bibnamefont
  {Takano}}, \bibinfo {author} {\bibfnamefont {M.}~\bibnamefont {Fuyama}},
  \bibinfo {author} {\bibfnamefont {R.}~\bibnamefont {Namiki}}, \ and\ \bibinfo
  {author} {\bibfnamefont {Y.}~\bibnamefont {Takahashi}},\ }\href
  {http://link.aps.org/doi/10.1103/PhysRevLett.102.033601} {\bibfield
  {journal} {\bibinfo  {journal} {Phys. Rev. Lett.}\ }\textbf {\bibinfo
  {volume} {102}},\ \bibinfo {pages} {033601} (\bibinfo {year}
  {2009})}\BibitemShut {NoStop}%
\bibitem [{\citenamefont {Kupriyanov}\ \emph {et~al.}(2005)\citenamefont
  {Kupriyanov}, \citenamefont {Mishina}, \citenamefont {Sokolov}, \citenamefont
  {Julsgaard},\ and\ \citenamefont {Polzik}}]{KupPol05}%
  \BibitemOpen
  \bibfield  {author} {\bibinfo {author} {\bibfnamefont {D.~V.}\ \bibnamefont
  {Kupriyanov}}, \bibinfo {author} {\bibfnamefont {O.~S.}\ \bibnamefont
  {Mishina}}, \bibinfo {author} {\bibfnamefont {I.}~\bibnamefont {Sokolov}},
  \bibinfo {author} {\bibfnamefont {B.}~\bibnamefont {Julsgaard}}, \ and\
  \bibinfo {author} {\bibfnamefont {E.~S.}\ \bibnamefont {Polzik}},\ }\href
  {http://link.aps.org/doi/10.1103/PhysRevA.71.032348} {\bibfield  {journal}
  {\bibinfo  {journal} {Phys. Rev. A}\ }\textbf {\bibinfo {volume} {71}}
  (\bibinfo {year} {2005})}\BibitemShut {NoStop}%
\bibitem [{\citenamefont {Vasilyev}\ \emph {et~al.}(2012)\citenamefont
  {Vasilyev}, \citenamefont {Hammerer}, \citenamefont {Korolev},\ and\
  \citenamefont {S\o{}rensen}}]{VasSor12}%
  \BibitemOpen
  \bibfield  {author} {\bibinfo {author} {\bibfnamefont {D.}~\bibnamefont
  {Vasilyev}}, \bibinfo {author} {\bibfnamefont {K.}~\bibnamefont {Hammerer}},
  \bibinfo {author} {\bibfnamefont {N.}~\bibnamefont {Korolev}}, \ and\
  \bibinfo {author} {\bibfnamefont {A.~S.}\ \bibnamefont {S\o{}rensen}},\
  }\href {http://stacks.iop.org/0953-4075/45/i=12/a=124007} {\bibfield
  {journal} {\bibinfo  {journal} {J. Phys. B}\ }\textbf {\bibinfo {volume}
  {45}},\ \bibinfo {pages} {124007} (\bibinfo {year} {2012})}\BibitemShut
  {NoStop}%
\bibitem [{\citenamefont {Newton}(1982)}]{Newton1982}%
  \BibitemOpen
  \bibfield  {author} {\bibinfo {author} {\bibfnamefont {R.}~\bibnamefont
  {Newton}},\ }\href@noop {} {\emph {\bibinfo {title} {Scattering Theory of
  Waves and Particles}}}\ (\bibinfo  {publisher} {Dover},\ \bibinfo {year}
  {1982})\BibitemShut {NoStop}%
\bibitem [{\citenamefont {M\"{u}ller}\ \emph {et~al.}(2005)\citenamefont
  {M\"{u}ller}, \citenamefont {Petrov}, \citenamefont {Oblak}, \citenamefont
  {Alzar}, \citenamefont {de~Echaniz},\ and\ \citenamefont
  {Polzik}}]{MulPol05}%
  \BibitemOpen
  \bibfield  {author} {\bibinfo {author} {\bibfnamefont {J.~H.}\ \bibnamefont
  {M\"{u}ller}}, \bibinfo {author} {\bibfnamefont {P.}~\bibnamefont {Petrov}},
  \bibinfo {author} {\bibfnamefont {D.}~\bibnamefont {Oblak}}, \bibinfo
  {author} {\bibfnamefont {C.~L.~G.}\ \bibnamefont {Alzar}}, \bibinfo {author}
  {\bibfnamefont {S.~R.}\ \bibnamefont {de~Echaniz}}, \ and\ \bibinfo {author}
  {\bibfnamefont {E.~S.}\ \bibnamefont {Polzik}},\ }\href
  {http://link.aps.org/doi/10.1103/PhysRevA.71.033803} {\bibfield  {journal}
  {\bibinfo  {journal} {Phys. Rev. A}\ }\textbf {\bibinfo {volume} {71}},\
  \bibinfo {pages} {033803} (\bibinfo {year} {2005})}\BibitemShut {NoStop}%
\bibitem [{\citenamefont {Deutsch}\ and\ \citenamefont
  {Jessen}(2009)}]{DeuJes09}%
  \BibitemOpen
  \bibfield  {author} {\bibinfo {author} {\bibfnamefont {I.~H.}\ \bibnamefont
  {Deutsch}}\ and\ \bibinfo {author} {\bibfnamefont {P.~S.}\ \bibnamefont
  {Jessen}},\ }\href {http://dx.doi.org/10.1016/j.optcom.2009.10.059}
  {\bibfield  {journal} {\bibinfo  {journal} {Opt. Comm.}\ }\textbf {\bibinfo
  {volume} {283}},\ \bibinfo {pages} {681} (\bibinfo {year}
  {2009})}\BibitemShut {NoStop}%
\bibitem [{\citenamefont {Koschorreck}\ \emph
  {et~al.}(2010{\natexlab{c}})\citenamefont {Koschorreck}, \citenamefont
  {Napolitano}, \citenamefont {Dubost},\ and\ \citenamefont
  {Mitchell}}]{Koschorreck2010}%
  \BibitemOpen
  \bibfield  {author} {\bibinfo {author} {\bibfnamefont {M.}~\bibnamefont
  {Koschorreck}}, \bibinfo {author} {\bibfnamefont {M.}~\bibnamefont
  {Napolitano}}, \bibinfo {author} {\bibfnamefont {B.}~\bibnamefont {Dubost}},
  \ and\ \bibinfo {author} {\bibfnamefont {M.~W.}\ \bibnamefont {Mitchell}},\
  }\href {\doibase 10.1103/PhysRevLett.105.093602} {\bibfield  {journal}
  {\bibinfo  {journal} {Phys. Rev. Lett.}\ }\textbf {\bibinfo {volume} {105}},\
  \bibinfo {pages} {093602} (\bibinfo {year} {2010}{\natexlab{c}})}\BibitemShut
  {NoStop}%
\bibitem [{\citenamefont {Norris}\ \emph {et~al.}(2012)\citenamefont {Norris},
  \citenamefont {Trail}, \citenamefont {Jessen},\ and\ \citenamefont
  {Deutsch}}]{NorDeu12}%
  \BibitemOpen
  \bibfield  {author} {\bibinfo {author} {\bibfnamefont {L.~M.}\ \bibnamefont
  {Norris}}, \bibinfo {author} {\bibfnamefont {C.~M.}\ \bibnamefont {Trail}},
  \bibinfo {author} {\bibfnamefont {P.~S.}\ \bibnamefont {Jessen}}, \ and\
  \bibinfo {author} {\bibfnamefont {I.~H.}\ \bibnamefont {Deutsch}},\ }\href
  {\doibase 10.1103/PhysRevLett.109.173603} {\bibfield  {journal} {\bibinfo
  {journal} {Phys. Rev. Lett.}\ }\textbf {\bibinfo {volume} {109}},\ \bibinfo
  {pages} {173603} (\bibinfo {year} {2012})}\BibitemShut {NoStop}%
\bibitem [{\citenamefont {Hammerer}(2006)}]{Ham06}%
  \BibitemOpen
  \bibfield  {author} {\bibinfo {author} {\bibfnamefont {K.}~\bibnamefont
  {Hammerer}},\ }\emph {\bibinfo {title} {Quantum Information Processing with
  Atomic Ensembles and Light}},\ \href
  {http://www.mpq.mpg.de/Theorygroup/CIRAC/wiki/images/e/e1/Hammerer_Klemens_-_PhD2006.pdf}
  {Ph.D. thesis},\ \bibinfo  {school} {Technische Universit\"{a}t M\"{u}nchen}
  (\bibinfo {year} {2006})\BibitemShut {NoStop}%
\bibitem [{\citenamefont {Deutsch}\ and\ \citenamefont
  {Garrison}(1991)}]{DeuGar91}%
  \BibitemOpen
  \bibfield  {author} {\bibinfo {author} {\bibfnamefont {I.~H.}\ \bibnamefont
  {Deutsch}}\ and\ \bibinfo {author} {\bibfnamefont {J.~C.}\ \bibnamefont
  {Garrison}},\ }\href {http://link.aps.org/doi/10.1103/PhysRevA.43.2498}
  {\bibfield  {journal} {\bibinfo  {journal} {Phys. Rev. A}\ }\textbf {\bibinfo
  {volume} {43}},\ \bibinfo {pages} {2498} (\bibinfo {year}
  {1991})}\BibitemShut {NoStop}%
\bibitem [{\citenamefont {Gardiner}\ and\ \citenamefont
  {Zoller}(2004)}]{GardZoller}%
  \BibitemOpen
  \bibfield  {author} {\bibinfo {author} {\bibfnamefont {C.}~\bibnamefont
  {Gardiner}}\ and\ \bibinfo {author} {\bibfnamefont {P.}~\bibnamefont
  {Zoller}},\ }\href@noop {} {\emph {\bibinfo {title} {Quantum Noise}}}\
  (\bibinfo  {publisher} {Springer},\ \bibinfo {year} {2004})\BibitemShut
  {NoStop}%
\bibitem [{\citenamefont {Cohen-Tannoudji}(1977)}]{CohenTannoudji}%
  \BibitemOpen
  \bibfield  {author} {\bibinfo {author} {\bibfnamefont {C.}~\bibnamefont
  {Cohen-Tannoudji}},\ }\bibinfo {organization} {1975 Les Houches Lectures,
  Session xxvii}\ (\bibinfo  {publisher} {North-Holland},\ \bibinfo {year}
  {1977})\ pp.\ \bibinfo {pages} {1--104}\BibitemShut {NoStop}%
\bibitem [{\citenamefont {Jacobs}\ and\ \citenamefont
  {Steck}(2006)}]{JacSte06}%
  \BibitemOpen
  \bibfield  {author} {\bibinfo {author} {\bibfnamefont {K.}~\bibnamefont
  {Jacobs}}\ and\ \bibinfo {author} {\bibfnamefont {D.~A.}\ \bibnamefont
  {Steck}},\ }\href
  {http://www.tandfonline.com/doi/abs/10.1080/00107510601101934#.Unl2BZGkSEA}
  {\bibfield  {journal} {\bibinfo  {journal} {Contemporary Physics}\ }\textbf
  {\bibinfo {volume} {47}},\ \bibinfo {pages} {279} (\bibinfo {year}
  {2006})}\BibitemShut {NoStop}%
\bibitem [{\citenamefont {Wiseman}\ and\ \citenamefont
  {Milburn}(2010)}]{WieMilBook}%
  \BibitemOpen
  \bibfield  {author} {\bibinfo {author} {\bibfnamefont {H.~M.}\ \bibnamefont
  {Wiseman}}\ and\ \bibinfo {author} {\bibfnamefont {G.~J.}\ \bibnamefont
  {Milburn}},\ }\href@noop {} {\emph {\bibinfo {title} {Quantum Measurement and
  Control}}}\ (\bibinfo  {publisher} {Cambridge University Press},\ \bibinfo
  {year} {2010})\BibitemShut {NoStop}%
\bibitem [{\citenamefont {Wineland}\ \emph {et~al.}(1992)\citenamefont
  {Wineland}, \citenamefont {Bollinger}, \citenamefont {Itano}, \citenamefont
  {Moore},\ and\ \citenamefont {Heinzen}}]{Wineland94}%
  \BibitemOpen
  \bibfield  {author} {\bibinfo {author} {\bibfnamefont {D.~J.}\ \bibnamefont
  {Wineland}}, \bibinfo {author} {\bibfnamefont {J.~J.}\ \bibnamefont
  {Bollinger}}, \bibinfo {author} {\bibfnamefont {W.~M.}\ \bibnamefont
  {Itano}}, \bibinfo {author} {\bibfnamefont {F.~L.}\ \bibnamefont {Moore}}, \
  and\ \bibinfo {author} {\bibfnamefont {D.~J.}\ \bibnamefont {Heinzen}},\
  }\href {\doibase 10.1103/PhysRevA.46.R6797} {\bibfield  {journal} {\bibinfo
  {journal} {Phys. Rev. A}\ }\textbf {\bibinfo {volume} {46}},\ \bibinfo
  {pages} {R6797} (\bibinfo {year} {1992})}\BibitemShut {NoStop}%
\bibitem [{\citenamefont {Budker}\ and\ \citenamefont
  {Romalis}(2007)}]{BudRom07}%
  \BibitemOpen
  \bibfield  {author} {\bibinfo {author} {\bibfnamefont {D.}~\bibnamefont
  {Budker}}\ and\ \bibinfo {author} {\bibfnamefont {M.}~\bibnamefont
  {Romalis}},\ }\href {http://dx.doi.org/10.1038/nphys566} {\bibfield
  {journal} {\bibinfo  {journal} {Nat. Phys.}\ }\textbf {\bibinfo {volume} {3}}
  (\bibinfo {year} {2007})}\BibitemShut {NoStop}%
\bibitem [{\citenamefont {Sewell}\ \emph {et~al.}(2012)\citenamefont {Sewell},
  \citenamefont {Koschorreck}, \citenamefont {Napolitano}, \citenamefont
  {Dubost}, \citenamefont {Behbood},\ and\ \citenamefont
  {Mitchell}}]{SewMit12}%
  \BibitemOpen
  \bibfield  {author} {\bibinfo {author} {\bibfnamefont {R.~J.}\ \bibnamefont
  {Sewell}}, \bibinfo {author} {\bibfnamefont {M.}~\bibnamefont {Koschorreck}},
  \bibinfo {author} {\bibfnamefont {M.}~\bibnamefont {Napolitano}}, \bibinfo
  {author} {\bibfnamefont {B.}~\bibnamefont {Dubost}}, \bibinfo {author}
  {\bibfnamefont {N.}~\bibnamefont {Behbood}}, \ and\ \bibinfo {author}
  {\bibfnamefont {M.~W.}\ \bibnamefont {Mitchell}},\ }\href
  {http://link.aps.org/doi/10.1103/PhysRevLett.109.253605} {\bibfield
  {journal} {\bibinfo  {journal} {Phys. Rev. Lett.}\ }\textbf {\bibinfo
  {volume} {109}} (\bibinfo {year} {2012})}\BibitemShut {NoStop}%
\bibitem [{\citenamefont {Hammerer}\ \emph {et~al.}(2010)\citenamefont
  {Hammerer}, \citenamefont {S\o{}rensen},\ and\ \citenamefont
  {Polzik}}]{PolzikRMP}%
  \BibitemOpen
  \bibfield  {author} {\bibinfo {author} {\bibfnamefont {K.}~\bibnamefont
  {Hammerer}}, \bibinfo {author} {\bibfnamefont {A.~S.}\ \bibnamefont
  {S\o{}rensen}}, \ and\ \bibinfo {author} {\bibfnamefont {E.~S.}\ \bibnamefont
  {Polzik}},\ }\href {http://link.aps.org/doi/10.1103/RevModPhys.82.1041}
  {\bibfield  {journal} {\bibinfo  {journal} {Rev. Mod. Phys.}\ }\textbf
  {\bibinfo {volume} {82}},\ \bibinfo {pages} {1041} (\bibinfo {year}
  {2010})}\BibitemShut {NoStop}%
\bibitem [{\citenamefont {Madsen}\ and\ \citenamefont
  {M\o{}lmer}(2004)}]{MadMol04}%
  \BibitemOpen
  \bibfield  {author} {\bibinfo {author} {\bibfnamefont {L.~B.}\ \bibnamefont
  {Madsen}}\ and\ \bibinfo {author} {\bibfnamefont {K.}~\bibnamefont
  {M\o{}lmer}},\ }\href {http://link.aps.org/doi/10.1103/PhysRevA.70.052324}
  {\bibfield  {journal} {\bibinfo  {journal} {Phys. Rev. A}\ }\textbf {\bibinfo
  {volume} {70}},\ \bibinfo {pages} {052324} (\bibinfo {year}
  {2004})}\BibitemShut {NoStop}%
\bibitem [{\citenamefont {Trail}\ \emph {et~al.}(2010)\citenamefont {Trail},
  \citenamefont {Jessen},\ and\ \citenamefont {Deutsch}}]{TraDeu10}%
  \BibitemOpen
  \bibfield  {author} {\bibinfo {author} {\bibfnamefont {C.~M.}\ \bibnamefont
  {Trail}}, \bibinfo {author} {\bibfnamefont {P.~S.}\ \bibnamefont {Jessen}}, \
  and\ \bibinfo {author} {\bibfnamefont {I.~H.}\ \bibnamefont {Deutsch}},\
  }\href {http://link.aps.org/doi/10.1103/PhysRevLett.105.193602} {\bibfield
  {journal} {\bibinfo  {journal} {Phys. Rev. Lett.}\ }\textbf {\bibinfo
  {volume} {105}},\ \bibinfo {pages} {193602} (\bibinfo {year}
  {2010})}\BibitemShut {NoStop}%
\bibitem [{\citenamefont {Takeuchi}\ \emph {et~al.}(2005)\citenamefont
  {Takeuchi}, \citenamefont {Ichihara}, \citenamefont {Takano}, \citenamefont
  {Kumakura}, \citenamefont {Yabuzaki},\ and\ \citenamefont
  {Takahashi}}]{TakTak05}%
  \BibitemOpen
  \bibfield  {author} {\bibinfo {author} {\bibfnamefont {M.}~\bibnamefont
  {Takeuchi}}, \bibinfo {author} {\bibfnamefont {S.}~\bibnamefont {Ichihara}},
  \bibinfo {author} {\bibfnamefont {T.}~\bibnamefont {Takano}}, \bibinfo
  {author} {\bibfnamefont {M.}~\bibnamefont {Kumakura}}, \bibinfo {author}
  {\bibfnamefont {T.}~\bibnamefont {Yabuzaki}}, \ and\ \bibinfo {author}
  {\bibfnamefont {Y.}~\bibnamefont {Takahashi}},\ }\href
  {http://link.aps.org/doi/10.1103/PhysRevLett.94.023003} {\bibfield  {journal}
  {\bibinfo  {journal} {Phys. Rev. Lett.}\ }\textbf {\bibinfo {volume} {94}},\
  \bibinfo {pages} {023003} (\bibinfo {year} {2005})}\BibitemShut {NoStop}%
\bibitem [{\citenamefont {Puentes}\ \emph {et~al.}(2013)\citenamefont
  {Puentes}, \citenamefont {Colangelo}, \citenamefont {Sewell},\ and\
  \citenamefont {Mitchell}}]{PueMit13}%
  \BibitemOpen
  \bibfield  {author} {\bibinfo {author} {\bibfnamefont {G.}~\bibnamefont
  {Puentes}}, \bibinfo {author} {\bibfnamefont {G.}~\bibnamefont {Colangelo}},
  \bibinfo {author} {\bibfnamefont {R.~J.}\ \bibnamefont {Sewell}}, \ and\
  \bibinfo {author} {\bibfnamefont {M.~W.}\ \bibnamefont {Mitchell}},\ }\href
  {http://stacks.iop.org/1367-2630/15/i=10/a=103031} {\bibfield  {journal}
  {\bibinfo  {journal} {N. J. Phys.}\ }\textbf {\bibinfo {volume} {15}}
  (\bibinfo {year} {2013})}\BibitemShut {NoStop}%
\bibitem [{\citenamefont {Cerf}\ \emph {et~al.}(2007)\citenamefont {Cerf},
  \citenamefont {Leuchs},\ and\ \citenamefont {Polzik}}]{CV2007}%
  \BibitemOpen
  \bibinfo {editor} {\bibfnamefont {N.}~\bibnamefont {Cerf}}, \bibinfo {editor}
  {\bibfnamefont {G.}~\bibnamefont {Leuchs}}, \ and\ \bibinfo {editor}
  {\bibfnamefont {E.}~\bibnamefont {Polzik}},\ eds.,\ \href@noop {} {\emph
  {\bibinfo {title} {Quantum Information with Continuous Variables of Atoms and
  Light}}}\ (\bibinfo  {publisher} {World Scientific},\ \bibinfo {year}
  {2007})\BibitemShut {NoStop}%
\bibitem [{\citenamefont {Garrison}\ and\ \citenamefont
  {Chiao}(2008)}]{GarChi08}%
  \BibitemOpen
  \bibfield  {author} {\bibinfo {author} {\bibfnamefont {J.~C.}\ \bibnamefont
  {Garrison}}\ and\ \bibinfo {author} {\bibfnamefont {R.~Y.}\ \bibnamefont
  {Chiao}},\ }\href@noop {} {\emph {\bibinfo {title} {Quantum Optics}}}\
  (\bibinfo  {publisher} {Oxford University Press},\ \bibinfo {year}
  {2008})\BibitemShut {NoStop}%
\end{thebibliography}%

\begin{appendix}

\begin{widetext}
\section{Paraxial scattering: classical theory} \label{Appendix::ParaxialScattering}

	The complex electric field is paraxial and quasimonchromatic, $ \vec{\mathcal{E}} (\rp , z,t) e^{i(k_0z-\omega_0 t)} $.  Thus, the slowly varying envelope is governed by the time-dependent paraxial wave equation~\cite{GarChi08},
	\begin{align}
		& i\bigg(\frac{\partial}{\partial z}  + \frac{1}{c}\frac{\partial}{\partial t} \bigg)  \vec{\mathcal{E}}(\rp , z, t) = -\frac{1}{2 k_0}\grad_\perp^2 \vec{\mathcal{E}}(\rp , z, t)-2 \pi k_0 \tensor{\chi}(\rp, z) \cdot \vec{\mathcal{E}}(\rp , z, t),
	\end{align}
where $\tensor{\chi}(\rp, z)$ is the spatially averaged dielectric susceptibility.  Defining $\vec{\mathcal{E}} (\rp , z, t) = \mathcal{A}(t-z/c) \vec{\mathcal{U}} (\rp, z)$,
\begin{equation}
\label{paraxEq}
i\frac{\partial}{\partial z}\vec{\mathcal{U}}(\rp , z)  = -\frac{1}{2 k_0}\grad_\perp^2 \vec{\mathcal{U}}(\rp , z) - 2 \pi k_0 \tensor{\chi}(\rp, z) \cdot \vec{\mathcal{U}}(\rp , z).
\end{equation}
This equation is isomorphic to the time-dependent Schr\"{o}dinger equation with the propagation distance $z$ playing the role of time and the susceptibility playing the role of the potential.  

We can define a Hilbert space of square-integrable functions on the transverse plane, and use Dirac notation to express the evolution of the transverse mode as a function of $z$, $\mathcal{U}(\rp , z)=\langle \rp | \mathcal{U}(z)\rangle $.  The free-space propagator ($z$-evolution operator), satisfies the free-particle Schr\"{o}dinger equation in two dimensions,
\begin{equation}
i\frac{\partial }{\partial z} \hat{K}=\frac{\hat{\mbf{p}}_\perp^2}{2 k_0} \hat{K},
\end{equation}
where $\hat{\mbf{p}}_\perp = -i \nabla_\perp$ in the position representation.  The solution, $\hat{K}(z-z')= \exp [{-i\frac{\hat{\mbf{p}}_\perp^2}{2 k_0} (z-z')} ]$, has the familiar position-space representation for the spreading of a wavepacket and Fraunhofer diffraction~\cite{Newton1982},

\begin{equation} \label{Eq::Propagator}
		K(\rp-\rp',z-z') =\langle \rp | \hat{K}(z-z') | \rp' \rangle = \frac{-i k_0}{2\pi (z-z')}\exp\left[ \frac{ik_0 |\rp - \rp'|^2}{2(z-z')} \right].
\end{equation}
The $z$-evolution of a freely propagating beam is given by  
\begin{equation}
\mathcal{U}(\rp , z) = \langle \rp | \mathcal{U}(z)\rangle  =\langle \rp | \hat{K}(z-z') |\mathcal{U}(z')\rangle =\int d^2\rp'\; K(\rp-\rp',z-z') \mathcal{U}(\rp' , z').
\end{equation}
Other properties of the propagator follow from unitarity, $\hat{K}^\dag(z-z')=\hat{K}(z'-z)$, and thus
\begin{eqnarray}
\label{backprop}
\mathcal{U}^* (\rp', z') \nonumber \nonumber&=& \langle \rp' | \hat{K}(z'-z)| \mathcal{U} (z) \rangle^* =   \langle \mathcal{U} (z)  | \hat{K}(z-z')| \rp' \rangle \nonumber\\
& =& \int  d^2 \rp  \mathcal{U}^* (\rp, z) \, K(\rp-\rp', z-z')   .
\end{eqnarray}
We define a complete orthogonal basis $\{\ket{u_{pl}(z)}\}$, normalized to a fixed transverse area $A$,
\begin{equation}
\langle u_{p'l'}(z) |u_{pl}(z) \rangle =  \int d^2\rp u^*_{p'l'}(z) u_{pl}(z) = A \, \delta_{p,p'} \delta_{l,l'}.
\end{equation}
The propagator can then be written as
\begin{equation}
\hat{K}(z-z')=\frac{1}{A}\sum_{p,l} \ket{u_{pl}(z)} \bra{u_{pl}(z')} \Rightarrow K(\rp-\rp', z-z') = \frac{1}{A}\sum_{p,l}u^*_{pl}(\rp', z') u_{pl}(\rp, z),
\end{equation}
with the boundary condition
$
K(\rp-\rp', t_0) = \delta^{(2)} (\rp'-\rp) 
$
that follows from completeness.

The scattering of paraxial fields thus follows in complete analogy to the scattering of nonrealistic Schr\"{o}dinger waves~\cite{Newton1982}, where the time-dependent formulation of scattering translates into the $z$-dependence.  The retarded Green's function for free propagation is defined as 
	\begin{eqnarray}
		G_+ (\rp-\rp',z-z') &=& -i \Theta(z-z') K(\rp-\rp',z-z'), \\
\left( i\frac{\partial}{\partial z}+\frac{1}{2 k_0}\grad_\perp^2 \right) G_+(\rp-\rp',z-z') &=&  \delta (z-z') \delta^{(2)} ( \rp - \rp'), \nonumber 
	\end{eqnarray}
where $\Theta(z)$ is the Heaviside step function.  In the first Born approximation, i.e. for dilute samples where multiple scattering is negligible, given an incident field (free propagating solution) $\vec{\mathcal{U}}_{\rm in} (\rp,z)$, the total scattering solution is
	\begin{align}
		\vec{\mathcal{U}}  (\rp, & z)   = \vec{\mathcal{U}}_{\rm in} (\rp,z) + \int_{-\infty}^\infty dz' \int d^2 \rp' G_+(\rp-\rp',z-z') (-2 \pi k_0) \tensor{\chi}(\rp, z) \cdot \vec{\mathcal{U}}_{\rm in} (\rp',z') \nonumber \\
  		& = \vec{\mathcal{U}}_{\rm in} (\rp,z) + i 2 \pi k_0 \int_{-\infty}^z dz' \int d^2 \rp' K(\rp-\rp', z-z') \tensor{\chi}(\rp, z')  \cdot \vec{\mathcal{U}}_{\rm in} (\rp',z') \label{Eq::ClassicalScattering}
	\end{align}
corresponding to the superposition of incident and reradiated fields.


\section{Quantization of the paraxial field} \label{Appendix::ParaxialQuantization}

Paraxial quantization follows from the slowly varying envelope approximation~\cite{DeuGar91}.  For these modes, we define the positive-frequency component of the electric field analogous to a classical beam
\begin{equation}
\hat{\mbf{E}}^{(+)}(\mbf{r}, t) = \sqrt{2 \pi \hbar \omega_0} \sum_\alpha \mbf{e _\alpha} \hat{\Psi}_\alpha (\rp,z,t) e^{i(k_0 z - \omega_0 t)},
\end{equation}
where $\alpha$ labels transverse polarization, and the slowly varying envelope satisfies the equal-time commutation relations of a nonrelativistic bosonic field,
\begin{equation}
\left[ \hat{\Psi}_\alpha(\rp,z,t), \hat{\Psi}^\dag_\beta(\rp',z',t) \right] = \delta_{\alpha, \beta} \delta^{(2)} (\rp - \rp') \delta (z-z').
\end{equation}
The free field satisfies the paraxial wave equation
\begin{equation}
i\frac{\partial}{\partial t} \hat{\Psi}_\alpha = -ic \frac{\partial}{\partial z} \hat{\Psi}_\alpha -\frac{1}{2 k_0} \grad^2_\perp \hat{\Psi}_\alpha,
\end{equation}
which is the Heisenberg equation of motion for an envelope governed by the free paraxial Hamiltonian
\begin{equation}
\hat{H}_{\text{free}}=   \sum_\alpha \int d^3 \mbf{r} \, \hat{\Psi}^\dag_\alpha  \left(- i\hbar c \frac{\partial}{\partial z}  -\frac{\hbar}{2 k_0} \grad^2_\perp  \right) \hat{\Psi}_\alpha .
\end{equation}
The free field solution is thus determined by the classical propagator,
\begin{equation} \label{Eq::ParaxialFreeFieldSolution}
\hat{\Psi}_{\alpha}(\rp, z, t) = \int d^2 \rp' K(\rp-\rp',ct) \hat{\Psi}_\alpha (\rp', z-ct,0).
\end{equation}
It then follows that the free field satisfies the general commutation relations,
	\begin{align}
		\Big[ & \hat{\Psi}_\alpha(\rp,z,t), \hat{\Psi}^\dag_\beta(\rp',z',t')  \Big] =   K(\rp - \rp',z-z')  \delta_{\alpha, \beta} \delta \left(z-z'-c(t-t')\right) , 
	\end{align}
and thus equal-$z$, unequal-$t$ commutation relations,
\begin{equation}
\left[ \hat{\Psi}_\alpha(\rp,z,t), \hat{\Psi}^\dag_\beta(\rp',z,t') \right] =\frac{1}{c} \delta_{\alpha, \beta} \delta^{(2)} (\rp - \rp') \delta (t-t') .
\end{equation}

The paraxial field is naturally decomposed into an orthonormal set of dimensionless transverse mode functions.  Here we use the Laguerre-Gauss modes $\{ u_{pl} (\rp, z)\}$ in cylindrical coordinates,
	\begin{align} \label{Eq::LGModes}
		u_{pl}(\rp,z)= & \, \mathcal{N}_{pl} \frac{w_0}{w(z)} \left( \frac{\sqrt{2} \rho}{w(z)} \right)^{|l|} L_p^{|l|}\left( \frac{2 \rho^2}{\big[ w(z)\big]^2}\right) e^{-\frac{ \rho^2}{[w(z)]^2}}e^{\frac{ik_0 \rho^2}{2 R(z)}}e^{-i (2p + l + 1)\Phi(z)}e^{-il\phi}, 
	\end{align}
where $\mathcal{N}_{pl} = \sqrt{p!/ (|l| + p)!}$ is the normalization constant, $L_p^{|l|}(x)$ indicates an associated Laguerre polynomial, and parameters $w(z)$, $R(z)$, and $\Phi(z)$ are given in \erf{Eq::GaussianParameters}.  These modes satisfy 
\begin{align} 
\int d^2 \rp u^*_{pl} (\rp , z) u_{p'l'} &(\rp , z) = A \, \delta_{p, p'} \delta_{l, l'},\label{Eq::TransverseOrthogonality} \\
\sum_{p,l} u_{pl}(\rp , z) u^*_{pl}(\rp' , z')  &=  A \, K(\rp-\rp', z-z'), \label{Eq::TransverseCompleteness}\\
\sum_{p,l} u_{pl}(\rp , z) u^*_{pl}(\rp' , z)  &=  A \, \delta^{(2)}(\rp-\rp')\label{Eq::TransverseCompSameZ},
\end{align}
where we have defined a quantization area, $A = \pi w_0^2/2$, as the natural scale for Gaussian beams of waist $w_0$.  Using \erf{Eq::ParaxialFreeFieldSolution} and the completeness relation, \erf{Eq::TransverseCompSameZ}, we define local mode creation and annihilation operators,
\begin{equation}
\hat{a}_{pl,\alpha}(z,t) = \int d^2\rp \sqrt{\frac{c}{A}} \hat{\Psi}_\alpha(\rp,z,t) u_{pl}^*(\rp,z), 
\end{equation}
that evolve under the free-field Hamiltonian according to $\hat{a}_{pl,\alpha}(z,t)=\hat{a}_{pl,\alpha}(0,t-z/c)=\hat{a}_{pl,\alpha}(z-ct,0)$, and satisfy free-field commutation relations
\begin{equation}
	\big[ \hat{a}_{pl,\alpha}(z,t),\hat{a}^{\dag}_{p'l',\beta}(z',t')\big] = \delta_{p, p'} \delta_{l, l'} \delta_{\alpha, \beta} \, \delta(t-t'-(z-z')/c).
\end{equation}
The positive frequency component of the electric field expanded in these modes is
\begin{align}
	\hat{\mbf{E}}^{(+)}&(\mbf{r},t) = \sum_{p,l,\alpha} \sqrt{\frac{2 \pi \hbar \omega_0}{c A}} \, \mbf{e}_\alpha  \, \hat{a}_{pl,\alpha}(z,t)  \,u_{pl}(\rp,z)  \, e^{i(k_0 z - \omega_0 t)}.
\end{align}

\section{Multimode master equation}
\label{Appendix::MultimodeJumpOperators}

	The joint dynamics of the collective atomic system and the paraxial field can be expressed using a master equation formalism~\cite{GardZoller},
	\begin{align}
		\frac{d}{dt} \hat{\rho} = - \frac{i}{\hbar} \big[ \hat{H}_{\rm eff} \hat{\rho} - \hat{\rho} \hat{H}_{\rm eff}^\dagger \big] + \Gamma \sum_{i, \alpha} \hat{W}_\alpha^{(i)} \hat{\rho} \hat{W}_\alpha^{(i) \dagger}.
	\end{align}
The effective Hamiltonian has a real part that drives coherent dynamics and an imaginary part describing loss,
	\begin{align}
		\hat{H}_{\rm eff} = \hat{H}_{\rm int} + \hat{H}_{\rm loss}.
	\end{align}
For a probe laser with polarization $\vec{\epsilon}$, the jump operators are \cite{DeuJes09}
	\begin{align} \label{Eq::JumpOperators}
		\hat{W}_\alpha^{(i)} = \sum_{f'} \frac{\Omega(\mathbf{r}_i)/2}{ \Delta_{f'} + i \Gamma/2} \mathbf{e^*_\alpha} \cdot \hat{\mathbf{d}}_{f'}^{(i)} \hat{\mathbf{d}}_{f'}^{(i)\dagger} \cdot \vec{\epsilon} 
	\end{align}
with local Rabi frequency $\Omega(\mathbf{r}_i) = \bra{J'}| d | \ket{J} \mathcal{E}_L(\mathbf{r}_i)$ .  

	For our spin-squeezing protocol, the probe is prepared with linear polarization along $x$ and photon flux $\dot{N}_L$.  The Hermitian part of the effective Hamiltonian, $\hat{H}_{\rm int}$, is given by the multimode Faraday Hamiltonian, \erf{Eq::TheFaradayInteraction}. For detuning large compared to the excited state hyperfine splitting and neglecting terms that describe vacuum-vacuum scattering, the anti-Hermitian part is
	\begin{align}  \label{Eq::LossHamiltonian}
		\hat{H}_{\rm loss} =& -i \hbar  \frac{ C^{(0)} }{2}  \sum_i \gamma_s(\mathbf{r}_i)  \hat{\mathbbm{1}}^{(i)}.
	\end{align}
Using \erf{Eq::LossHamiltonian}, \erf{Eq::IrreducibleDecomp}, and \erf{Eq::LocalScatRate} we can write down the part of master equation from diffuse scattering,
	\begin{align}\label{Eq::NonRotMaster}
		\frac{d \hat{\rho}}{dt} \Big|_{\rm diff} =  \sum_{i} \gamma_s & (\mathbf{r}_i) \Big[ \big( |C^{(0)}|^2 - C^{(0)} \big) \hat{\rho} + |C^{(1)}|^2 \big( \hat{f}_z^{(i)} \hat{\rho} \hat{f}_z^{(i)} +  \hat{f}_y^{(i)} \hat{\rho} \hat{f}_y^{(i)} \big) \Big]. 
	\end{align} 	
For the case of an atom driven on an $S_{1/2} \rightarrow P_{J}$ transition, $|C^{(0)}|^2 - C^{(0)} = -2/9$ and $|C^{(1)}|^2 = g^2_f/9$.  To define the quantization axis, a large bias magnetic field $B_0$ is applied along the $z$-axis, causing rapid Larmor precession at frequency $\Omega_0 = g_f \mu_B B_0$. Transforming to the rotating frame,  $\hat{f}_z^{(i)} \rightarrow \hat{f}_z^{(i)}$ and $\hat{f}_y^{(i)} \rightarrow \cos(\Omega_0 t )\hat{f}_y^{(i)}- \sin(\Omega_0 t) \hat{f}_x^{(i)}$. Substituting these relations into Eq. (\ref{Eq::NonRotMaster}) and time-averaging according to the rotating wave approximation yields the map for local decoherence in \erf{Eq::DiffuseME}.
	
\section{Derivation of the multimode homodyne polarimetry stochastic master equation}  \label{Appendix::SMEDerivation}

	The stochastic master equation (SME) describes the evolution of the atomic ensemble as continuous homodyne polarimetry measurements are performed on the output light.  Although we showed in Sec. \ref{Sec::SemiclassicalTheory} that only light in the probe mode is measured, we present here the SME that results from independent measurements of the position quadratures $\hat{X}_{pl}$ in each mode $pl$, following the standard prescription given in Refs. \cite{JacSte06, WieMilBook}. A more general SME arises from such a continuous polarimetry measurement than the case where solely the $\hat{X}_{00}$ quadrature is measured and measurement records on all modes $pl\neq00$ are discarded.
	
	Prior to measurement, the time evolution operator $\hat{U}(\Delta t) = \exp ( -i \Delta t \hat{H}/\hbar )$ describing the interaction of the light and spin waves over a time interval $\Delta t$ is
	\begin{equation}
		\hat{U}(\Delta t)  = \prod_{p,l} \hat{U}_{pl}(\Delta t). 
	\end{equation}
The interaction in a single spatial mode is generated by the multimode Faraday Hamiltonian in \erf{Eq::TheFaradayInteraction},
	\begin{align}
			\hat{U}_{pl}(\Delta t) =\exp \! \bigg[  \! - \! i \Delta t & \sqrt{\frac{ \kappa }{2}} \Big(  \mbox{Re}\big\{ \hat{F}^{pl}_z \big\} \hat{P}_{pl}  -\mbox{Im} \big\{ \hat{F}^{pl}_z \big\} \hat{X}_{pl} \Big)\! \bigg] \nonumber.
	\end{align}
After this interaction the light and spin waves are entangled so that a polarimetry measurement of the $\hat{X}_{pl}$ is correlated with quantum backaction on the atomic ensemble.  The evolution of the system conditioned on independent measurements of each mode is determined by the Kraus operator,
	\begin{equation} \label{Eq::GenKrausOp}
		\hat{A}(\Delta t)= \prod_{p,l} \hat{A}_{pl}(\Delta t).
	\end{equation} 
Here, $\hat{A}_{pl}(\Delta t)$ is the Kraus component for measurement outcome $x_{pl}$ in the spatial mode $pl$: 
	\begin{align}
		\hat{A}_{pl} (\Delta t)  \equiv & \bra{\hat{X}_{pl} = x_{pl}} \hat{U}_{pl}(\Delta t) \ket{0} \nn \\
		=&  \exp \bigg[ \Delta t \sqrt{\frac{ \kappa }{2 }} \hat{F}^{pl}_z \hat{X}_{pl} - \frac{\kappa \Delta t}{4}  \left(\mbox{Re}\big\{ \hat{F}^{pl}_z \big\}^2 + i \mbox{Im} \big\{ \hat{F}^{pl}_z \big\} \mbox{Re}\big\{ \hat{F}^{pl}_z \big\} \right) \bigg]  .
	\end{align}
The measurement photocurrent is a Gaussian stochastic process with mean $\langle \mbox{Re}\{ \hat{F}^{pl}_z \}\rangle \Delta t$ and variance 1/$\kappa$, 
	\begin{align}
		dy_{pl}=\big\langle \mbox{Re} \{ \hat{F}^{pl}_z \} \big\rangle \Delta t+\frac{\Delta W_{pl}}{\sqrt{\kappa} },
	\end{align}
where $\Delta W_{pl}$ is a Wiener increment with zero mean and variance $\Delta t$.  In the infinitesimal limit, $\Delta t \rightarrow dt$ and $\Delta W_{pl} \rightarrow dW_{pl}$, we expand the Kraus component to first order in $dt$,
	\begin{align} \label{Eq::KrausOpmn}
		\hat{A}_{pl}(d t) =  \hat{\mathbbm{1}}& +\frac{\kappa}{4}\hat{F}^{pl}_z \big\langle \hat{F}^{pl}_z + \hat{F}_z^{pl\dagger} \big\rangle dt -\frac{\kappa}{8}\hat{F}^{pl}_z \hat{F}_z^{pl\dagger} d t+ \sqrt{ \frac{\kappa}{4} } \hat{F}^{pl}_z dW_{pl}. 
	\end{align}
We have used the statistical independence of the stochastic Wiener processes, $d W_{pl} dW_{p'l'} = \delta_{p, p'}\delta_{l, l'}d t$.

	After the measurements are performed, the conditional collective atomic state is updated via the map
	\begin{align} \label{Eq::KrausUpdate}
		\hat{\rho}(t+d t)&=\frac{\hat{A}(d t)\hat{\rho}(t)\hat{A}^\dagger(d t)}{\Tr{ \hat{A}^\dagger (d t)\hat{A}(d t) \hat{\rho}(t) }}.
	\end{align}
Using Eqs. (\ref{Eq::GenKrausOp}) and (\ref{Eq::KrausOpmn})  with Eq. (\ref{Eq::KrausUpdate}), we derive the conditional atomic state. In differential form, the SME is
	\begin{align}\label{Eq::FullSME}
		d \hat{\rho}&= \sum_{p,l} \left( \sqrt{ \frac{\kappa}{4} } \mathcal{H}_{pl}[\hat{\rho} ] \, dW_{pl} + \frac{\kappa}{4} \mathcal{L}_{pl}[\hat{\rho}] \, dt \right),
	\end{align}
where the measurement-update superoperator $\mathcal{H}_{pl}[\hat{\rho} ]$ is defined in \erf{Eq::HSuperoperator} and the Lindblad superoperator $\mathcal{L}_{pl}[\hat{\rho}]$ in \erf{Eq::LSuperoperator}. 

For measurements of the fundamental mode only, we ignore the measurement records for $pl\neq00$, which is equivalent to averaging over measurement records or tracing over these modes.  The result is the stochastic master equation
	\begin{align} \label{ApEq::HomodyneSME}
		d \hat{\rho}&= \sqrt{ \frac{\kappa}{4} } \mathcal{H}_{00}[\hat{\rho}]  \, dW + \frac{\kappa}{4} \sum_{p,l} \mathcal{L}_{pl}[\hat{\rho}] \, dt,
	\end{align}
where $dW=dW_{00}.$  Since the modes $pl\neq00$ are unmeasured, information about the ensemble is lost and \erf{ApEq::HomodyneSME} does not preserve purity.

\section{Derivation of the mean spin and covariance equations of motion} \label{Appendix::SpinCovariances}	

While the squeezing parameter, \erf{Eq::SqueezingParam}, depends solely upon the mean and variance of the fundamental spin wave defined by the spatial mode of the laser probe, the diffuse scattering by individual atoms is not collective in nature and acts to couple the different spin waves to one another.  In order to model the dynamical evolution of the squeezing, including decoherence, we must track the evolution of  a hierarchy of differential equations coupling the means and covariances of spin waves in all spatial modes. This appendix provides a detailed derivation of these equations and the numerical methods used in their solution for the case of an ensemble of spin-1/2 atoms.

	We first consider the evolution of $\expects{\hat{F}_x^{pl}}$, the mean of a spin wave in spatial mode $pl$, where $\hat{F}_x^{pl} = \sum_i  \beta_{pl}(\mbf{r}_i) \hat{f}_x^{(i)}$. Because collective scattering and measurement backaction negligibly affect the dynamics of the mean spin, to good approximation the evolution of $\expects{\hat{F}_x^{pl}}$ is dominated by diffuse scattering and is described by \erf{Eq::1stOrderEvol}.  For spin-1/2, the local map, \erf{Eq::DiffuseME}, simplifies to ${\mathcal{D}_i}[\hat{f}_x^{(i)}] =-\hat{f}_x^{(i)}/3$.  Using \erf{Eq::LocalScatRate}, we get an equation of motion,
\begin{align}\label{Eq::SimplifiedMeanFx}
\frac{d}{dt}\expect{\hat{F}_x^{pl}}=-\frac{\gamma_0}{3}\sum_i  \beta_{00}(\mbf{r}_i) \beta_{pl}(\mbf{r}_i)\expect{\hat{f}_x^{(i)}}.
\end{align} 
By decomposing $\beta_{00}(\mbf{r}) \beta_{pl}(\mbf{r})$ in terms of orthogonal mode functions, the right hand side of \erf{Eq::SimplifiedMeanFx} can be expressed in terms of spin wave operators. In terms of the mode functions,
	\begin{align}
		 \beta_{00}(\mbf{r}_\perp,z)  \beta_{pl}(\mbf{r}_\perp,z)& =|u_{00}(\mbf{r}_\perp,z)|^2u_{pl}^*(\mbf{r}_\perp,z)u_{00}(\mbf{r}_\perp,z) 
		= \sum_{p',l'} c^{pl}_{p'l'}(z) \beta_{p'l'}(\mbf{r}_\perp,z), \label{Eq::ProjCoeff_proto}
\end{align}
where we have made use of orthogonality and completeness in Eqs. (\ref{Eq::TransverseOrthogonality}) and (\ref{Eq::TransverseCompSameZ}) to define projection coefficients,
	\begin{align} \label{Eq::ProjCoeff}
		c^{pl}_{p'l'}(z)  \equiv  \frac{1}{A} \int d^2 \mathbf{r}_\perp \left[ u_{00}(\mathbf{r}_\perp, z)\right]^2 u^*_{pl}(\mathbf{r}_\perp, z) u_{p'l'}(\mathbf{r}_\perp, z).
	\end{align}
By restricting \erf{Eq::SimplifiedMeanFx} to a coarse-grained slice of thickness $\delta z$ at longitudinal coordinate $z_k$ and performing the projection in \erf{Eq::ProjCoeff_proto}, we obtain an infinite hierarchy of differential equations that couple mean spin waves in a given slice to one another,
\begin{align}\label{Eq::ZkSliceFx}
\frac{d}{dt}\expect{\hat{F}_x^{pl}(z_k)}=-\frac{\gamma_0}{3}\sum_{p',l'}c^{pl}_{p'l'}(z_k)\expect{\hat{F}_x^{p'l'}(z_k)}.
\end{align} 

Solving the resulting finite system of coupled differential equations requires initial conditions of the mean spin waves in each slice. Using $\expect{\hat{f}_x(t_0)}=1/2$ for the initial SCS state of the ensemble,
	\begin{align}\label{Eq::meanSlice}
		\expect{\hat{F}_x^{pl}(z_k, t_0)} = \sum_{i_k}\beta_{pl}(\mbf{r}_{i_k})\expect{\hat{f}_x^{(i_k)}(t_0)}  =\frac{1}{2}\sum_{i_k}\beta_{pl}(\mbf{r}_{i_k}),
	\end{align} 
where $i_k$ is an index  over all atoms in slice $z_k$. For a average atomic density, $\eta(\mathbf{r})$, the sum becomes an integral, 
	\begin{align}
		\expect{\hat{F}_x^{pl}(z_k, t_0)} \rightarrow \frac{\delta z}{2}\int d^2 \rp \eta(\rp,z_k)\beta_{pl}(\rp,z_k).
	\end{align} 
An approximate solution to \erf{Eq::ZkSliceFx} is found for each slice by choosing $\delta z$ and truncating the number of spin waves at some index $p_{max}, \,l_{max}$.  Summing over the solutions at each slice gives the mean of the fundamental spin wave,
\begin{align}\label{Eq::FundSpinwave}
\expect{\hat{F}_x^{00}(t)}=\sum_{k}\expect{\hat{F}_x^{00}(z_k,t)}.
\end{align} 
Equation (\ref{Eq::FundSpinwave}) is the mean spin in the definition of the squeezing parameter.  

	To solve for the variance of the fundamental spin wave, we follow a similar procedure. As shown in \erf{Eq::SpinHalfVariance}, the fundamental variance couples through diffuse scattering to covariances between spin waves in slices $z_k$ and $z_{k'}$:
	\begin{align} \label{Eq::GenCovariance}
		\expect{\Delta\hat{F}_z^{pl}(z_k)\Delta\hat{F}_z^{p'l'}(z_{k'})} =\expect{\hat{F}_z^{pl}(z_k)\hat{F}_z^{p'l'}(z_{k'})}-\expect{\hat{F}_z^{pl}(z_k)}\expect{\hat{F}_z^{p'l'}(z_{k'})} .
		\end{align}
From the SME in \erf{Eq::HomodyneSME}, we find the equations of motion for these covariances.  Unlike the mean spin, the effects of continuous measurement must be included along with diffuse scattering. However, decoherence from collective scattering, described by the map $\mathcal{L}_{pl}$ in \erf{Eq::LSuperoperator}, does not affect these covariances since the $\hat{F}_z^{pl}$ commute with one another.

	First, we examine the contribution of continuous measurement. From the SME in \erf{ApEq::HomodyneSMEtext} and the rule of It\={o} calculus that differentials must be taken to second order \cite{JacSte06}, i.e. $d(XY) = (dX) Y + X (dY) + (dX)(dY)$, we find
	\begin{align} 
		 d\expect{ \Delta\hat{F}_z^{pl}(z_k)   \Delta\hat{F}_z^{p'l'}(z_{k'})}\Big|_{\text{meas}} =  
		\sqrt{ \frac{\kappa}{4} } \bigg\{ & \big\langle \mathcal{H}_{00}[\hat{F}_z^{pl}(z_k)\hat{F}_z^{p'l'}(z_{k'})]\big\rangle- \big\langle\mathcal{H}_{00}[\hat{F}_z^{pl}(z_k)]\big\rangle \big\langle\hat{F}_z^{p'l'}(z_{k'})\big\rangle  \label{Eq::CovHomodyne} \\
& - \big\langle\hat{F}_z^{pl}(z_k)\big\rangle \big\langle\mathcal{H}_{00}[\hat{F}_z^{p'l'}(z_{k'})]\big\rangle
 \bigg\} dW -\frac{\kappa}{4}\big\langle\mathcal{H}_{00}[\hat{F}_z^{pl}(z_k)]\big\rangle\big\langle\mathcal{H}_{00}[\hat{F}_z^{p'l'}(z_{k'})]\big\rangle dt.\nn
	\end{align} 
The map $\mathcal{H}_{00}$, \erf{Eq::HSuperoperator}, couples the first- and second-order moments of the spin waves to higher-order moments.
For the initial SCS along $x$ and during its subsequent evolution, the spin waves $\hat{F}_z^{pl}$ are Gaussian distributed, both over the entire cloud and within each coarse-grained slice $z_k$. Thus, third-order moments of commuting observables can be expressed in terms of first- and second-order moments with the relation, $\expects{XYZ} = \expects{XY}\expects{Z} + \expects{XZ}\expects{Y} + \expects{YZ}\expects{X} - 2\expects{X}\expects{Y}\expects{Z}$ \cite{JacSte06}.
In this regime, all stochastic terms in \erf{Eq::CovHomodyne} cancel, leaving the deterministic equation: 
	\begin{align}
		\frac{d}{dt}\expect{\Delta\hat{F}_z^{pl}(z_k)  \Delta\hat{F}_z^{p'l'} (z_{k'})}\Big|_{\text{meas}} &= -\kappa\big\langle\Delta\hat{F}_z^{pl}(z_k)\Delta\hat{F}_z^{00}\big\rangle\big\langle\Delta\hat{F}_z^{p'l'}(z_{k'})\Delta\hat{F}_z^{00}\big\rangle \nn \\
		&=-\kappa\sum_{k'',k'''}\big\langle\Delta\hat{F}_z^{pl}(z_k)\Delta\hat{F}_z^{00}(z_{k''})\big\rangle\big\langle\Delta\hat{F}_z^{p'l'}(z_{k'})\Delta\hat{F}_z^{00}(z_{k'''})\big\rangle. \label{Eq::CovarianceBackaction}
	\end{align}
These dynamics, which arise from continuous polarimetry measurements, serve to generate the correlations that produce spin squeezing.  Note that if we take $l,\,l',\,p,\,p'=0$ and sum over all $k$ and $k'$, we recover the familiar case derived in \cite{JacSte06}. 

	We now turn our attention to diffuse scattering.  The evolution of the first-order terms in the covariance, \erf{Eq::GenCovariance}, is
	\begin{align} \label{Eq::FzMeanSpin}
		\frac{d}{dt} \expect{\hat{F}_z^{pl}(z_k)} \Big|_{\rm diff} = -\frac{2 \gamma_0}{9} \sum_{p',l'}c^{pl}_{p'l'}(z_k) \expect{\hat{F}_z^{p'l'}(z_{k'})},
	\end{align}
where we have used the fact that ${\mathcal{D}_i}[\hat{f}_z^{(i)}] = -2\hat{f}_z^{(i)}/9$.  The evolution of the second-order term in \erf{Eq::GenCovariance} is governed entirely by atomic pairwise correlations,
	\begin{align}\label{eq::CovCorrelations}
		\frac{d}{dt}\expect{& \hat{F}_z^{pl}(z_k) \hat{F}_z^{p'l'}(z_{k'})}\Big|_{\text{diff}}=  \sum_{i_k\neq j_{k'}} \beta_{pl} (\mbf{r}_{i_k}) \beta_{p'l'} (\mbf{r}_{j_{k'}})\frac{d}{dt}\expect{\hat{f}_z^{(i_{k})} \hat{f}_z^{(j_{k'})}}\big|_{\text{diff}},
	\end{align}
where $i_k$ and $j_{k'}$ label atoms in coarse-grained slices at $z_k$ and $z_{k'}$, respectively. Using Eq. (\ref{Eq::2ObservableEOM}), the pairwise correlations evolve according to
	\begin{equation}
		\frac{d}{dt}\expect{\hat{f}_z^{(i_{k})} \hat{f}_z^{(j_{k'})}}\Big|_{\text{diff}} = -\frac{2 \gamma_0}{9} \big[ \beta_{00}(\mbf{r}_{i_k}) +\beta_{00}(\mbf{r}_{j_{k'}}) \big] \expect{\hat{f}_z^{(i_{k})} \hat{f}_z^{(j_{k'})}}.
	\end{equation}
Substituting this into Eq. (\ref{eq::CovCorrelations}) yields,
	\begin{align}
		\frac{d}{dt}\expect{\hat{F}_z^{pl}(z_k)\hat{F}_z^{p'l'}(z_{k'})}\Big|_{\text{diff}}  = & - \frac{2 \gamma_0}{9}\sum_{i_k\neq j_{k'}} \beta_{pl} (\mbf{r}_{i_k}) \beta_{p'l'} (\mbf{r}_{j_{k'}})\big[ \beta_{00}(\mbf{r}_{i_k}) +\beta_{00}(\mbf{r}_{j_{k'}}) \big] \expect{\hat{f}_z^{(i_{k})} \hat{f}_z^{(j_{k'})}} \nn \\
		  = & -\frac{2 \gamma_0}{9} \bigg[ \sum_{i_k}\beta_{00}(\mbf{r}_{i_k})\beta_{pl}(\mbf{r}_{i_k})\expect{\hat{f}_z^{(i_k)}\hat{F}_z^{p'l'}(z_{k'})}+\sum_{i_{k'}}\beta_{00}(\mbf{r}_{i_{k'}})\beta_{pl}(\mbf{r}_{i_{k'}})\expect{\hat{F}_z^{pl}(z_k)\hat{f}_z^{(i_{k'})}} \bigg]  \nn \\
		&+\delta_{k,k'} \frac{\gamma_0}{9}\sum_{i_k}\beta_{00}(\mbf{r}_{i_k})\beta_{pl}(\mbf{r}_{i_k})\beta_{p'l'}(\mbf{r}_{i_k}) \label{Eq::TwoPointEvolution}.
	\end{align}
where the term in the last line comes from adding and subtracting the $i_k = j_k$ terms, which allowed us to perform the sum over one of the atom indices in the previous line.  The sum in the final term can be expressed as an integral over the density of the atomic cloud,
\begin{align}
N_{p'l'}^{pl}(z_k)= \delta z \int d^2\rp\eta(\rp,z_k)\beta_{00}(\rp,z_k)\beta_{pl}(\rp,z_k)\beta_{p'l'}(\rp,z_k).
\end{align}
Note that for the fundamental mode, $p,p',l,l'=0$, $N_{p'l'}^{pl}(z_k)$ is $N^{(3)}_{\text{eff}}$ at slice $z_k$. 

	Combining \erf{Eq::FzMeanSpin} and \erf{Eq::TwoPointEvolution} and projecting the covariances into the spin waves using \erf{Eq::ProjCoeff}, we arrive at a differential equation that couples spin wave covariances.  
Including the dynamics due to continuous measurement \erf{Eq::CovarianceBackaction}, the full equation of motion for the covariances is
	\begin{align}\label{eq::FullEOMCOV}
		 \frac{d}{dt}  \expect{\Delta\hat{F}_z^{pl}(z_k) & \Delta\hat{F}_z^{p'l'}(z_{k'})}= -\kappa\sum_{k'',k'''}\big\langle\Delta\hat{F}_z^{pl}(z_k)\Delta\hat{F}_z^{00}(z_{k''})\big\rangle\big\langle\Delta\hat{F}_z^{p'l'}(z_{k'})\Delta\hat{F}_z^{00}(z_{k'''})\big\rangle  \\
		&-\frac{2\gamma_s}{9}\sum_{p''l''}\Big[ c^{pl}_{p''l''}(z_k)\expect{\Delta\hat{F}_z^{p''l''}(z_k)\Delta\hat{F}_z^{p'l'}(z_{k'})} +c_{p''l''}^{p'l'}(z_{k'})\expect{\Delta\hat{F}_z^{pl}(z_k)\Delta\hat{F}_z^{p''l''}(z_{k'})}\Big] +\frac{\gamma_s}{9}N_{p'l'}^{pl}(z_k)\delta_{k,k'}. \nonumber
	\end{align}
As in the case of the mean spin waves, this set of equations is solved by truncating \erf{eq::FullEOMCOV} at some $p_{\text{max}}$ and  $l_{\text{max}}$. Following Eq. (\ref{eq::CovCorrelations}), using $\expect{\hat{f}_z^{(i_{k})} \hat{f}_z^{(j_{k'})}}_{i_k \neq j_{k'}} =0$ for the initial SCS, the initial spin wave covariances are
	\begin{align}
		&\expect{\Delta\hat{F}_z^{pl}(z_k)\Delta\hat{F}_z^{p'l'}(z_{k'})}(t_0)=  \delta_{k,k'}\frac{\delta z}{4}\int d^2\rp \, \eta(\rp,z_k) \beta_{pl}(\rp,z_k)\beta_{p'l'}(\rp,z_k).
	\end{align}
With these initial conditions and the equations of motion, we can solve for the evolution of all covariances in the presence of both QND measurement backaction and decoherence by optical pumping. Summing the covariances in the fundamental spin wave over all slices yields the variance in the fundamental spin wave:
\begin{align}
(\Delta F_z^{00}(t))^2=\sum_{k,k'}\expect{\Delta\hat{F}_z^{00}(z_k)\Delta\hat{F}_z^{00}(z_{k'})}(t).
\end{align}
From $(\Delta F_z^{00}(t))^2$ and $\expect{\hat{F}_x^{00}(t)}$, we calculate the dynamics of the squeezing parameter, \erf{Eq::SqueezingParam}. 

For spin-$f$ alkali atoms, the derivation of the mean spin equations, \erf{Eq::MeanSpins}, and fundamental spin wave variance equation, \erf{Eq::GenCovarianceEvolution}, follows a similar prescription as for spin-1/2, but with more general processes that include transfer of coherences, spontaneous emission, and hyperfine optical pumping.  This makes the dynamics substantially more complex and will be treated in detail in a future publication.

\end{widetext}
	
\end{appendix}

\end{document}